\documentclass[letter,11pt]{article}
 \pdfoutput=1
 \usepackage[font=small,labelfont=bf]{caption}
 \usepackage{jheppub} 
 \usepackage{colortbl}
 \usepackage{graphicx}
 \usepackage{tikz}
\usepackage[sort&compress]{natbib}
 \usepackage{amssymb}
 \usepackage{amsmath}
\usepackage{dcolumn}
\usepackage{bm}
\usepackage{color}
\usepackage{multirow}

\newcommand{\met}{{\:/\!\!\!\! E_T}} 
\newcommand{\mpt}{{\;/\!\!\!\! \vec{P}_T}} 

\DeclareMathOperator\sign{sign}

\newcommand{\dk}[1]{{#1}}
\newcommand{\amc}{{\sc MadGraph5\textunderscore}a{\sc MC@NLO}}
 \newcommand{\lsim}{{\;\raise0.3ex\hbox{$<$\kern-0.75em\raise-1.1ex\hbox{$\sim$}}\;}}
\newcommand{\gsim}{{\;\raise0.3ex\hbox{$>$\kern-0.75em\raise-1.1ex\hbox{$\sim$}}\;}}
\newcommand{\beq}{\begin{equation}}
\newcommand{\eeq}{\end{equation}}
\newcommand{\bea}{\begin{eqnarray}}
\newcommand{\eea}{\end{eqnarray}}

\def\baa{\begin{array}}
\def\eaa{\end{array}}

\mathchardef\minus="002D

\def\met{E_T\hspace{-0.45cm}/\hspace{0.25cm}}

\title{\boldmath Resolving Combinatorial Ambiguities in Dilepton $t\bar t$ Event Topologies
with Constrained $M_2$ Variables }
 
\author[a]{Dipsikha Debnath,}
\author[b]{Doojin Kim,}
\author[c]{Jeong Han Kim,}
\author[c,d]{Kyoungchul Kong,}
\author[a]{Konstantin T.~Matchev,} 

\affiliation[a]{Physics Department, University of Florida, Gainesville, FL 32611, USA}
\affiliation[b]{Theory Department, CERN, CH-1211 Geneva 23, Switzerland}
\affiliation[c]{Department of Physics and Astronomy, University of Kansas, Lawrence, KS 66045, USA}
\affiliation[d]{Pittsburgh Particle physics, Astrophysics, and Cosmology Center, Department of Physics and Astronomy, University of Pittsburgh, Pittsburgh, PA 15260, USA.}

\abstract{We advocate the use of on-shell constrained $M_2$ variables in order to mitigate the combinatorial problem 
in SUSY-like events with two invisible particles at the LHC. We show that in comparison to other approaches in the literature, 
the constrained $M_2$ variables provide superior ansatze for the unmeasured invisible momenta and therefore
can be usefully applied to discriminate combinatorial ambiguities. We illustrate our procedure with the example of
dilepton $t\bar{t}$ events. We critically review the existing methods based on the Cambridge $M_{T2}$ variable
and MAOS-reconstruction of invisible momenta, and show that their algorithm can be simplified without loss of sensitivity,
due to a perfect correlation between events with complex solutions for the invisible momenta and events exhibiting a kinematic endpoint violation.
Then we demonstrate that the efficiency for selecting the correct partition is further improved by utilizing the $M_2$ variables instead.
Finally, we also consider the general case when the underlying mass spectrum is unknown, and no kinematic endpoint information is available.
}
 
\preprint{CERN-TH-2017-130\\
\hspace*{12.5cm}PITT-PACC-1704} 
\date{June 14, 2017}

\begin{document} 
\maketitle
\flushbottom

\section{Introduction}
\label{sec:introduction}

Events with missing transverse energy\footnote{$/\!\!\!\! E_T$ is an unfortunate misnomer which stands for 
the magnitude of the missing transverse momentum $\mpt$.} ($\met$) are arguably the most exciting class of events at the 
Large Hadron Collider (LHC). They offer the tantalizing possibility of discovering the elusive dark matter --- 
if dark matter particles were produced in the LHC collisions, they would leave the detector without a trace, 
and the only sign of their presence would be the imbalance in the total transverse momentum of the event.
Unfortunately, events with $\met$ are also notoriously difficult to interpret and analyze: 
\begin{itemize}
\item {\em Instrumental effects.} Since the missing transverse momentum $\mpt$ is measured only 
as the recoil against {\em all other} visible objects in the event, it can be easily faked by mismeasurement and the 
finite detector resolution \cite{Chatrchyan:2011tn}. This problem becomes more severe if the signature involves QCD jets, 
whose energies and momenta are poorly measured in comparison to leptons and photons. 
\item {\em Unknown nature of the invisible particles.} A priori, we do not know the nature of the invisible particles --- 
they could be new particles, or simply the Standard Model (SM) neutrinos \cite{Chang:2009dh}. 
\item {\em Incomplete kinematic information.} We do not know how many invisible particles were present in the event
to begin with \dk{\cite{Agashe:2010gt,Agashe:2010tu,Giudice:2011ib,Cho:2012er,Agashe:2012fs}}. We also do not know their individual momenta, and
only the net sum $\mpt$ of their transverse components is available.
\end{itemize}

The first step in the analysis of any sample of $\met$ events is to hypothesize a certain event topology,
and design suitable variables adapted to this interpretation \cite{Barr:2011xt}. It is already at this stage that 
one is facing a combinatorial problem, namely, how to associate the various reconstructed objects 
in the event to the elementary particles in the final state of the event topology. Only in very special cases 
does the problem not arise --- if the event topology is very simple and/or all final state particles are distinct. 
In general, a typical $\met$ event at the LHC does suffer from a combinatorics problem, for the following two reasons:
\begin{itemize}
\item At hadron colliders like the LHC, strong production of colored particles is the dominant production mechanism.
When those colored particles decay to the invisible dark matter candidates, the color is shed in the form of QCD jets, 
which can be confused with jets from initial state radiation (ISR) \cite{Plehn:2005cq,Alwall:2008qv,Papaefstathiou:2009hp,Jackson:2016mfb}. In fact, the ISR combinatorics problem is very general and
affects any multijet events at hadron colliders, regardless of the presence of $\met$ in the event.
\item The lifetime of the dark matter particles is typically protected by some new symmetry. This is often chosen to be a discrete $Z_2$ parity,
under which the SM particles are even, while the new physics particles are odd. In that case, the new particles are necessarily pair produced,
so that each event contains two independent decay chains. This creates a partitioning ambiguity, since the experimenter 
has to decide whether to assign each reconstructed object to the first or the second decay chain \cite{Barr:2010zj}. 
Wrong assignments would tend to wash out the desired kinematic features and degrade the measurements.
\end{itemize}

In principle, the combinatorial problem can be addressed in two different ways:
\begin{itemize}
\item {\em Sidestep the combinatorial problem.} The idea here is to design the analysis in such a way that the combinatorial problem does not become an issue. Two possibilities are:
\begin{itemize}
\item {\em Use global inclusive variables which do not suffer from a combinatorics problem.} 
These variables treat the event as a whole and thus do not depend on the exact event topology, and the combinatorics problem does not arise in the first place. Some well known examples are $M_{eff}$ \cite{Hinchliffe:1996iu,Tovey:2000wk}, $\hat s_{min}$ \cite{Konar:2008ei}, $\met$ \cite{Hubisz:2008gg}, etc. 
The disadvantage is that such variables are suboptimal when compared to more exclusive variables which take advantage 
of the individual characteristics of the event topology.
\item {\em Use variables which optimize over all possible combinatorial assignments.} In this case,
instead of trying to figure out the correct assignment in a given event, one considers all possibilities, 
then chooses the one\footnote{The chosen option does not necessarily have to be the correct one.} 
which preserves the relevant useful property of the kinematic variable used in the analysis.
As an example, consider an attempt to measure the {\em upper} kinematic endpoint of some relevant distribution, 
such as a two-body invariant mass or the Cambridge $M_{T2}$ variable \cite{Lester:1999tx}. One could simply compute the 
value of the variable under all possible assignments, then choose the {\em smallest} among them to be used in the 
analysis \dk{\cite{Hinchliffe:1996iu,Allanach:2000kt,Lester:2007fq,Alwall:2009zu,Bai:2010hd,Wiesler:2012rkl,Dev:2015kca,Kim:2015bnd,Klimek:2016axq,Debnath:2016gwz}}.%
\footnote{A similar idea can be applied 
to measure a {\em lower} kinematic endpoint --- in that case one would choose the {\em largest} value among all possibilities.}
While this procedure is guaranteed to preserve the kinematic endpoint, it also adversely distorts the shape of the kinematic distribution
in the vicinity of the endpoint, making it more difficult to observe in the presence of SM background. 
\end{itemize}
\item {\em Resolve the combinatorial problem by choosing the ``best" assignment event by event.} In this case one tries 
to design an algorithm which will single out one (or maybe several) among the many possible assignments as the most likely ``correct" assignment, 
then use the value of the kinematic variable obtained with this specific choice. Ideally, the algorithm should return a unique selection, 
which would be correct 100\% of the time. Unfortunately, this is rarely achievable in practice, and an important measure quantifying the
success of the algorithm is the purity of the resulting sample, i.e., the fraction of events in which the combinatorics was successfully resolved. 
In principle, there can be different approaches to designing such an algorithm, from the use of a single exclusive variable to 
a multivariate technique like a neural network analysis \cite{Shim:2014aua}. For example, depending on the process at hand, 
one can attempt to tag ISR jets by a suitable combination of cuts on the jet rapidity and transverse momentum \cite{Krohn:2011zp} 
or on the invariant mass and $M_{T2}$ \cite{Kim:2015uea}. The partitioning problem into two decay chains
is usually addressed by the so-called ``hemisphere" algorithm, developed originally within CMS \cite{Ball:2007zza} and later 
adopted in many phenomenological studies \cite{Matsumoto:2006ws,Cho:2007dh,Nojiri:2008hy}. 
There have been attempts to further improve on the hemisphere algorithm by suitable cuts on the invariant mass and either the jet $p_T$
\cite{Rajaraman:2010hy} or $M_{T2}$ \cite{Baringer:2011nh}, by excluding certain reconstructed objects from the clustering algorithm \cite{Alwall:2009zu,Nojiri:2008vq},
or by recursive jigsaw reconstruction \cite{Jackson:2017gcy}. In general, 
methods which invoke fewer assumptions, are robust and model independent, but lead to rather vague conclusions,
while methods with more assumptions give better results, but are not generally applicable.
\end{itemize}

In the case of $\met$ events, the combinatorics problem is exacerbated by the fact that the momenta of the invisible particles are unknown. 
If the decay chains are sufficiently long, so that there are enough kinematic constraints, one can attempt to compute the 
individual invisible particle momenta on an event per event basis \cite{Nojiri:2003tu,Kawagoe:2004rz,Cheng:2007xv,Cheng:2008mg,Cheng:2009fw}.
Unfortunately, this procedure itself suffers from a combinatorics problem, which only becomes worse as the decay chains get 
longer (as required for the method to work). For shorter decay chains, like the ones considered in this paper, the method does not apply.
 
Since the invisible momenta cannot be reconstructed exactly, the next best thing to do is to use some sort of an approximation for them \cite{Kim:2017awi}.
Again, different approaches are possible. For example, one could use a matrix element method (MEM) to select the most likely values
of the invisible momenta. However, the MEM itself suffers from combinatorics, and is rather model dependent since it requires us to fully specify the underlying physics. 
A better approach would be to rely only on kinematics and obtain the invisible momenta by optimizing a suitable kinematic function.
But what constitutes a good target function for such optimization? Initially, the focus was placed on {\em transverse} mass variables
like $M_{T2}$ \cite{Lester:1999tx,Barr:2003rg} and its variants \cite{Burns:2008va,Barr:2009jv,Konar:2009qr,Konar:2009wn}.
While transverse quantities are Lorentz invariant under longitudinal boosts, they only provide an ansatz for the {\em transverse} 
components of the individual invisible momenta, and one still needs to provide a supplementary procedure for calculating the 
{\em longitudinal} components of the invisible momenta. One such complementary technique is the 
MAOS\footnote{MAOS stands for $M_{T2}$-Assisted On-Shell reconstruction.} 
reconstruction \cite{Cho:2008tj}, where one imposes an additional on-shell kinematic constraint which can be solved for the
longitudinal momentum component of each invisible particle. It has been shown that the MAOS approach provides a reasonably good 
approximation to the true values of the invisible momenta, and can be usefully applied for mass and spin measurements 
\cite{Cho:2008tj,Park:2011uz,Guadagnoli:2013xia}. The MAOS technique was then used to 
design a novel algorithm \cite{Choi:2011ys} for resolving the combinatorial ambiguity in dilepton $t\bar{t}$ events, further expanding on the ideas from
Refs.~\cite{Rajaraman:2010hy,Baringer:2011nh}. The algorithm aims to resolve the two-fold\footnote{In the case of dilepton $t\bar{t}$ events, 
the two jets originating from the top decays can be distinguished from ISR jets by b-tagging.} ambiguity in selecting the correct 
lepton-jet pairing and involves the following three steps:
\begin{itemize}
\item {\bf Step I.} Following the proposal of Ref.~\cite{Baringer:2011nh}, some number of wrong lepton-jet combinations can be eliminated if they violate the expected endpoints in the distributions of the invariant mass $m_{b\ell}$ and $M_{T2}$.
\item {\bf Step II.} Utilizing the ansatz found in Step I for the transverse components of the invisible momenta, attempt a MAOS reconstruction of the longitudinal components in two cases: 
\begin{enumerate}
\item using the known value of the top quark mass $m_t$;
\item using the known value of the $W$-boson mass $m_W$.
\end{enumerate}
Eliminate additional wrong combinations if the solutions for the longitudinal momenta in either case turn out to be complex.
\item {\bf Step III.} In this final step, one uses the reconstructed masses for the $W$-boson (in case II.1) and the top quark (in case II.2) in conjunction with $M_{T2}$ to 
decide which of the two lepton-jet pairings is the likelier one.
\end{itemize}
While this algorithm was originally designed to handle the two-fold ambiguity in $t\bar{t}$ events, where the mass spectrum is known,
with suitable modifications it can also be applied to new physics searches, as advertised in Ref. \cite{Choi:2011ys}. For instance, in the MAOS reconstruction Step II, instead of using the fixed values of the known masses $m_t$ and $m_W$, one could use the measured endpoints in the respective $M_{T2}$ subsystems \cite{Burns:2008va}.

Recently it has been pointed out that the $M_{T2}$ approach has a $(3+1)$-dimensional analogue in terms of a general class
of on-shell constrained invariant mass variables $M_2$ \cite{Barr:2011xt,Mahbubani:2012kx,Cho:2014naa}.
Compared to $M_{T2}$, the $M_2$ variables have several advantages:
\begin{itemize}
\item Being defined in (3+1) dimensions, they allow us to easily and directly enforce all relevant on-shell constraints in a given event topology
\cite{Ross:2007rm,Barr:2011xt}.
\item Unlike the case of $M_{T2}$, the optimization procedure required to compute the value of $M_2$ automatically provides an ansatz for both the 
transverse {\em and} the longitudinal components of the invisible momenta. In this sense, once one commits to using $M_2$ variables instead of $M_{T2}$, the MAOS reconstruction step for finding the longitudinal momentum components  is unnecessary.
\item The maximally constrained $M_2$ variable can be expected to provide the best possible ansatz for the individual invisible momenta, since it takes into account all relevant kinematic constraints in a given event topology \cite{Kim:2017awi}. 
\end{itemize}

The main goal of this paper is to utilize these advantages of the $M_2$ variables and design an improved algorithm 
for resolving the combinatorial ambiguity in SUSY-like events with two invisible particles at the LHC.
As our benchmark, we shall use the current state of the art algorithm which was proposed 
and tested for dilepton $t\bar{t}$ events in Ref.~\cite{Choi:2011ys}. Correspondingly, in section~\ref{sec:review} 
we shall first give a brief review of the relevant background information regarding the kinematics of the 
dilepton $t\bar{t}$ event topology. Then in section~\ref{sec:momentum} we shall carefully define the different 
options for kinematic reconstruction of the invisible momenta \cite{Kim:2017awi}. We shall see that in principle there can be
different ways of applying the ideas of MAOS reconstruction, $M_2$-assisted reconstruction, or some combination of both.
In section~\ref{sec:momentum} we shall also compare the accuracy of several representative methods for 
invisible momentum reconstruction. 

The next three sections will be devoted to the issue of resolving the combinatorial ambiguity.
First in section~\ref{sec:oldmethod} we critically review each of the three steps of the current state of the art method
based on the Cambridge $M_{T2}$ variable and MAOS-reconstruction of invisible momenta \cite{Baringer:2011nh,Choi:2011ys}. 
Our goal will be to improve the algorithm in two aspects:
\begin{itemize}
\item {\em Better performance.} By considering various modifications, e.g., utilizing the alternative set of $M_2$ variables, 
or alternative implementations of the MAOS method itself, we shall attempt to improve the efficiency\footnote{Throughout the paper, 
we shall use the terms ``efficiency" and ``purity" interchangeably to denote the same quantity --- the fraction of events in which
the algorithm is successful in identifying the correct partition.} 
of the algorithm in selecting the correct partition in dilepton $t\bar{t}$ events.
\item {\em Simplicity.} At the same time, we shall keep an eye on the relative performance of each algorithm component, 
and if we find components which underperform, we shall eliminate them from consideration, thus simplifying the algorithm.
For example, in section~\ref{sec:complex} we shall demonstrate that Step II can be safely disregarded since it is fully correlated
with Step I and does not give anything new.
\end{itemize}
Then in section~\ref{sec:newmethod} we consider several new ideas which go beyond the three steps of the current algorithm.
In section~\ref{sec:quadrants} we consider expanding the set of variables used in Step I from two to three, 
since the dilepton $t\bar{t}$ event topology allows not just two, but three independent kinematic endpoints 
\cite{Burns:2008va,Chatrchyan:2013boa}. Then in section~\ref{sec:M2Ct} we discuss a special class of 
maximally constrained $M_2$ variables where the knowledge of the top and $W$-boson masses can be
taken into account already during the optimization stage\footnote{Note that this is impossible in the case of purely transverse variables like $M_{T2}$.},
thus further improving the ansatz for the transverse invisible momenta. In section~\ref{sec:smintheta} 
we study the potential benefit from using a global inclusive variable such as $\sqrt{\hat{s}}$
or an angular variable such as the scattering angle of the parents in the center-of-mass frame.
Finally, in section \ref{sec:generalcase} we treat the general case when the underlying mass spectrum is unknown, 
and no kinematic endpoint information is available. We consider a simplified version of the algorithm 
which is suitably adapted to  this scenario, and investigate its performance in the general new physics mass parameter space.
We discuss future extensions of this work and summarize in section~\ref{sec:conclusions}.

\section{Dilepton $t\bar{t}$ kinematics and mass-constraining variables}
\label{sec:review}

\begin{figure}[t]
\centering
\includegraphics[scale=0.9]{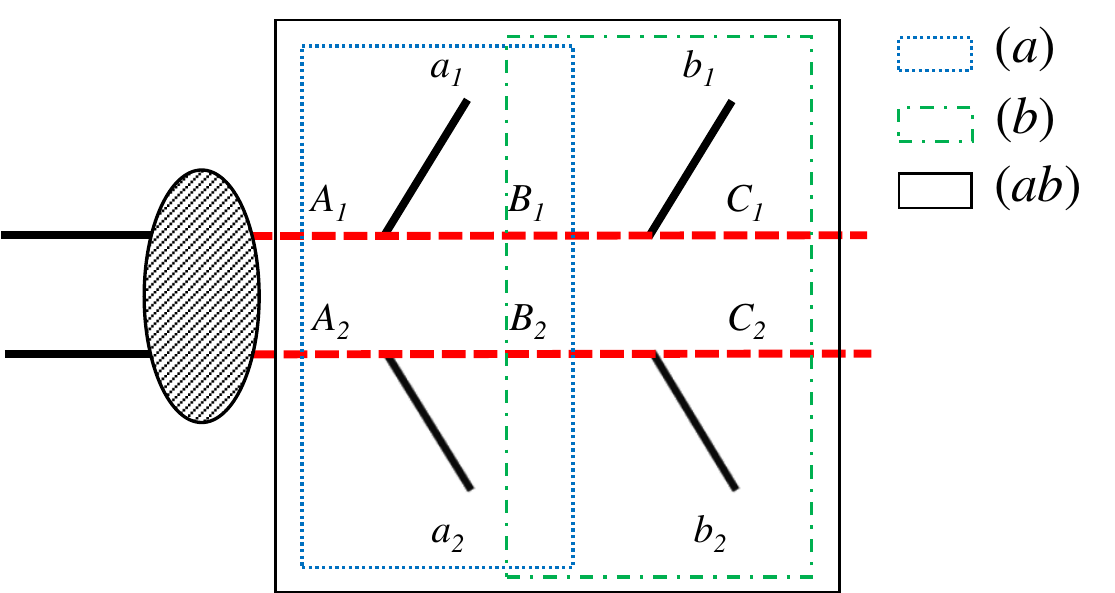}
\caption{The event topology considered in this paper, 
together with the three possible subsystems. 
The blue dotted, the green dot-dashed, and
  the black solid boxes indicate the subsystems $(a)$, $(b)$, and
  $(ab)$, respectively. The figure is taken from Ref. \cite{Cho:2014naa}.
  \label{fig:decaysubsystem}}
\end{figure}

In this section we shall introduce the basic notation and review the relevant class of mass-constraining variables which 
will be used later to obtain suitable ansatze for the invisible momenta. For the most part, we shall stick to the notation 
and terminology of Refs.~\cite{Kim:2017awi,Cho:2014naa}.
Following \cite{Baringer:2011nh,Choi:2011ys}, we focus primarily on the ``dilepton $t\bar{t}$" event topology 
depicted in Fig.~\ref{fig:decaysubsystem}. This choice is motivated by several factors:
\begin{itemize}
\item As far as the combinatorial problem is concerned, this is the simplest example which is not trivial --- if we were 
to consider a \dk{single-step} two-body decay on each side, there would be no combinatorial issue to begin with, 
and if we were to consider longer decay chains, the problem would become more difficult.
\item This event topology is realized in the SM production of $t\bar{t}$ events, providing a useful toy playground for testing 
new ideas for studying new physics \cite{Chatrchyan:2013boa,ATLAS:2012poa,Phan:2013trw,Sirunyan:2017idq}.
\item Several new physics models can lead to this event topology, including stop-pair production in supersymmetry
\cite{Cho:2014yma} and pair-production of leptoquarkinos \cite{Reuter:2010nx}.
\end{itemize}

Thus the general event topology considered in this paper is the pair-production of two identical parent particles $A_i$, followed by 
a 2-step 2-body decay for each one (see Fig.~\ref{fig:decaysubsystem}): 
\begin{equation}
p p \to A_1 A_2\, , \quad A_i \to a_i B_i \,, \quad B_i \to b_i C_i \, , \quad (i=1, 2)\, .
\end{equation}
In principle, $A_i$, $B_i$ and $C_i$ should be thought of as some unknown BSM particles, 
while $a_i$ and $b_i$ are SM particles whose four-momenta are measured. The particles $C_i$ are invisible in the detector, 
and their momenta $q_i$  are constrained only by the $\mpt$ measurement and their (a priori unknown) masses, $\tilde m_{C_i}$, with $q_i^2 = \tilde m_{C_i}^2$. 

The 2-step 2-body event topology of Fig.~\ref{fig:decaysubsystem} allows for three different subsystems, as indicated
by the colored rectangular boxes \cite{Burns:2008va}. Each subsystem is labelled by the visible particles in it, 
and defined by a choice of parent and daughter particles, leaving the third type of particles as ``relatives":
in subsystem $(ab)$ the parents are $A_i$, the daughters are $C_i$ and the relatives are $B_i$;
in subsystem $(a)$ the parents are $A_i$, the daughters are $B_i$ and the relatives are $C_i$, while
in subsystem $(b)$ the parents are $B_i$, the daughters are $C_i$ and the relatives are $A_i$.
The mass-constraining kinematic variables defined below can be applied to any of the three subsystems, 
thus each variable has three different versions, depending on the chosen subsystem.
For simplicity, in what follows we shall assume that the event topology of Fig.~\ref{fig:decaysubsystem} is symmetric,
i.e., $A_1=A_2$, $B_1=B_2$,  and $C_1=C_2$ (see \cite{Barr:2009jv,Konar:2009qr} for generalizing to the asymmetric case).

We first consider the traditional transverse variable $M_{T2}$ \cite{Lester:1999tx}. Let
the two transverse masses of the parent particles be $M_{TP_i}(\vec{q}_{iT},\tilde m)$,
where $\vec{q}_{iT}$ is the transverse momentum of $C_i$ and $\tilde m$ is a test mass for the daughter particles, 
which is $\tilde m_{C_i}$ for the case of subsystems $(ab)$ and $(b)$ and
$\tilde m_{B_i}$ for the case of subsystem $(a)$. The kinematic variable
$M_{T2}$ is now defined as the absolute minimum of the larger of these two transverse masses, 
with respect to all possible partitions of the individual invisible transverse momenta
$\vec{q}_{iT}$,
\bea
M_{T2} (\tilde m) &\equiv& \min_{\vec{q}_{1T},\vec{q}_{2T}}\left\{\max\left[M_{TP_1}(\vec{q}_{1T},\tilde m),\;M_{TP_2} (\vec{q}_{2T},\tilde m)\right] \right\} \, .
\label{eq:mt2def}\\
\vec{q}_{1T}+\vec{q}_{2T} &=& \mpt    \nonumber
\eea 

Alternatively, one could apply the same procedure to the {\em actual} parent masses, $M_{P_i}$, 
and define the (3+1)-dimensional analogue of Eq.~(\ref{eq:mt2def}) as
\bea
M_{2} (\tilde m) &\equiv& \min_{\vec{q}_{1},\vec{q}_{2}}\left\{\max\left[M_{P_1}(\vec{q}_{1},\tilde m),\;M_{P_2} (\vec{q}_{2},\tilde m)\right] \right\} \, ,
\label{eq:m2def}\\
\vec{q}_{1T}+\vec{q}_{2T} &=& \mpt   \nonumber
\eea 
where now the minimization is performed over the 3-component momentum vectors $\vec{q}_{1}$ and $\vec{q}_{2}$ \cite{Barr:2011xt}.
As shown in Refs. \cite{Ross:2007rm,Barr:2011xt,Cho:2014naa}, at this point the two definitions (\ref{eq:mt2def}) and (\ref{eq:m2def})
are equivalent, in the sense that the resulting two variables, $M_{T2}$ and $M_2$, will have the same numerical value. 

The case when $M_2$ begins to differ from $M_{T2}$ is when we start to apply additional kinematic constraints 
beyond the $\mpt$ condition $\vec{q}_{1T}+\vec{q}_{2T} = \mpt $. Then the $M_2$ variable can be further refined 
and one can obtain non-trivial variations \cite{Cho:2014naa}:
\bea
M_{2XX} &\equiv& \min_{\vec{q}_{1},\vec{q}_{2}}\left\{\max\left[M_{P_1}(\vec{q}_{1},\tilde m),\;M_{P_2} (\vec{q}_{2},\tilde m)\right] \right\},  
\label{eq:m2XXdef}
\\
\vec{q}_{1T}+\vec{q}_{2T} &=& \mpt   \nonumber
\eea 
\bea
M_{2CX} &\equiv& \min_{\vec{q}_{1},\vec{q}_{2}}\left\{\max\left[M_{P_1}(\vec{q}_{1},\tilde m),\;M_{P_2} (\vec{q}_{2},\tilde m)\right] \right\},  
\label{eq:m2CXdef}\\
\vec{q}_{1T}+\vec{q}_{2T} &=& \mpt   \nonumber \\
M_{P_1}&=& M_{P_2} \nonumber 
\eea 
\bea
M_{2XC} &\equiv& \min_{\vec{q}_{1},\vec{q}_{2}}\left\{\max\left[M_{P_1}(\vec{q}_{1},\tilde m),\;M_{P_2} (\vec{q}_{2},\tilde m)\right] \right\},  
\label{eq:m2XCdef}\\
\vec{q}_{1T}+\vec{q}_{2T} &=& \mpt   \nonumber \\
M_{R_1}^2&=& M_{R_2}^2 \nonumber 
\eea 
\bea
M_{2CC} &\equiv& \min_{\vec{q}_{1},\vec{q}_{2}}\left\{\max\left[M_{P_1}(\vec{q}_{1},\tilde m),\;M_{P_2} (\vec{q}_{2},\tilde m)\right] \right\}.
\label{eq:m2CCdef}\\
\vec{q}_{1T}+\vec{q}_{2T} &=& \mpt   \nonumber \\
M_{P_1}&=& M_{P_2} \nonumber  \\
M_{R_1}^2&=& M_{R_2}^2 \nonumber 
\eea 
Here $M_{P_i}$ ($M_{R_i}$) is the \dk{reconstructed} mass of the parent (relative) particle in the $i$-th decay chain \dk{during the associated minimization procedure} and a subscript ``$C$" 
indicates that an equal mass constraint is applied for the two parents (when ``$C$" is in the first position) or for the relatives
(when ``$C$" is in the second position). A subscript ``$X$" simply means that no such constraint is applied.
In any given subsystem, the variables (\ref{eq:mt2def}-\ref{eq:m2CCdef}) are related \dk{event-by-event} in the following way \cite{Cho:2014naa}
\bea
M_{T2} = M_{2XX}=M_{2CX} \leq M_{2XC} \leq M_{2CC}. 
\label{eq:hierarchy}
\eea

Until now, we have treated the event topology of Fig.~\ref{fig:decaysubsystem} in very general terms. In particular, we have not made any 
assumptions about the nature of the visible particles $a_i$ and $b_i$. If they are all indistinguishable, e.g., jets from 
gluino pair-production events, $p p \to \tilde g \tilde g \to j j \tilde q \tilde q \to j j j j+ \met$, the resulting
combinatorial issues are rather severe, and one should perhaps first focus on testing the hypothesis for the event topology \cite{Bai:2010hd}.
Here we would like to start with a more tractable problem, where some of the final state particles are distinguishable. 
Keeping in mind the dilepton $t\bar{t}$ example and the analogous BSM signatures, we shall take particles $a_i$ to be $b$-jets, and particles $b_i$ to be leptons,
i.e., $a_1=b$, $a_2 = \bar b$, $b_1=\ell^+$ and $b_2=\ell^-$, where $\ell=\{e, \mu\}$ and $b$ is the bottom quark.
Since the charge of the $b$-jet is difficult to determine, there is a two-fold partitioning ambiguity: the correct partition is
\beq
P_C: \{b, \ell^+\}\oplus \{\bar{b}, \ell^-\},
\label{eq:PC}
\eeq
while the wrong partition is 
\beq
P_W: \{\bar b, \ell^+\}\oplus \{b, \ell^-\}.
\label{eq:PW}
\eeq

In the rest of this paper, we shall be concerned with designing algorithms which would preferentially select 
the correct pairing (\ref{eq:PC}) over the wrong one (\ref{eq:PW}). For this purpose, we shall mostly utilize the 
Cambridge $M_{T2}$ variable (\ref{eq:mt2def}) and the constrained $M_{2CC}$ variable (\ref{eq:m2CCdef}).
Each of these two variables can be applied to one of the three possible $t\bar{t}$ subsystems, $(b\ell)$, $(\ell)$ and $(b)$.
Notice, however, that in the ``smaller" subsystems $(b)$ and $(\ell)$, the two partitions (\ref{eq:PC}) and (\ref{eq:PW})
give identical values of $M_{T2}$, thus the corresponding subsystem $M_{T2}$ variables $M_{T2}^{(b)}$ and $M_{T2}^{(\ell)}$
will not be useful to us for the purposes of resolving the combinatorial issue. In contrast, all three subsystem $M_{2CC}$ variables,
$M_{2CC}^{(b\ell)}$, $M_{2CC}^{(b)}$,  and $M_{2CC}^{(\ell)}$, depend on the partitioning --- 
either directly, or through the relative constraint $M_{R_1}=M_{R_2}$.

Recently, Ref.~\cite{Kim:2017awi} introduced another interesting variation of the $M_{2CC}$ variable, which 
takes advantage of the potentially known mass for a relative particle. For example, if the mass of the $B_i$ particles is known,
we can enforce it as an additional constraint during the minimization in the $(ab)$ subsystem. 
Specifying to the $t\bar{t}$ case, where $A_i$ are the top quarks $t_i$ and $B_i$ are the $W$-bosons $W_i$, we can write 
\bea
M_{2CW}^{(b\ell)} &\equiv& \min_{\vec{q}_{1},\vec{q}_{2}}\left\{\max\left[M_{t_1}(\vec{q}_{1},\tilde m),\;M_{t_2} (\vec{q}_{2},\tilde m)\right] \right\} \, ,
\label{eq:m2CWdef}\\
\vec{q}_{1T}+\vec{q}_{2T} &=& \mpt   \nonumber \\
M_{t_1}&=& M_{t_2} \nonumber  \\
M_{W_1}&=& M_{W_2} = m_W \nonumber 
\eea 
where $m_W$ is the experimentally measured $W$-boson mass. Similarly, if we take the mass $m_t$ of the top quarks to be known, 
there is a new variable in the $(\ell)$ subsystem:
\bea
M_{2Ct}^{(\ell)} &\equiv& \min_{\vec{q}_{1},\vec{q}_{2}}\left\{\max\left[M_{W_1}(\vec{q}_{1},\tilde m),\;M_{W_2} (\vec{q}_{2},\tilde m)\right] \right\}.
\label{eq:m2Ctdef}\\
\vec{q}_{1T}+\vec{q}_{2T} &=& \mpt   \nonumber \\
M_{W_1}&=& M_{W_2} \nonumber  \\
M_{t_1}&=& M_{t_2} = m_t \nonumber 
\eea

\section{Reconstruction schemes for invisible momenta}
\label{sec:momentum}

All of the kinematic variables introduced in the previous section are defined in terms of an optimization procedure
over all possible values of the individual invisible momenta. The procedure then singles out one particular choice of 
the invisible momenta, which is used to calculate the corresponding variable. We can also use this choice as a
useful ansatz for the invisible momenta, and then apply standard analysis techniques as if the momenta of the invisible
particles were known \cite{Cho:2008tj,Kim:2017awi}.

\begin{figure}[t]
\centering
\includegraphics[width=5.1cm]{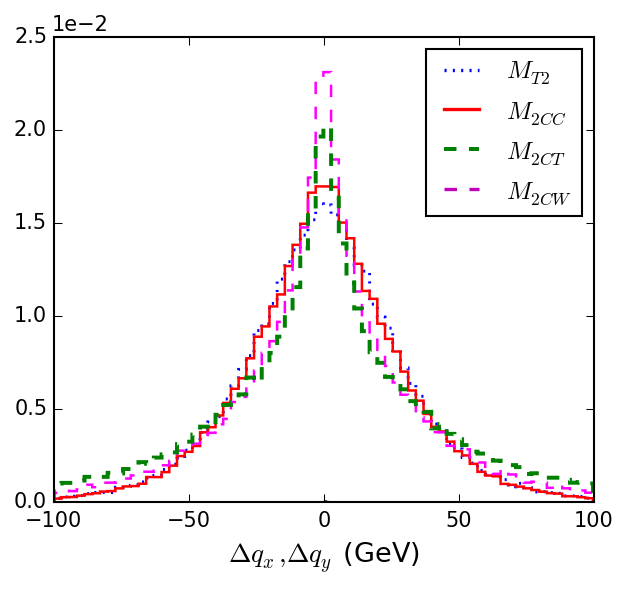} 
\includegraphics[width=5.1cm]{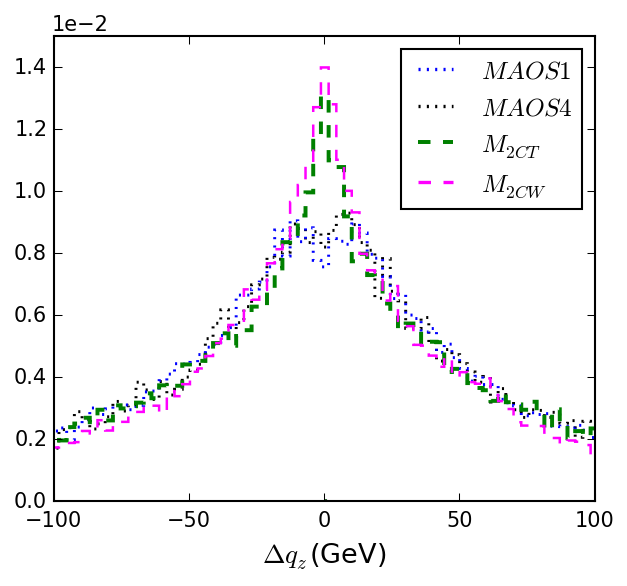}
\includegraphics[width=5.1cm]{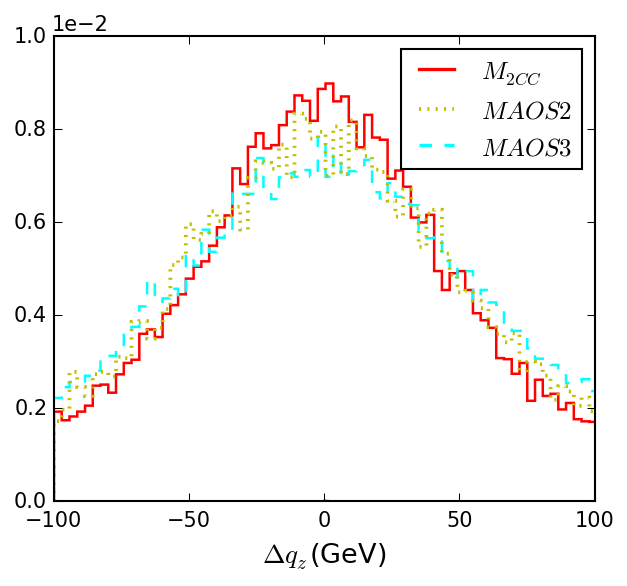}
\caption{\label{fig:momentum1d} Difference between the reconstructed and the true values for the invisible momentum components in
dilepton $t\bar{t}$ events. The left panel shows results from several methods for fixing the transverse components $q_{ix}$ and $q_{iy}$  
by minimizing an invariant mass variable: $M_{T2}^{(b\ell)}$ (blue dotted line), $M_{2CC}^{(b\ell)}$ (red solid line),
$M_{2Ct}^{(\ell)}$ (green dashed line) and $M_{2CW}^{(b\ell)}$ (magenta dashed line).
The middle (right) panel shows corresponding results for the longitudinal components, 
obtained with methods which use (do not use) additional mass information:
MAOS1($b\ell$;$m_t$) (blue dotted line),
MAOS4($b\ell$;$m_W$) (black dotted line),
$M_{2Ct}^{(\ell)}$ (green dashed line),
$M_{2CW}^{(b\ell)}$ (magenta dashed line),
M${}_2$A($b\ell$) (red solid line),
MAOS2($b\ell$;$b\ell$) (yellow dotted line) and 
MAOS3($b\ell$;$b\ell$) (cyan dashed line).
}
\end{figure}

\begin{figure}[tbp]
\centering
\includegraphics[width=6.5cm]{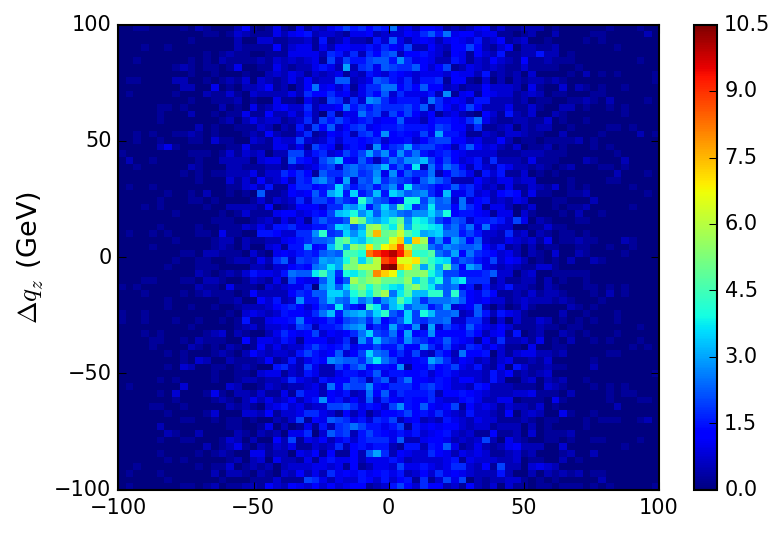} 
\includegraphics[width=6.5cm]{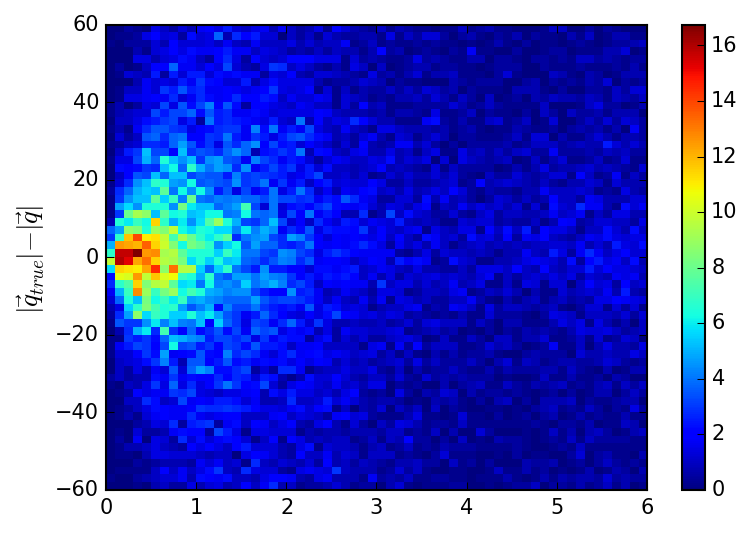} \\ \vspace*{-0.1cm}
\includegraphics[width=6.5cm]{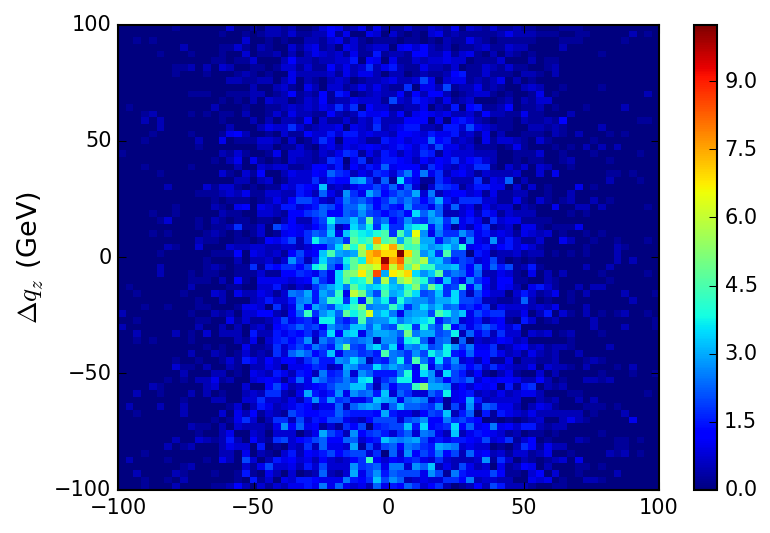}  
\includegraphics[width=6.5cm]{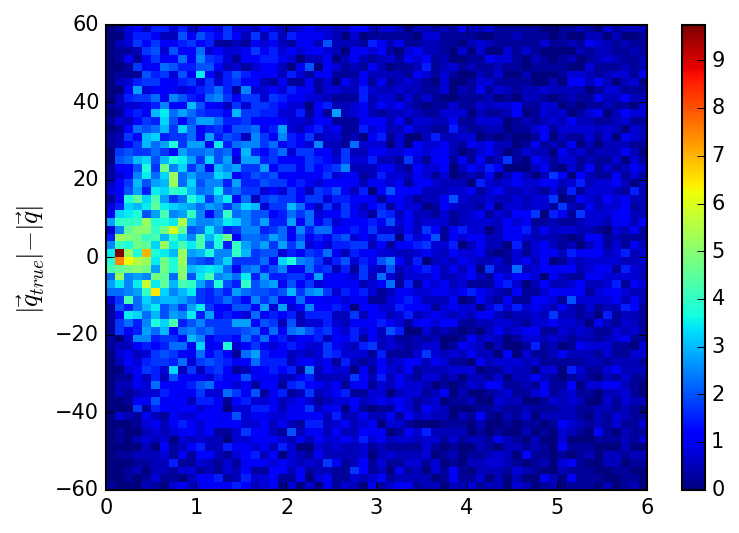} \\ \vspace*{-0.1cm}
\includegraphics[width=6.5cm]{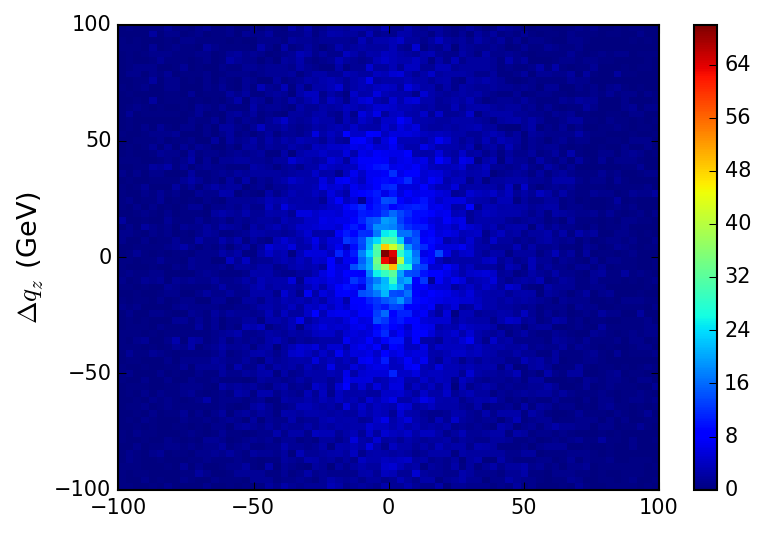} 
\includegraphics[width=6.5cm]{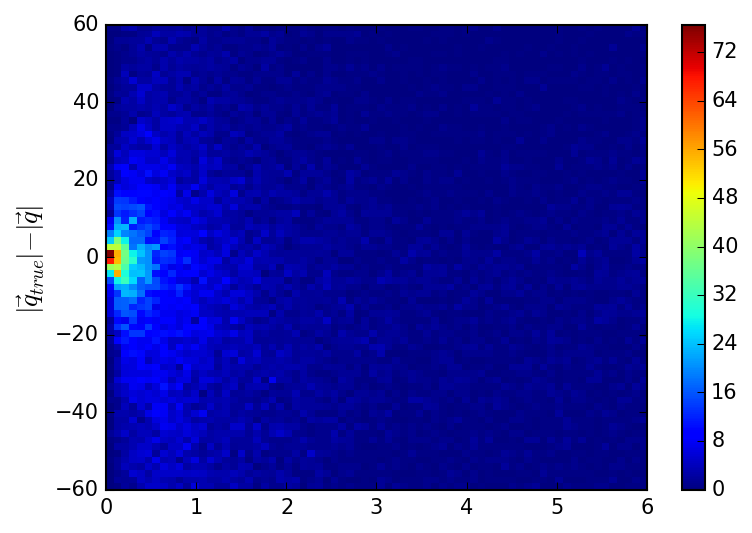}  \\ \vspace*{-0.1cm}
\includegraphics[width=6.5cm]{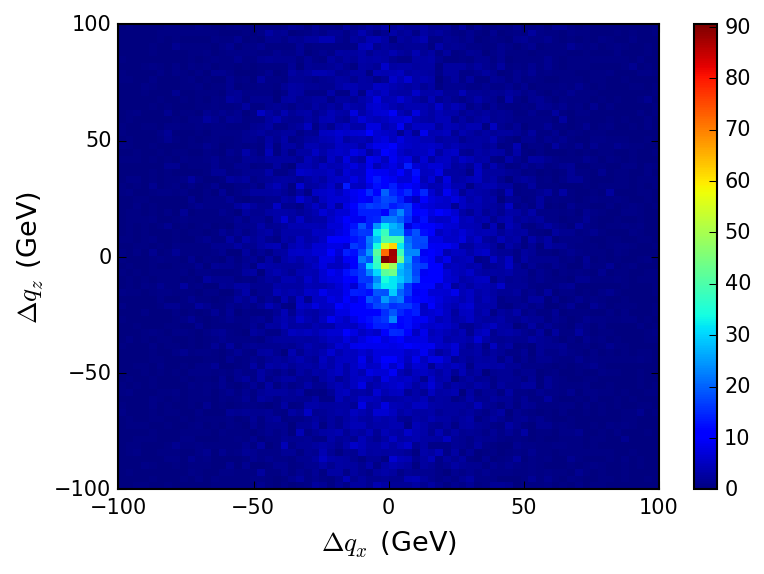} 
\includegraphics[width=6.5cm]{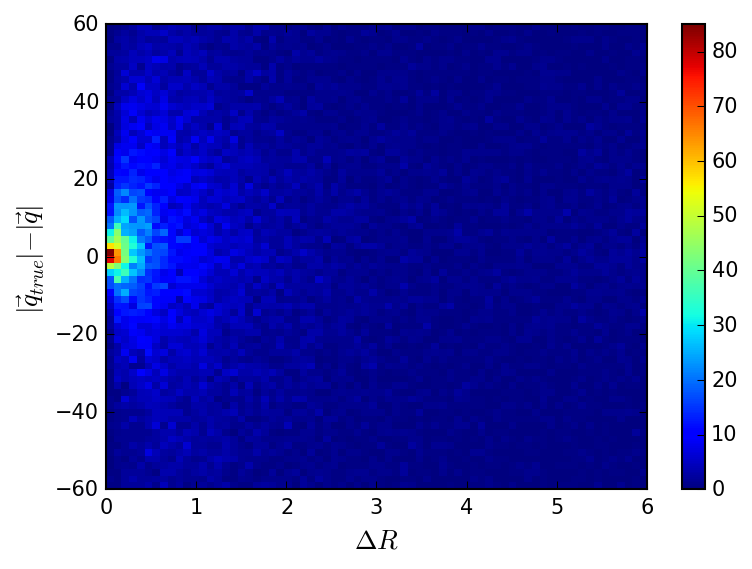} \\ \vspace*{-0.4cm}
\caption{\label{fig:momentum2dmass} Correlations between $\Delta q_z$ and $\Delta q_x$ (left) and 
$ |\vec q_{true}|-|\vec q \, |$ and $\Delta R(\vec q_{true},\vec q \,)$ (right) for four different 
schemes (from top to bottom): MAOS1($b\ell$;$m_t$), MAOS4($b\ell$;$m_W$), $M_{2Ct}^{(\ell)}$ and $M_{2CW}^{(b\ell)}$.}
\end{figure}

\begin{figure}[t]
\centering
\includegraphics[width=6.5cm]{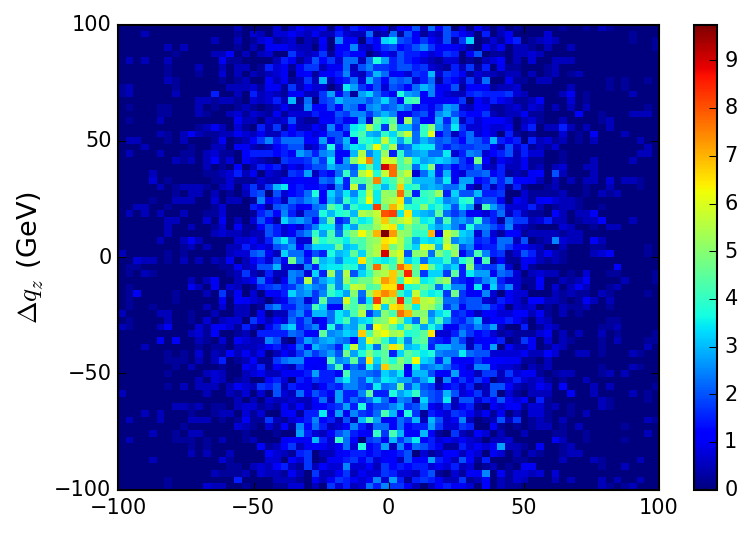} 
~\includegraphics[width=6.5cm]{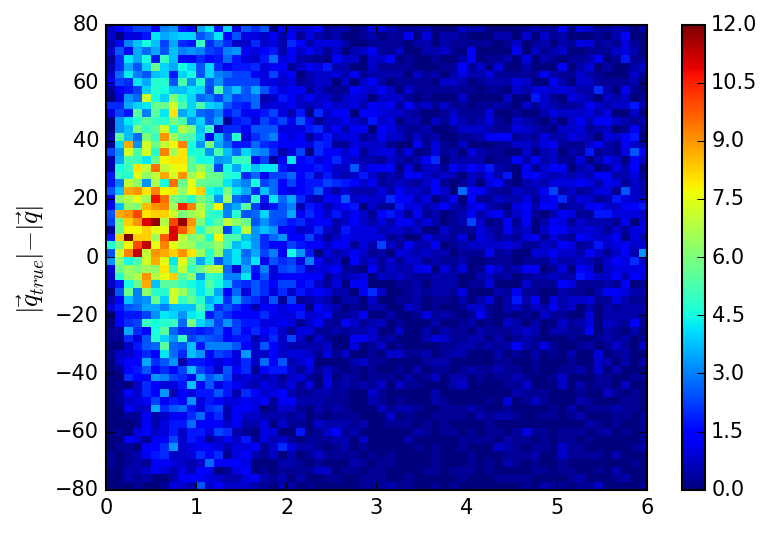} \\ 
\includegraphics[width=6.7cm]{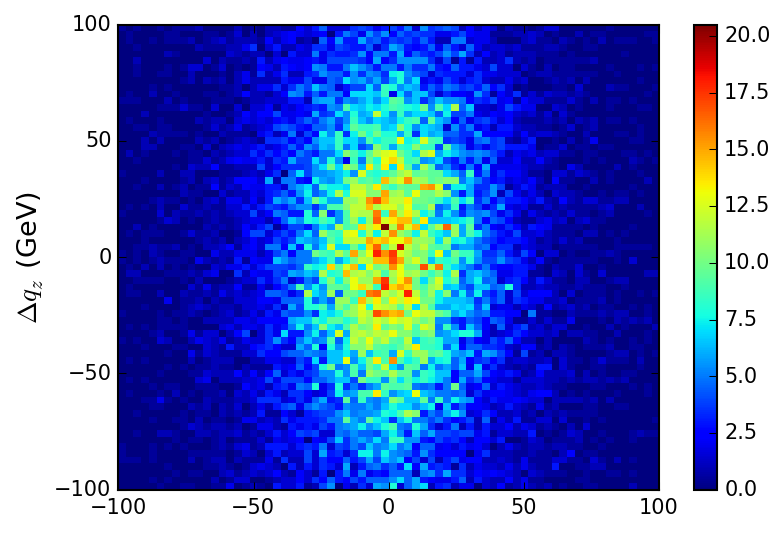}  
\includegraphics[width=6.6cm]{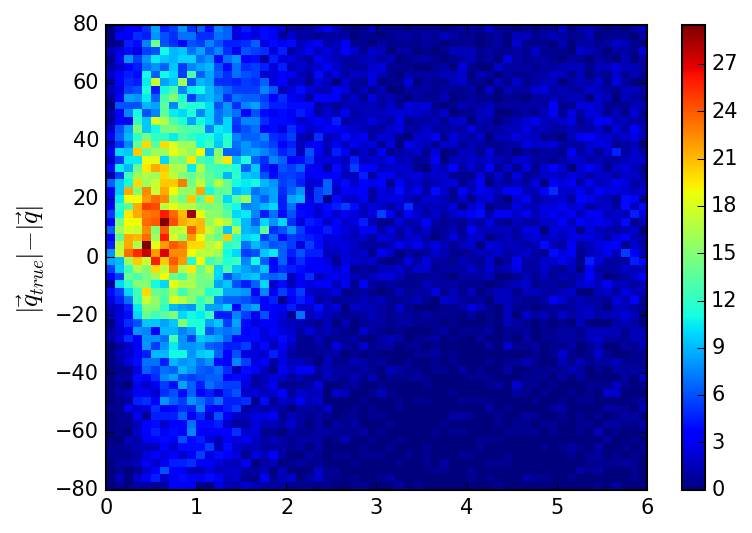} \\ 
\includegraphics[width=6.5cm]{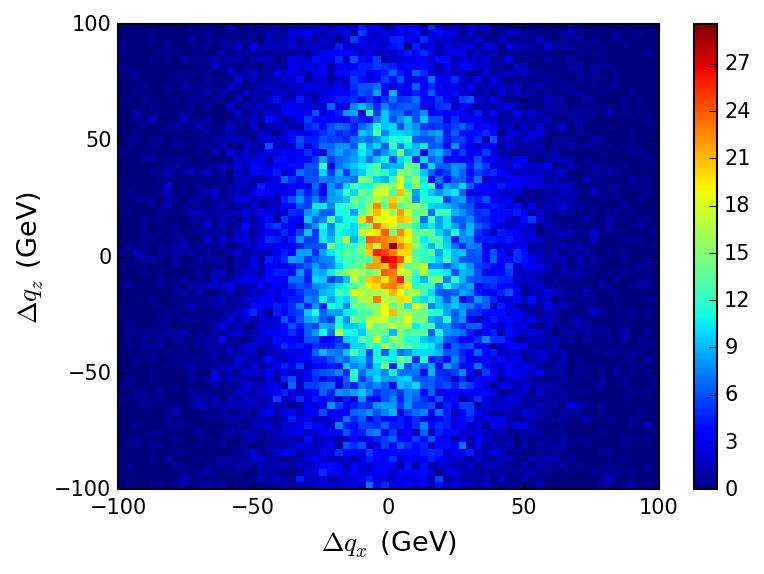} 
~\includegraphics[width=6.5cm]{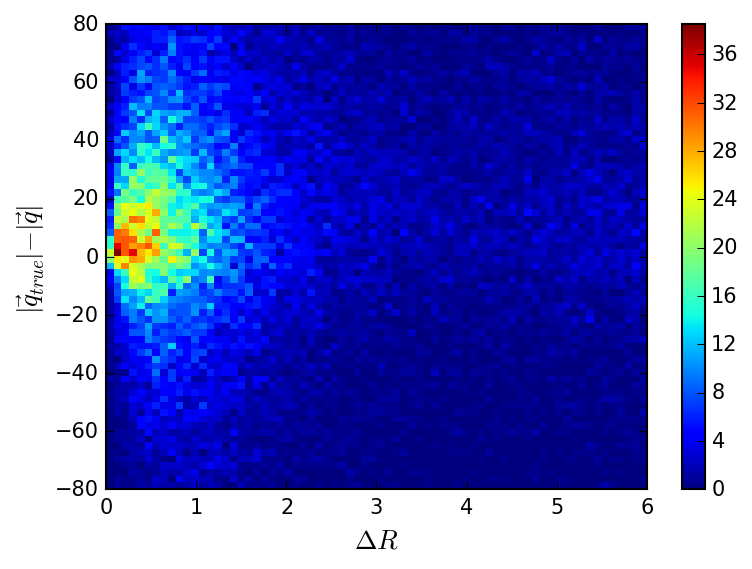} \\  
\caption{\label{fig:momentum2d} The same as Fig.~\ref{fig:momentum2dmass}, but for methods which do not use an extra mass input (from top to bottom):
MAOS2($b\ell$;$b\ell$),
MAOS3($b\ell$;$b\ell$) and 
M${}_2$A($b\ell$).
}
\end{figure}

The two main goals of this section are:
\begin{itemize}
\item \dk{to} list systematically the different ways in which the variables from the previous section can be used 
(sometimes in combination) to obtain an ansatz for the invisible momenta (see Table~\ref{tab:methods});
\item \dk{to} compare the accuracy of several representative schemes for invisible momentum reconstruction
(see Figs.~\ref{fig:momentum1d}-\ref{fig:momentum2d}).
\end{itemize}

The ansatz for the invisible momenta is generally obtained in two steps\footnote{In all cases, one must specify a test mass for 
the lightest particle (the neutrino in the case of dilepton $t\bar{t}$ events.}:
\begin{enumerate}
\item {\em Fixing the transverse components $\vec{q}_{iT}$ of the invisible momenta.} In principle, 
there are several possible options here: one can use either an $M_{T2}$ variable, or an $M_2$ variable, which
can then be applied to any of the three possible subsystems in Fig.~\ref{fig:decaysubsystem}. In addition, if one wished to use the 
mass information for a relative particle, one could also consider the maximally constrained 
variables (\ref{eq:m2CWdef}) and (\ref{eq:m2Ctdef}). The four columns of Table~\ref{tab:methods} list four representative examples, 
illustrating both the use of different variables ($M_{T2}$ versus $M_{2CC}$) and the use of different subsystems ($(b\ell)$ versus $(\ell)$).
\begin{table}[t]
\centering
\def\arraystretch{1.3}
\scalebox{0.91}{
\begin{tabular}{||c||c|c||c|c||}
\hline
 \multicolumn{5}{||c||}{Schemes for fixing the components of the invisible momenta}\\
\hline
\multicolumn{1}{||c||}{longitudinal}          &  \multicolumn{4}{c||}{transverse}  \\  \hline
input               &	$M_{T2}^{(b\ell)}=M_{2CX}^{(b\ell)}$       &	$M_{T2}^{(\ell)}=M_{2CX}^{(\ell)}$       & $M_{2CC}^{(b\ell)}$   & $M_{2CC}^{(\ell)}$   \\
\hline   
\hline
  $m_t$   & \cellcolor{yellow}MAOS1($b\ell$;$m_t$)   & \cellcolor{yellow}MAOS4($\ell$;$m_t$)  &  \cellcolor{yellow}CMAOS1($b\ell$;$m_t$)  & \cellcolor{yellow}CMAOS4($\ell$;$m_t$)   \\
\hline
 $m_W$  &  \cellcolor{yellow}MAOS4($b\ell$;$m_W$) &  \cellcolor{yellow}MAOS1($\ell$;$m_W$) &  \cellcolor{yellow}CMAOS4($b\ell$;$m_W$)  & \cellcolor{yellow}CMAOS1($\ell$;$m_W$)  \\ \hline
\hline
 $M_{T2}^{(b\ell)}$   &  MAOS2($b\ell$;$b\ell$)    & MAOS2($\ell$;$b\ell$)  &  CMAOS2($b\ell$;$b\ell$)   &  CMAOS2($\ell$;$b\ell$) \\
\hline
$M_{T2}^{(\ell)}$   & MAOS2($b\ell$;$\ell$)      & MAOS2($\ell$;$\ell$)    & CMAOS2($b\ell$;$\ell$)   &  CMAOS2($\ell$;$\ell$)   \\ \hline
$M_T^{(b\ell)}$   & \cellcolor{orange}MAOS3($b\ell$;$b\ell$)     & MAOS3($\ell$;$b\ell$)   & \cellcolor{orange}CMAOS3($b\ell$;$b\ell$)    &  CMAOS3($\ell$;$b\ell$) \\
\hline
$M_T^{(\ell)}$   & MAOS3($b\ell$;$\ell$)   & MAOS3($\ell$;$\ell$)    & CMAOS3($b\ell$;$\ell$)  & CMAOS3($\ell$;$\ell$)   \\
\hline\hline
$M_{2CC}^{(b\ell)}$  &  ---   &  --- & \cellcolor{orange}M${}_2$A($b\ell$) &   --- \\
\hline
$M_{2CC}^{(\ell)}$     & ---   &  --- &   ---  & \cellcolor{orange}M${}_2$A($\ell$) \\
\hline
\end{tabular}
}
\caption{\label{tab:methods} Various methods for reconstructing the transverse and longitudinal momenta of invisible particles 
in the dilepton $t\bar{t}$ event topology of Fig.~\ref{fig:decaysubsystem}. In all cases, one must specify a test mass for the lightest particle (the neutrino), then 
superscripts $(b \ell)$ and $(\ell)$ are used to denote respectively the $(ab)$ and $(b)$ subsystems of Fig.~\ref{fig:decaysubsystem}
(or alternatively, the subsystems $(2,2,0)$ and $(2,1,0)$ in the notation of Ref.~\cite{Burns:2008va}). 
The methods in the yellow (orange) cells will be investigated in detail in Table~\ref{table:efficiency} (Table~\ref{table:efficiencyM2CC}) below.}
\end{table}
\item {\em Fixing the longitudinal components $q_{iz}$ of the invisible momenta.} Having thus determined the transverse invisible components, 
the second step is to obtain values for the longitudinal components $q_{iz}$ of the invisible momenta. There are several possibilities (refer to Table~\ref{tab:methods}):
\begin{itemize}
\item {\em Classic MAOS with mass information (MAOS1 and MAOS4).} In the original MAOS approach
\cite{Cho:2008tj}, a mass shell constraint for an intermediate resonance is imposed on each side of the event.
Following the notation of \cite{Kim:2017awi}, we shall make the distinction between cases where the resonance is a parent particle
(MAOS1) and a relative particle (MAOS4). 
In the classic MAOS reconstruction, the transverse invisible components are obtained from $M_{T2}$, but this can be done
for one of several possible subsystems, so we need to implement some notation to indicate which subsystem was used.
For example, the abbreviation MAOS1($b\ell$;$m_t$) in Table~\ref{tab:methods} implies that the transverse invisible momenta 
were obtained from $M_{T2}^{(b\ell)}$, while the longitudinal invisible momenta were computed from the on-shell conditions
for the parent particles (thus MAOS1) with mass $m_t$. Similarly, the abbreviation MAOS4($\ell$;$m_t$) indicates
the use of $M_{T2}^{(\ell)}$ for fixing the transverse invisible momenta, then applying on-shell conditions for the top quarks, which
in subsystem $(\ell)$ are relative particles (thus the name MAOS4). In both MAOS1 and MAOS4, the longitudinal momenta are obtained up to 
a four-fold ambiguity, as one has to solve a quadratic equation for each decay side.
\item  {\em Classic MAOS without mass information (MAOS2 and MAOS3).} 
There are two other MAOS schemes, which are applicable in the absence of any mass information about the parent or relative particles
\cite{Choi:2009hn,Cho:2009wh,Park:2011uz,Choi:2010dw}.
In MAOS2 one forces each parent mass to be equal to the computed $M_{T2}$ value, i.e.,
$M_{P_i}(\vec{q}_i)=M_{T2}$, $i=1,2$, while in MAOS3 one demands that the parent mass be equal to the 
corresponding {\em transverse} parent mass obtained during the $M_{T2}$ calculation:
$M_{P_i}(\vec{q}_i)=M_{TP_i}(\vec{q}_{iT})$, $i=1,2$. Once again, each of these two MAOS schemes 
can be applied to any of the three possible subsystems \cite{Kim:2017awi}. Furthermore, in the previous step 1 we could in principle
use a different subsystem for the determination of the transverse invisible components, therefore now we need {\em two}
subsystem labels to completely define the procedure. We shall employ the notation where the first subsystem label refers to the 
determination of the transverse invisible momenta, while the second subsystem label refers to the computation of the respective 
longitudinal components. For example, the abbreviation MAOS3($\ell$;$b\ell$) implies that the transverse components were obtained from
$M_{T2}^{(\ell)}$, and then the longitudinal components were calculated from the MAOS3 condition for the parents in the 
$(b\ell)$ subsystem, i.e., the top quarks: $M_{t_i}(\vec{q}_i)=M_{Tt_i}(\vec{q}_{iT})$, $i=1,2$. The MAOS3 procedure
always results in an unique ansatz, while MAOS2 is unique only for balanced events, i.e., events with 
$M_{TP_1}=M_{TP_2}$; for unbalanced events, MAOS2 gives exactly two solutions \cite{Cho:2014naa}. 
\item {\em $M_2$-assisted invisible momentum reconstruction.} Another alternative is to use an $M_2$ variable --- recall that the 
$M_2$ optimization procedure provides an ansatz for the full 3-vectors $\vec{q}_i$ of the invisible momenta. As indicated in
Table~\ref{tab:methods}, such $M_2$-Assisted reconstructions will be denoted with M${}_2$A and will carry a corresponding
subsystem label as well.
\item {\em Hybrid methods.} The remaining methods in Table~\ref{tab:methods} are hybrid in the sense that they 
rely on a constrained $M_{2CC}$ variable for obtaining the transverse components of the invisible momenta and on one of the MAOS 
methods for the determination of the longitudinal components. We shall call such methods CMAOS for ``constrained" MAOS.
The rationale for considering these methods is that, as we shall see below, the constrained $M_{2CC}$ variables often provide
superior ansatze for the transverse invisible momenta. Once again, one can ``mix and match" the subsystems, which necessitates the
use of two subsystem arguments for the CMAOS procedures listed in Table~\ref{tab:methods}.
\end{itemize}
\end{enumerate}

Note that Table \ref{tab:methods} does not include all logical possibilities --- for example, in order to keep the table compact, 
we did not list the the maximally constrained variables (\ref{eq:m2CWdef}) and (\ref{eq:m2Ctdef}), which represent another 
M${}_2$A option for simultaneously computing the transverse and longitudinal invisible components.
One should also distinguish between methods which use additional mass inputs ($m_t$ or $m_W$) and methods which do not --- 
in what follows, we shall be careful to compare the performance of those two categories of methods separately.
For example, the methods in the yellow-shaded cells of Table~\ref{tab:methods} require an additional mass input and as such
they will be discussed and contrasted in Table~\ref{table:efficiency} of section \ref{sec:step3}. 
On the other hand, the orange-shaded cells of Table~\ref{tab:methods} highlight a few representative methods which do not require additional mass inputs --- 
those methods will be compared separately in Table~\ref{table:efficiencyM2CC} of section \ref{sec:step3}. 
Note that the M${}_2$A methods from Table~\ref{tab:methods} do not use extra mass information. 

Having defined the different momentum reconstruction schemes, we are now in position to compare their performance.
Following \cite{Cho:2008tj,Guadagnoli:2013xia}, we shall first ask, \dk{how close each scheme gets} to reproducing 
the {\em actual} values for the invisible momenta. Fig.~\ref{fig:momentum1d} shows a comparison of the true values $\vec{q}_{true}$ of the 
transverse components (left panel) and the longitudinal components (middle and right panels) of the invisible momenta 
to the corresponding reconstructed values $\vec{q}$ obtained with different methods from Table~\ref{tab:methods}.
The left panel in Fig.~\ref{fig:momentum1d} contains the combined distributions of the transverse momentum differences
$\Delta q_x\equiv q_{x,true}-q_x$ and $\Delta q_y\equiv q_{y,true}-q_y$ resulting from four different transverse momentum 
reconstruction schemes: $M_{T2}^{(b\ell)}$ (blue dotted line), $M_{2CC}^{(b\ell)}$ (red solid line),
$M_{2Ct}^{(\ell)}$ (green dashed line) and $M_{2CW}^{(b\ell)}$ (magenta dashed line). In all four cases, the
distributions are peaked at $\Delta q=0$, which indicates that on average all four methods work rather well.
We also observe that the distributions for $M_{2Ct}^{(\ell)}$ and $M_{2CW}^{(b\ell)}$, which utilize an extra mass input, 
are more sharply peaked, leading to much smaller errors. Among the two remaining distributions, $M_{2CC}^{(b\ell)}$ 
appears to perform slightly better than $M_{T2}^{(b\ell)}$.

The middle and right panels of Fig.~\ref{fig:momentum1d} show similar plots for the longitudinal momentum difference
$\Delta q_z\equiv q_{z,true}-q_z$, obtained with various methods for reconstructing the longitudinal invisible momenta:
MAOS1($b\ell$;$m_t$) (blue dotted line),
MAOS4($b\ell$;$m_W$) (black dotted line),
$M_{2Ct}^{(\ell)}$ (green dashed line),
$M_{2CW}^{(b\ell)}$ (magenta dashed line),
M${}_2$A($b\ell$) (red solid line),
MAOS2($b\ell$;$b\ell$) (yellow dotted line) and 
MAOS3($b\ell$;$b\ell$) (cyan dashed line).
Among the methods requiring an additional mass input (middle panel), $M_{2Ct}^{(\ell)}$ and $M_{2CW}^{(b\ell)}$ 
again work best, while among the more conservative methods (right panel), M${}_2$A appears to 
outperform MAOS2 and MAOS3 (see also \cite{Kim:2017awi}).

Figs.~\ref{fig:momentum2dmass} and \ref{fig:momentum2d} provide a more detailed view of the results from 
Fig.~\ref{fig:momentum1d} by showing the correlations between $\Delta q_z$ and $\Delta q_x$ (left panels)
and between the difference in magnitudes $ |\vec q_{true}|-|\vec q \, |$
and the direction mismatch $\Delta R(\vec q_{true},\vec q \,)\equiv \sqrt{ (\Delta \eta)^2+(\Delta\varphi)^2}$
(right panels).
Figs.~\ref{fig:momentum2dmass} and \ref{fig:momentum2d} reveal that in general, the transverse components of the invisible momenta are 
reconstructed more accurately than the longitudinal components, and that having additional mass information at one's disposal 
definitely helps. The right panels of Fig.~\ref{fig:momentum2d} also show that for those methods, it is more likely to underestimate 
(than to overestimate) the magnitude of the invisible momentum --- 
this is easy to understand for the case of events in which the two transverse invisible momenta partially cancel each other out in the $\mpt$ sum.

In conclusion of this section, we note that it is known that the performance of the methods with respect to invisible momentum reconstruction 
can be further improved by selecting only events near the kinematic endpoint of the respective invariant mass variable from 
which the ansatz originated \cite{Cho:2008tj,Kim:2017awi}. However, this benefit comes with a significant loss in statistics, 
and we shall not pursue this idea further here.

\section{Critical review of the standard method}
\label{sec:oldmethod}

In this section, we analyze the standard method outlined in Refs.~\cite{Baringer:2011nh,Choi:2011ys}
for resolving the combinatorics problem in dilepton $t\bar{t}$ events. The method involves three steps, which were 
briefly reviewed in the Introduction, and will be now examined in detail in the following three subsections.
For our numerical studies, we generate a \dk{partonic} $t \bar{t}$ dilepton sample
with 50k events, using the \amc~framework at the LHC with $\sqrt{s} = 14$ TeV center of mass energy and the default set of 
parton distribution functions \cite{Alwall:2011uj}. The masses of the top quark and the $W$-boson are set to 173 GeV and \dk{80.419} GeV, respectively, 
and we also take into account the proper finite widths --- as we shall see below, this leads to the presence of events for which 
the top quarks and/or the $W$-bosons can be significantly off-shell. In order to reduce the background, we apply the same basic cuts as those 
used in Ref.~\cite{Choi:2011ys}. This leaves us with 18,456 events after cuts, for a cut efficiency of 37\%. 
The different versions of the $M_{T2}$ and $M_2$ kinematic variables will be computed with the {\sc OPTIMASS} package \cite{Cho:2015laa}.

\subsection{Step I: $M_{T2}^{(b\ell)}$ and $m_{b\ell}$ cuts}
\label{sec:step1}

The first step of the algorithm relies on the fact that in the event topology of Fig.~\ref{fig:decaysubsystem}, there exist several invariant mass variables,
whose distributions exhibit an upper kinematic endpoint. If we choose the correct partition (\ref{eq:PC}), all of these endpoints should be satisfied
(barring off-shell effects). On the other hand, the wrong partition (\ref{eq:PW}) may lead to one (or more) endpoint violations. The art of designing
a good method for resolving the combinatorics lies in choosing the optimal invariant mass variables which will maximize the number of events 
for which the wrong partition (\ref{eq:PW}) results in endpoint violations.

In principle, there are two types of invariant mass variables which can have kinematic endpoints:
\begin{itemize}
\item {\em Using visible particles from the same decay chain.} One can study the invariant mass of a collection of visible particles 
emerging from the same decay chain. For a long decay chain, there are many possible combinations \cite{Kim:2015bnd}, 
but for a short decay chain like the one in Fig.~\ref{fig:decaysubsystem}, the choice is unique - we can only form the two-body 
invariant mass of the $b$-jet and the lepton on each side. This gives us two values, $m_{b\ell^+}$ and $m_{\bar{b}\ell^-}$,
each of which should obey the kinematic endpoint $m_{b\ell}^{max}$, as illustrated in the left panel of Fig.~\ref{fig:step1}.
\begin{figure}[t]
\centering
\includegraphics[width=4.8cm]{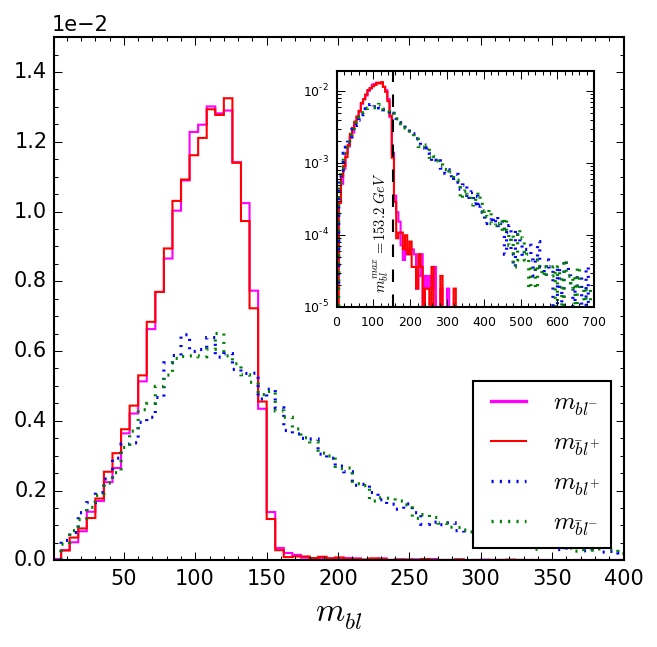}  \hspace*{-0.1cm}
\includegraphics[width=5.2cm]{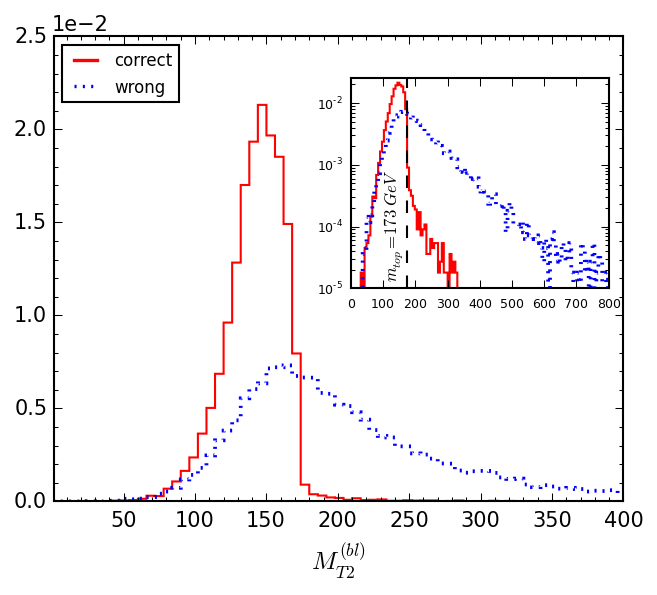}\hspace*{-0.1cm} 
\includegraphics[width=5.2cm]{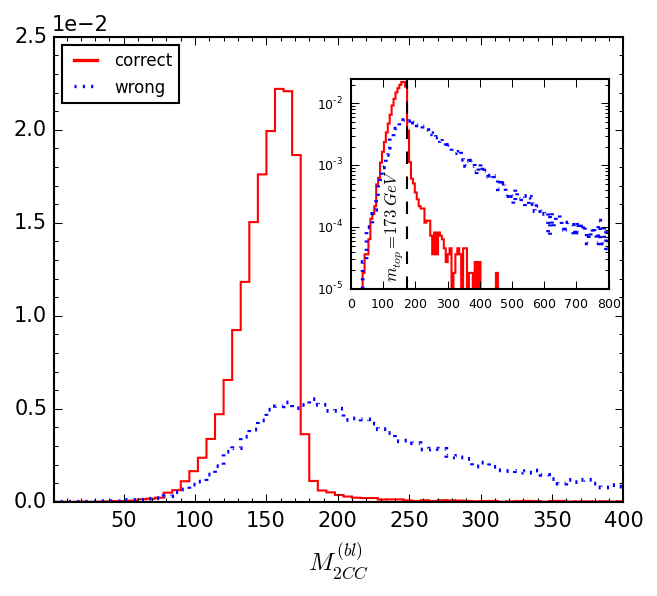}  \\
\caption{\label{fig:step1} The distribution of  $m_{b\ell}$  (left), $M_{T2}^{(b\ell)}$ (middle) and $M_{2CC}^{(b\ell)}$ (right)  for the correct 
partition (\ref{eq:PC}) (solid lines) and the wrong partition (\ref{eq:PW}) (dotted lines). The inserts show a wider range of the $x$-axis and
use a log scale for the $y$-axis. The corresponding $M_{T2}$ and $M_{2CC}$ distributions for the $(b)$ and $(\ell)$ subsystems are
shown in Fig.~\ref{fig:subsystem} below.
}
\end{figure}
Following Refs.~\cite{Baringer:2011nh,Choi:2011ys}, we shall apply the stronger condition that the {\em larger} of these two 
values should also obey the upper kinematic endpoint:
\beq
\max \{ m_{b\ell^+} \, , m_{\bar{b}\ell^-}  \} ~\leqslant ~  m_{b\ell}^{max} ~ \equiv ~  \sqrt{\frac{ \big ( m_{t}^2 - m_W^2\big ) \big ( m_W^2 - m_\nu^2 \big )}{ m_W^2}   }\, , 
\label{eq:mbl}
\eeq
where $m_t$, $m_W$ and $m_\nu$ are respectively the masses of the top quark, the $W$-boson and the neutrino (we neglect the masses of the $b$-quark and the lepton).
With their nominal values from the standard model, the endpoint is located at $m_{b\ell}^{max}=153.2$ GeV (see the insert in the left panel of Fig.~\ref{fig:step1}).
\item {\em Using visible particles from both decay chains.} The other possibility is to use the measured momenta of visible particles from both decay chains
in order to construct invariant mass variables which also exhibit upper kinematic endpoints \cite{Barr:2011xt}.
The prototypical example of such a variable is the Cambridge variable $M_{T2}$ (see the middle panel in Fig.~\ref{fig:step1}), but there are 
other possibilities as well, e.g., $M_{CT}$ \cite{Tovey:2008ui,Matchev:2009ad}, $M_{CT2}$ \cite{Cho:2009ve}, and more recently, 
$M_{2CC}$ \cite{Cho:2014naa} (see the right panel in Fig.~\ref{fig:step1}). Following Refs.~\cite{Baringer:2011nh,Choi:2011ys}, 
we shall continue to consider $M_{T2}$, but we shall also entertain the possibility of using $M_{2CC}$ instead.
For the correct partition, the distributions of $M_{T2}$ and $M_{2CC}$ have common kinematic endpoints, and so 
the values of $M_{T2}$ and $M_{2CC}$ obey the hierarchy
\begin{eqnarray}
M_{T2}^{(b\ell)} ~ \leqslant ~ M_{2CC}^{(b\ell)} ~ & \leqslant&  ~ m_{t} \, ,  \label{eq:mt2bl} \\ [2mm]
M_{T2}^{(\ell)}   ~ \leqslant ~ M_{2CC}^{(\ell)}   ~ & \leqslant&  ~ m_{W} \, ,   \label{eq:mt2l}  
\end{eqnarray}
where the endpoint values correspond to using the true\footnote{If the neutrino mass were unknown, one could use any 
arbitrary value for the test daughter particle mass, and then extract the endpoint values in (\ref{eq:mt2bl}) and (\ref{eq:mt2l}) from the data.} 
value of the neutrino mass $m_{\nu}=0$. More importantly, for the wrong partition, the shapes of the 
$M_{T2}$ and $M_{2CC}$ distributions are different (compare the blue dotted lines in the middle and right panels of Fig.~\ref{fig:step1}), 
which will affect the efficiency for selecting the correct partition. Due to the general property (\ref{eq:hierarchy}), 
the wrong partition will still preserve the hierarchy $M_{T2} \leqslant M_{2CC}$, and therefore, 
the chances of endpoint violations will be increased if we were to use $M_{2CC}$ in place of $M_{T2}$ \cite{Cho:2014yma}.
\end{itemize} 
 \begin{minipage}[t!]{\textwidth}
 \centering
  \begin{minipage}[b]{0.47\textwidth}
	\includegraphics[width=7.cm]{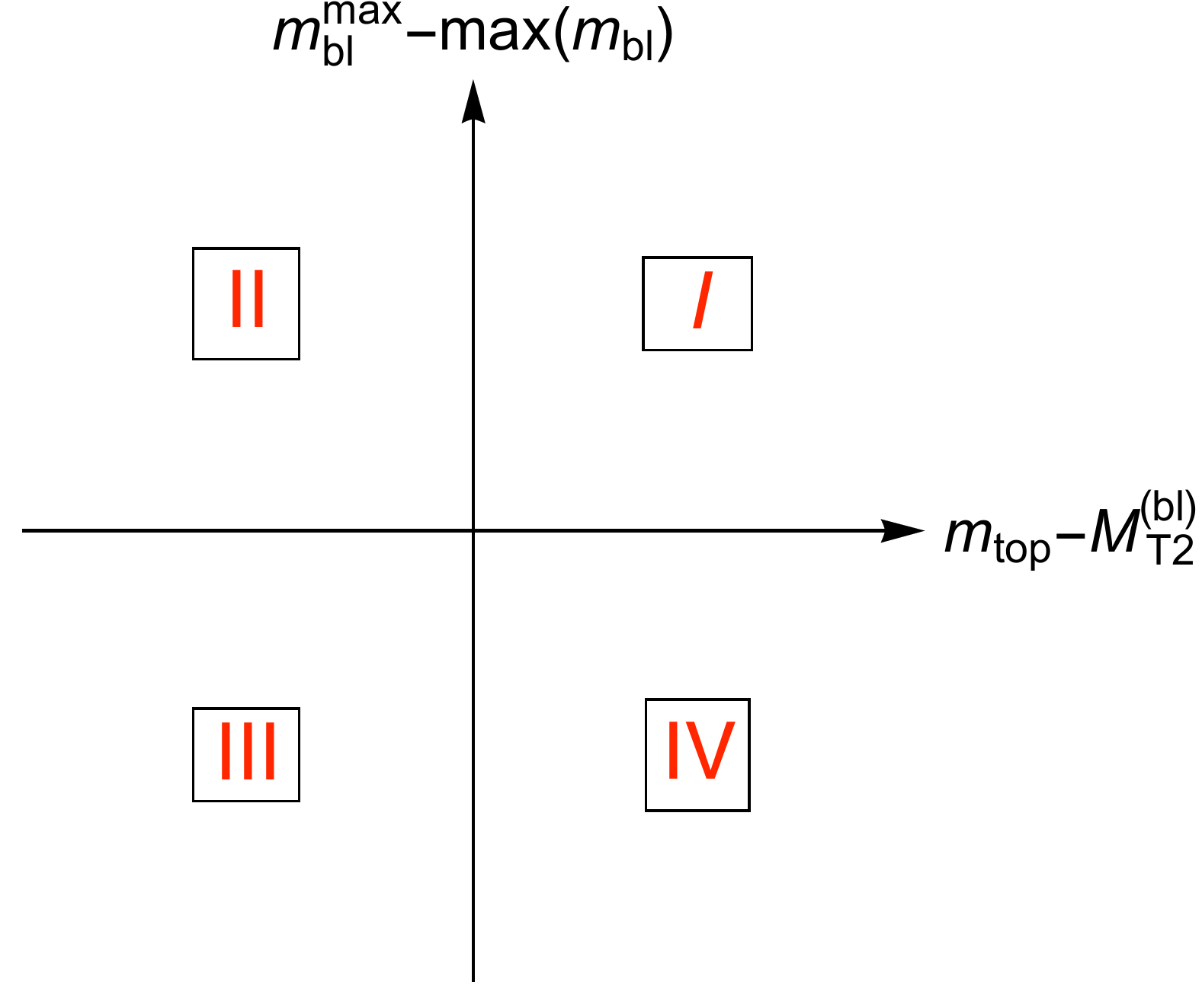} 
	\vspace*{-0.3cm}
	\captionof{figure}{\label{fig:quadrant} Definition of the four quadrants in the plane ($m_t-M_{T2}^{(b\ell)}$, $m_{b\ell}^{max}-\max_j\{m^{(j)}_{b\ell}\}$). 
	}
\vspace*{0.5cm}
  \end{minipage}
%
%
\hspace*{0.04\textwidth}
  \begin{minipage}[b]{0.47\textwidth}
    \centering 
\scalebox{1.1}{
\begin{tabular}{||c||c|c|c|c||}
\hline
Quadrant & \multicolumn{4}{c||}{Quadrant for $P_W$}   \\
\cline{2-5} 
for $P_C$ &  I  &  II  &  III  &  IV  \\ \hline \hline
I  &  ~~~~
   &  \cellcolor{green}   ~~~
    &  \cellcolor{green}   ~~~
    &  \cellcolor{green}   ~~~
 \\ \hline
II  &   \cellcolor{red}  
   &  
    &  \cellcolor{green}  
     & 
	 	 \\ \hline
III &   \cellcolor{red}    
   &   \cellcolor{red}  
    &
    &   \cellcolor{red}  
           	 	 \\ \hline  
IV  &   \cellcolor{red}  
   &  
    &  \cellcolor{green}  
    &  
           	 \\ \hline
          \hline
\end{tabular}
}
\captionof{table}{\label{table:quadrant} 
Resolving the combinatorial ambiguity at Step I. 
Each event is tagged with two quadrant numbers, one for each partition $P_i$. 
The quadrant number for the correct (wrong) combination is given in the `row' (`column') label of the 
4 by 4 matrix above. 
The green, \dk{red, and white} fields indicate correctly resolved, wrongly resolved, and unresolved cases, \dk{respectively}.}
\vspace*{0.5cm}
\end{minipage}
  \end{minipage}

The distributions in Fig.~\ref{fig:step1} clearly motivate the use of the variables $m_{b\ell}$ and $M_{T2}^{(b\ell)}$ (or perhaps $M_{2CC}^{(b\ell)}$ instead) 
to resolve the two-fold combinatorial ambiguity.\footnote{For now, as in Refs.~\cite{Baringer:2011nh,Choi:2011ys},
we shall focus on the $(b\ell)$ subsystem, where one would expect the largest number of endpoint violations for the wrong kinematics
\cite{Cho:2014yma}. The other two subsystems, $(b)$ and $(\ell)$, will be discussed later in section~\ref{sec:quadrants}.} 
This idea is implemented as Step I of the algorithm, by requiring that the invariant mass variables computed with a given
partition $P_i, (i=1,2)$, obey the two kinematic endpoints (\ref{eq:mbl}) and (\ref{eq:mt2bl}). If one of the partitions obeys both endpoints,
while the other does not, the former (latter) is declared to be the correct (wrong) partition $P_C$ ($P_W$).

In order to quantify the discussion in the rest of paper, we introduce a simple Cartesian coordinate system designed to keep track of the 
kinematic endpoint violations (see Fig.~\ref{fig:quadrant}). The $x$ and $y$ variables will be chosen so that their values are 
positive (negative) in the absence (presence) of a kinematic endpoint violation. To this end, we shall consider the difference between the 
value of the upper kinematic endpoint and the value of the variable itself --- this difference is expected to be positive for the correct
partition $P_C$, and conversely, if the difference is negative, it is likely that we have chosen the wrong partition $P_W$. 
Thus in Fig.~\ref{fig:quadrant} we choose the $x$-axis to be $m_t-M_{T2}^{(b\ell)}$ (later on we shall also consider $m_t-M_{2CC}^{(b\ell)}$),
while for the $y$-axis we take $m_{b\ell}^{max}-\max_j\{m^{(j)}_{b\ell}\}$, where $m^{(1)}_{b\ell}$ and $m^{(2)}_{b\ell}$ 
are the invariant masses of the two $b$-lepton pairs in a given partition. As usual, the plane in Fig.~\ref{fig:quadrant} is divided into four quadrants, 
labelled I, II, III and IV. With this setup, one would expect that the correct partition $P_C$ will be registered in the first quadrant I,
while the wrong partition $P_W$ can end up anywhere, including quadrants II, III and IV, which would indicate some sort of an endpoint violation.

\begin{figure}[t]
\centering
\includegraphics[width=7cm]{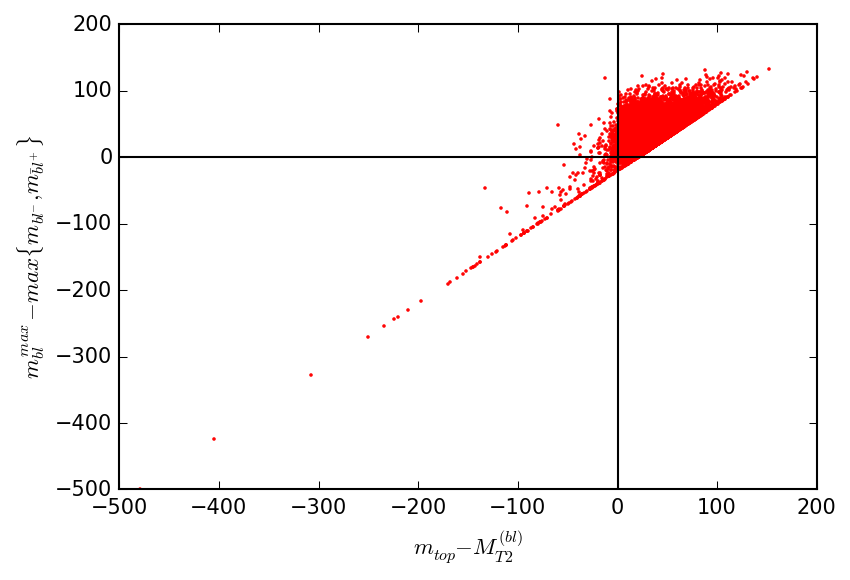} 
\includegraphics[width=7cm]{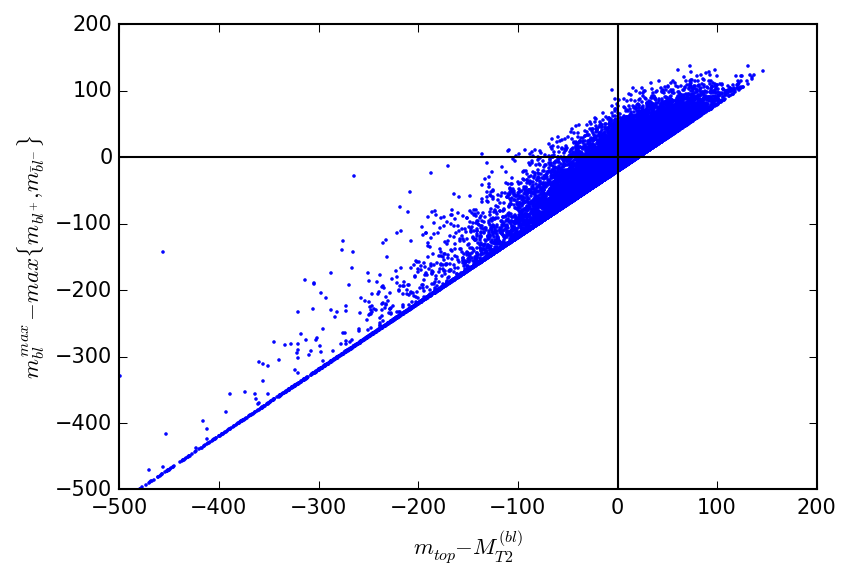} \\
\includegraphics[width=7cm]{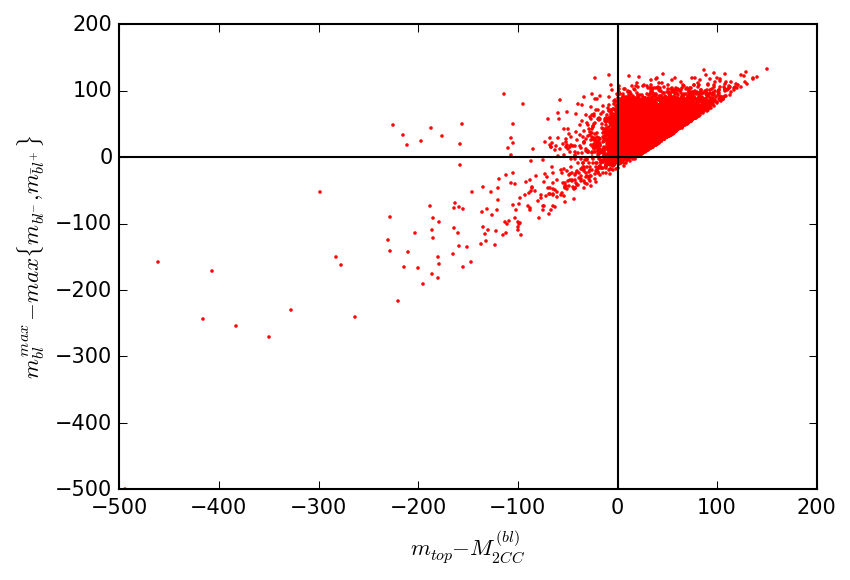} 
\includegraphics[width=7cm]{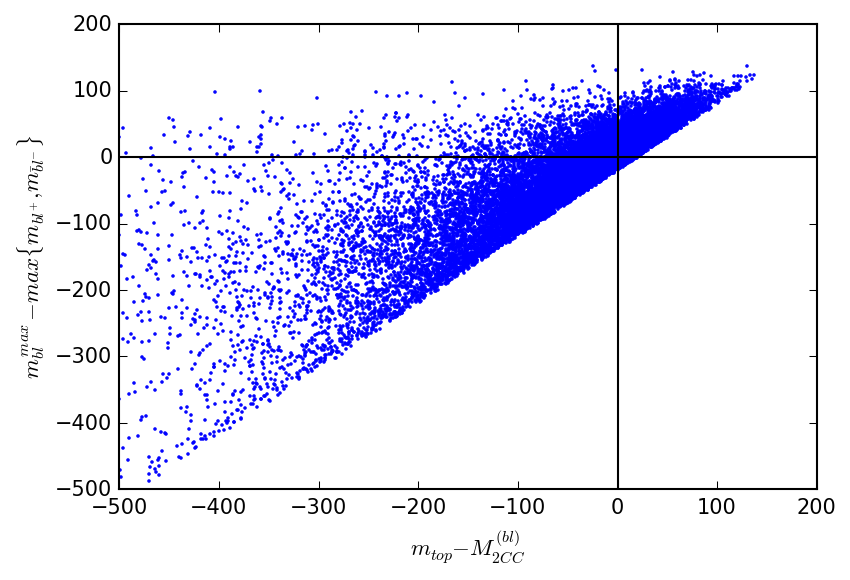}
\caption{\label{fig:scatterplot} Scatter plots in the plane of Fig.~\ref{fig:quadrant} for the correct partition $P_C$ (left panels, red points) 
and the wrong partition $P_W$ (right panels, blue points). In the top row the $x$-axis is chosen to be $m_t-M_{T2}^{(b\ell)}$,
while in the bottom row the $x$-axis is $m_t-M_{2CC}^{(b\ell)}$. 
}
\end{figure}
These expectations are confirmed in Fig.~\ref{fig:scatterplot}, which shows scatter plots in the plane of Fig.~\ref{fig:quadrant} 
for events with the correct partition $P_C$ (left panels, red points) and the wrong partition $P_W$ (right panels, blue points). 
We see that the correct partition mostly populates quadrant I, although there is some leakage into the other three quadrants due to 
off-shell effects. On the other hand, the wrong partition cases are significantly spread out, and the majority of the events live 
outside quadrant I. The effect is even more pronounced if we trade $M_{T2}^{(b\ell)}$ for $M_{2CC}^{(b\ell)}$ and consider
$m_t-M_{2CC}^{(b\ell)}$ as our $x$-axis variable (see the plots in the bottom row of Fig.~\ref{fig:scatterplot}).

We are now in position to define the action of Step I of the algorithm. For each event, there are two possible partitions,
$P_C$ from (\ref{eq:PC}) and $P_W$ from (\ref{eq:PW}). Since we do not know which is which, from now on we shall denote them 
with $P_k$, $(k=1,2)$. (It does not matter which partition is labelled first and which is labelled second.) Each partition
$P_k$ will produce a point in one of the four quadrants within the plane of Fig.~\ref{fig:quadrant}. We will then resolve 
the partitioning ambiguity according to Table~\ref{tab:decision}. 
%
\begin{table}[t]
\centering
\scalebox{1.01}{
\begin{tabular}{||c||c|c|c|c||}
\hline
\hline
Quadrant & \multicolumn{4}{c|}{Quadrant for $P_2$}   \\
\cline{2-5} 
for $P_1$ &  ~~I~~  & ~ II ~ &  ~III~  &  ~IV~ \\ \hline \hline
I  &  unresolved     &  $P_C=P_1$ & $P_C=P_1$  & $P_C=P_1$  \\ \hline
II & $P_C=P_2$    &  unresolved  & $P_C=P_1$  &  unresolved \\ \hline
III & $P_C=P_2$   & $P_C=P_2$   & unresolved  & $P_C=P_2$  \\ \hline
IV  & $P_C=P_2$ & unresolved & $P_C=P_1$ & unresolved\\ \hline  \hline
\end{tabular}
}
\caption{\label{tab:decision} 
Decision table for resolving the combinatorial ambiguity at Step I.
} 
\end{table}
Whenever one of the two partitions falls in quadrant I while the other does not, then the partition in quadrant I will be declared to be the correct one ($P_C$).
If both partitions fall in the same quadrant, then the event remains ``unresolved" at Step I and we need to wait for the next steps of the algorithm.
The situation becomes more complicated if both partitions fall outside quadrant I, and within different quadrants. In that case, we shall make the 
distinction between quadrant III, where both endpoints are violated, and quadrants II and IV, where there is a single endpoint violation.
Correspondingly, if one partition falls in quadrant III while the other does not, the partition in quadrant III will be declared as the wrong partition ($P_W$).
Finally, if one partition is in quadrant II while the other in quadrant IV, the event remains unresolved.

As a result of the application of Table~\ref{tab:decision}, each event will fall into one of two categories: resolved, for which one of the partitions $P_k$ has been 
declared to be correct, and unresolved, for which no decision has been reached at that stage. Furthermore, the resolved events will not always be identified
correctly --- on occasion, the algorithm will misidentify the wrong partition $P_W$ as being the correct one. In order to better understand the power of each method below, 
we shall find it convenient to quote our results in the form of $4\times 4$ tables like Table~\ref{table:quadrant}, where we separately keep track of the 
quadrant for the correct partition $P_C$ (indicated by the row label) and the quadrant for the wrong partition $P_W$ (indicated by the column label).
Each event will belong to one of the 16 boxes of Table~\ref{table:quadrant}, and we will be interested in the number of events $N_{(I,J)}$ within each box,
where the ``quadrant indices" $I$ (for the correct partition) and $J$ (for the wrong partition) take values in the set $\{I, II, III, IV\}$. The action by
Table~\ref{tab:decision} then causes all events within the green-shaded boxes of Table~\ref{table:quadrant} to be {\em correctly} resolved, the events within the 
red-shaded boxes of Table~\ref{table:quadrant} to be {\em wrongly} resolved, while the events within the unshaded boxes of Table~\ref{table:quadrant} 
to remain {\em unresolved}. Then, for any given event sample, the total number $N_C$ of correctly resolved events will be given by the total number of events 
within the green-shaded boxes of Table~\ref{table:quadrant}:
\beq
N_C \equiv N_{(I,II)} + N_{(I,III)} + N_{(I,IV)} + N_{(II,III)} + N_{(IV,III)} .
\label{eq:NC}
\eeq
Similarly, the total number of wrongly resolved events $N_W$ will be equal to the total number of events within the red-shaded boxes of Table~\ref{table:quadrant}:
\beq
N_W \equiv N_{(II,I)} + N_{(III,I)} + N_{(IV,I)} + N_{(III,II)} + N_{(III,IV)} .
\label{eq:NW}
\eeq
Finally, the total number of unresolved events $N_U$ is the sum of all events within the unshaded (white) boxes of Table~\ref{table:quadrant}:
\beq
N_U \equiv N_{(I,I)} + N_{(II,II)} + N_{(III,III)} + N_{(IV,IV)} + N_{(II,IV)} + N_{(IV,II)}.
\label{eq:NU}
\eeq

The algorithms below will be applied so that once an event is resolved, it does not get reclassified at a later stage, i.e., subsequent 
steps of the algorithm only affect the remaining unresolved events. Obviously, at different steps of the algorithms, the number of correctly resolved events 
($N_C$), wrongly resolved events ($N_W$) and unresolved events $(N_U)$ will vary, but those three numbers will always add up to the total number of events 
$N_T$ in the sample:
\beq
N_T \equiv \sum_{I,J} N_{(I,J)} = N_C + N_W + N_U.
\label{eq:norm}
\eeq

In order to compare different algorithms, we define the expected efficiency (sometimes called purity) as
\beq
\varepsilon = \frac{N_C+0.5N_U}{N_C+N_U+N_W} =  \frac{N_C+0.5N_U}{N_T}. 
\label{efficiency}
\eeq
For the purposes of calculating the efficiency, we shall assume that any unresolved events are eventually decided with a coin flip (50\% efficiency).

\begin{table}[t]
\centering
\scalebox{1.01}{
\begin{tabular}{|c||c|c|c|c|}
\hline
\multicolumn{5}{|c|}{Quadrant counts based on $M_{T2}^{(b\ell)}$ and $m_{b\ell}$} \\
\hline
\hline
Quadrant & \multicolumn{4}{c|}{Quadrant for $P_W$}   \\
\cline{2-5} 
for $P_C$ &  ~~I~~  & ~ II ~ &  ~III~  &  ~IV~ \\ \hline \hline
I  &  6194   &  \cellcolor{green}  910  &  \cellcolor{green}  9793 &  \cellcolor{green} 1020 \\ \hline
II &   \cellcolor{red}      22  &  31  &  \cellcolor{green} 106  &  5 \\ \hline
III &   \cellcolor{red}    29 &  \cellcolor{red} 10  &   191  & \cellcolor{red} 4 \\ \hline
IV  & \cellcolor{red} 41   & 3 & \cellcolor{green} 91 & 6\\ \hline  \hline
\end{tabular}
}
\hspace*{-0.4cm}
\centering
\scalebox{1.01}{
\begin{tabular}{|c||c|c|c|c|}
\hline
\multicolumn{5}{|c|}{Quadrant counts based on $M_{2CC}^{(b\ell)}$ and $m_{b\ell}$} \\
\hline
\hline
Quadrant & \multicolumn{4}{c|}{Quadrant for $P_W$}   \\
\cline{2-5} 
for $P_C$ &  ~~I~~  & ~II~  &  ~III ~ &  ~IV~ \\ \hline \hline
I  &    4494 &   \cellcolor{green} 2317  &   \cellcolor{green} 10222 &  \cellcolor{green} 217 \\ \hline
II &    \cellcolor{red}   168  &   178  &    \cellcolor{green} 480  &  5 \\ \hline
III &   \cellcolor{red}    37 &  \cellcolor{red}  26  &    251  &\cellcolor{red}  1 \\ \hline
IV  &  \cellcolor{red}  17   & 3 & \cellcolor{green} 38 & 2 \\ \hline
  \hline
\end{tabular}
}
\caption{\label{table:mt2_m2cc_mbl_cuts} 
Quadrant counts $N_{(I,J)}$ based on $M_{T2}^{(b\ell)}$ and $m_{b\ell}$ (left) and $M_{2CC}^{(b\ell)}$ and $m_{b\ell}$ (right) after the basic cuts. The corresponding efficiencies are 82\% (left) and 85.3\% (right).
} 
\end{table}

Our results for $N_{(I,J)}$ are shown in Table~\ref{table:mt2_m2cc_mbl_cuts}, where the quadrants from Fig.~\ref{fig:quadrant}
have been defined in terms of the variables $M_{T2}^{(b\ell)}$ and $m_{b\ell}$ (left $4\times 4$ table) and the variables
$M_{2CC}^{(b\ell)}$ and $m_{b\ell}$ (right $4\times 4$ table).
As expected, the most populated entries are found in the first rows, which confirms that in the case of the correct partition $P_C$, 
endpoint violations are relatively rare\footnote{As a sanity check, we have verified that if we turn off the width effects by hand, 
by forcing both the top quarks and the $W$-bosons on-shell, all entries in the second through fourth rows are exactly zero.}. 
We also find a handful of events in the off-diagonal boxes of the first column --- for those events, off-shell effects caused the correct partition $P_C$ to
violate one or both of the kinematic endpoints, while the wrong partition $P_W$ accidentally happened to satisfy both kinematic endpoints.  
Such events are problematic since they will be wrongly identified --- one should keep in mind that only the symmetric combination of
events $N_{(I,J)}+N_{(J,I)}$ is experimentally observable, since a priori we do not know which is the correct partition. 
Nevertheless, we observe that in the cases when one of the partitions ends up in quadrant I while the other does not,
the large majority of the events will be correctly identified, since
\beq
N_{(I,II)} + N_{(I,III)} +  N_{(I,IV)} \gg N_{(II,I)} + N_{(III,I)} + N_{(IV,I)}.
\label{firstrow}
\eeq
Another piece of good news is that whenever one of the partitions violates exactly one endpoint, while the other violates both, 
it is much more likely that the former (latter) is the correct (wrong) partition:
\beq
N_{(II,III)} \gg N_{(III,II)}; \quad N_{(IV,III)} \gg N_{(III,IV)}.
\label{23boxes}
\eeq  
Combining the relations (\ref{firstrow}) and (\ref{23boxes}) and using the definitions (\ref{eq:NC}) and (\ref{eq:NW}),
we conclude that $N_C \gg N_W$, and therefore Step I of the method works relatively well\footnote{This can also be seen 
directly by comparing the green-shaded and the red-shaded boxes of Table~\ref{table:mt2_m2cc_mbl_cuts}.}. 
With our definition for the efficiency (\ref{efficiency}), the classic method based on $M_{T2}^{(b\ell)}$ and $m_{b\ell}$
gives an efficiency of 82\%. Looking back at the left side of Table~\ref{table:mt2_m2cc_mbl_cuts}, we see that for the standard method,
the efficiency is hurt by the relatively large fraction of events which remain unresolved at this stage --- the 
unshaded boxes contain a total of $N_U=6430$ events, or about 35\% of the sample. The situation improves somewhat if we use the
$M_{2CC}^{(b\ell)}$ variable instead of $M_{T2}^{(b\ell)}$: in that case, the right $4\times 4$ table of Table~\ref{table:mt2_m2cc_mbl_cuts}
contains fewer unresolved events ($N_U=4933$, or only 27\%), while the desired relations (\ref{firstrow}) and (\ref{23boxes})
are further enhanced. 
Both of these effects are responsible for increasing the efficiency (\ref{efficiency}) of the new $M_{2CC}^{(b\ell)}$ method
to 85.3\%. Note that in both methods, the majority of the unresolved events are found in the very first diagonal box $(I,I)$,
where both partitions are fully consistent with the kinematics of the assumed event topology. This is why,
in what follows we shall focus our attention on the additional steps of the algorithm which can successfully
classify the remaining unresolved events, especially those in the $(I,I)$ box.

\subsection{Step II: The presence of complex solutions}
\label{sec:complex}





In the second step of the method, one attempts to reconstruct the longitudinal momenta of the invisible particles, by 
enforcing on-shell conditions for a parent particle (MAOS1) or for a relative particle (MAOS4). Since the on-shell conditions
result in quadratic equations, the solutions are not guaranteed to be real. The idea of Step II is to compare the two possible
partitions $P_k$ in terms of the number of complex solution pairs ${\mathbb C}$ for the longitudinal momenta. Since there is a separate calculation for each
decay chain, there are three possible outcomes:
\begin{itemize}
\item ${\mathbb C}=0$. Both decay chains result in real solutions.
\item ${\mathbb C}=1$. Exactly one decay chain gives a pair of complex solutions, while the other decay chain has real solutions.
\item ${\mathbb C}=2$. Both decay chains result in complex solutions.
\end{itemize}
For the purposes of applying Step II of the method, there is no need to distinguish between the cases of ${\mathbb C}=1$ and ${\mathbb C}=2$, 
since the important point is simply that ${\mathbb C}>0$. The action of Step II is the following: 
if one of the partitions $P_k$ gives ${\mathbb C}=0$, while the other has ${\mathbb C}>0$, then
the former (latter) partition is declared to be the correct (wrong) one.

\begin{table}[t]
\centering
\scalebox{1.0}{
\begin{tabular}{|c||c|c|c||c|c|c||c|c|c||c|c|c||c|}
\cline{2-13}
\multicolumn{1}{c||}{}  &\multicolumn{3}{c||}{I}     &  \multicolumn{3}{c||}{II}        & \multicolumn{3}{c||}{III}  & \multicolumn{3}{c||}{IV}  & \multicolumn{1}{c}{}  \\
\cline{2-14}
 \multicolumn{1}{c||}{} 
  &  0  &  1  &  2  &  0 &  1 & 2   & 0 & 1 & 2 & 0 & 1 & 2 & ${\mathbb C}$\\
  \hline
  \hline
  I   & 6194   & 0    & 0 & \cellcolor{green}  910   &  \cellcolor{green}  0  & \cellcolor{green}  0 & \cellcolor{green}  9793  & \cellcolor{green}  0  & \cellcolor{green}  0 & \cellcolor{green}  1020 & \cellcolor{green}  0 & \cellcolor{green}  0 & $P_C$\\
  \cline{2-14}
     & 6194   &  0 & 0 & \cellcolor{green}  0  & \cellcolor{green}  1  & \cellcolor{green}  909   & \cellcolor{green}  0  & \cellcolor{green}  7215 & \cellcolor{green}  2578 & \cellcolor{green}  1020 & \cellcolor{green}  0 & \cellcolor{green}  0 & $P_W$\\ 
   \hline
   \hline
   II   & \cellcolor{red}  0   & \cellcolor{red} 2   & \cellcolor{red} 20  & 0  &  0 & 31 & \cellcolor{green}  0 & \cellcolor{green}   2  & \cellcolor{green}  104  & 0 & 0 & 5  & $P_C$\\
  \cline{2-14}
     & \cellcolor{red} 22  &  \cellcolor{red} 0 & \cellcolor{red} 0  & 0 & 0 & 31  & \cellcolor{green}   0  & \cellcolor{green}  69 & \cellcolor{green}  37  & 5 & 0 & 0 & $P_W$\\ 
    \hline
    \hline
   III   & \cellcolor{red} 0   & \cellcolor{red} 22   & \cellcolor{red} 7  & \cellcolor{red} 0  & \cellcolor{red}  5 & \cellcolor{red} 5 & 0 & 124  & 67  & \cellcolor{red} 0 & \cellcolor{red} 4 & \cellcolor{red} 0 & $P_C$ \\
  \cline{2-14}
     & \cellcolor{red} 29  & \cellcolor{red}  0 & \cellcolor{red} 0  & \cellcolor{red} 0 & \cellcolor{red} 0  & \cellcolor{red} 10  & 0 & 138 & 53 & \cellcolor{red} 4 & \cellcolor{red} 0 & \cellcolor{red} 0& $P_W$\\
    \hline
    \hline
    IV   & \cellcolor{red} 41  & \cellcolor{red} 0 & \cellcolor{red} 0  & 3  & 0   &  0 & \cellcolor{green}  91 & \cellcolor{green}  0 & \cellcolor{green}  0  & 6  & 0 & 0  & $P_C$ \\
  \cline{2-14}
   & \cellcolor{red} 41   & \cellcolor{red}  0 & \cellcolor{red} 0  & 0 & 0  & 3  & \cellcolor{green}  0 & \cellcolor{green}  66 & \cellcolor{green}  25  & 6 & 0 & 0  & $P_W$ \\  
   \hline
   \hline
\end{tabular}
}
\caption{\label{table:tsub_cut} 
Classification of events according to the complexity of the solutions for the longitudinal invisible momenta in the case of 
MAOS1($b\ell$;$m_t$), i.e., 
the transverse momenta of the neutrinos are fixed during the minimization of $M_{T2}^{(b\ell)}$, while
the longitudinal components are obtained from the on-shell conditions for the top quarks on each side.
The table is organized by quadrant counts $N_{(I,J)}$ as in Table~\ref{table:mt2_m2cc_mbl_cuts}.  
For each $(I,J)$ quadrant pair, we further classify the events according to the number of decay sides ${\mathbb C}$ with
complex solutions for the longitudinal momenta, and the table lists the corresponding values of $N_{(I,J)}^{({\mathbb C})}$.  
This is done separately for the correct partition $P_C$ (upper rows) and for the wrong partition $P_W$ (lower rows).
}
\end{table}

\begin{table}[t]
\centering
\scalebox{1.0}{
\begin{tabular}{|c||c|c|c||c|c|c||c|c|c||c|c|c||c|}
\cline{2-13}
\multicolumn{1}{c||}{}   
 &\multicolumn{3}{c||}{I}     &  \multicolumn{3}{c||}{II}        & \multicolumn{3}{c||}{III}  & \multicolumn{3}{c||}{IV} & \multicolumn{1}{c}{}   \\
\cline{2-14}
\multicolumn{1}{c||}{}  &  0  &  1  &  2  &  0 &  1 & 2   & 0 & 1 & 2 & 0 & 1 & 2 & ${\mathbb C}$\\
  \hline
  \hline
  I   & 5005   & 1086    & 103  & \cellcolor{green} 582   &  \cellcolor{green} 299  &  \cellcolor{green}29 &  \cellcolor{green}8245  &  \cellcolor{green}1497  &  \cellcolor{green}51 &  \cellcolor{green}908 &  \cellcolor{green}109 &  \cellcolor{green}3 & $P_C$\\
  \cline{2-14}
     & 4094   &  2074 & 26 &  \cellcolor{green}136  &  \cellcolor{green}751  &  \cellcolor{green}23   &  \cellcolor{green}3893  &  \cellcolor{green}5817 &  \cellcolor{green}83  &  \cellcolor{green}661 &  \cellcolor{green}359 &  \cellcolor{green}0 & $P_W$\\ 
   \hline
   \hline
   II  & \cellcolor{red} 7   & \cellcolor{red} 14   & \cellcolor{red} 1  & 12 &  16 & 3 &  \cellcolor{green}58 &  \cellcolor{green} 46  &  \cellcolor{green}2  & 2 & 3 & 0 & $P_C$   \\
  \cline{2-14}
    & \cellcolor{red} 9   & \cellcolor{red}  13 & \cellcolor{red} 0  & 5 & 24  & 2  &  \cellcolor{green}26  &  \cellcolor{green}79 &  \cellcolor{green}1  & 1 & 4 & 0 & $P_W$ \\ 
    \hline
    \hline
   III  & \cellcolor{red} 12   & \cellcolor{red} 17   & \cellcolor{red} 0  & \cellcolor{red} 5   & \cellcolor{red}  5 & \cellcolor{red} 0 & 84 & 106  & 1  & \cellcolor{red} 3 & \cellcolor{red} 1 & \cellcolor{red} 0 & $P_C$  \\
  \cline{2-14}
    & \cellcolor{red} 22   & \cellcolor{red}  7 & \cellcolor{red} 0  & \cellcolor{red} 3 & \cellcolor{red} 6  & \cellcolor{red} 1  & 96 & 95 & 0  & \cellcolor{red} 3 & \cellcolor{red} 1 & \cellcolor{red} 0 & $P_W$ \\
    \hline
    \hline
    IV  & \cellcolor{red} 36  & \cellcolor{red} 5 & \cellcolor{red} 0  & 3  & 0   &  0 &  \cellcolor{green}78 &  \cellcolor{green}13 &  \cellcolor{green}0  & 6  & 0 & 0 & $P_C$   \\
  \cline{2-14}
     & \cellcolor{red} 31   & \cellcolor{red}  10 & \cellcolor{red} 0  & 1 & 2  & 0  &  \cellcolor{green}58 &  \cellcolor{green}33 &  \cellcolor{green}0  & 6 & 0 & 0 & $P_W$\\  
   \hline
   \hline
\end{tabular}
}
\caption{\label{table:wsub_cut} 
The same as Table \ref{table:tsub_cut}, but for the case of MAOS4($b\ell$;$m_W$), i.e., 
the transverse momenta of the neutrinos are still obtained from the minimization of $M_{T2}^{(b\ell)}$, 
but now the longitudinal components are computed from the on-shell conditions for the $W$-bosons instead.
} 
\end{table}

As already discussed in section~\ref{sec:momentum}, there are several ways to implement the MAOS idea and compute 
longitudinal invisible momenta. Tables~\ref{table:tsub_cut} and \ref{table:wsub_cut} show results for the cases of
MAOS1($b\ell$;$m_t$) and  MAOS4($b\ell$;$m_W$), respectively\footnote{The former was the method used in Ref.~\cite{Choi:2011ys}, 
but the latter is in principle a viable option as well.}. Similarly to Table~\ref{table:mt2_m2cc_mbl_cuts},
each table is organized by quadrants, and within each cell $(I,J)$ we show the number of events $N_{(I,J)}^{({\mathbb C})}$
with a given value of ${\mathbb C}$, for the correct partition (upper rows) and the wrong partition (lower rows).
 
Let us first focus on Table~\ref{table:tsub_cut}, which shows several interesting trends. First, recall the motivation behind Step II --- one was hoping to find that 
the correct partition $P_C$ would always give real solutions (${\mathbb C}=0$), while the wrong partition $P_W$ would always lead to complex solutions (${\mathbb C}>0$).
Table~\ref{table:tsub_cut} reveals that this expectation is indeed true, but only for certain quadrant pairs: $(I,II)$, $(I,III)$, $(IV,II)$ and $(IV,III)$. For those cases, Step II would be 
able to perfectly resolve the combinatorial ambiguity, and it seems that the method works as designed. Unfortunately, three out of these four 
quadrant pairs are already shaded in green (see Table~\ref{table:quadrant}), which means that those events were already perfectly resolved by Step I, 
thus the additional benefit from Step II for those three quadrant pairs is exactly zero. As for the fourth quadrant pair, $(IV,II)$, it is very sparsely populated, 
and furthermore, any benefit there would be offset by the negative effects from the symmetric case of $(II,IV)$, where the results are 
contrary to our expectations above --- now it is the correct partition $P_C$ which leads to complex solutions\footnote{The fact that the correct partition 
$P_C$ may result in complex momenta should not be surprising --- this can be due to finite width and off-shell effects. As a sanity check, we have verified that 
in the zero-width limit only the first row in Table~\ref{table:tsub_cut} has any non-zero entries, while the second, third and fourth rows are empty.}.
(The same phenomenon is observed for the other three symmetric pairs as well --- see the red-shaded boxes corresponding to 
$(II,I)$, $(III,I)$, and $(III,IV)$, where it is again the correct partition $P_C$ which has ${\mathbb C}>0$.)

Thus we conclude that for the events which were already resolved at Step I (the green-shaded and red-shaded cells in Table~\ref{table:tsub_cut}), 
Step II does not bring anything new --- its results are either fully correlated with Step I (as for the quadrant pairs $(I,II)$, $(I,III)$ and $(IV,III)$ 
and their red-shaded symmetric partners on the other side of the diagonal), or inconclusive, since the two partitions behave identically 
(e.g., for quadrant pairs $(I,IV)$ and $(II,III)$ and their partners). Therefore, we need to concentrate on the unresolved events in the unshaded cells in 
Table~\ref{table:tsub_cut}, since those were precisely the events which Step II was meant to address. Unfortunately, we observe 
that in the unshaded cells along the diagonal in Table~\ref{table:tsub_cut}, the two partitions lead to the same result and cannot be discriminated,
while the remaining two cells $(II,IV)$ and $(IV,II)$ were already discussed above --- their statistics is too low, and they tend to cancel each other out, 
thus they will not appreciably affect the overall efficiency.

Based on the results from Table~\ref{table:tsub_cut}, we conclude that if one were to apply Step II based on the MAOS1($b\ell$;$m_t$) method, 
which was used in Ref.~\cite{Choi:2011ys}, there would be no additional benefit beyond Step I, and therefore Step II is unnecessary and can be eliminated.  
However, this still leaves open the question whether some modified version of Step II can still be useful, e.g., applying a different MAOS scheme like  
MAOS4($b\ell$;$m_W$) (see Table~\ref{table:wsub_cut}), or perhaps using one of the CMAOS schemes based on the $M_{2CC}$ variables 
(see Tables~\ref{table:tsub_cut_m2cc} and \ref{table:wsub_cut_m2cc}). 
%
\begin{table}[t]
\centering
\scalebox{1.0}{
\begin{tabular}{|c||c|c|c||c|c|c||c|c|c||c|c|c||c|}
\cline{2-13}
 \multicolumn{1}{c||}{}   &\multicolumn{3}{c||}{I}     &  \multicolumn{3}{c||}{II}        & \multicolumn{3}{c||}{III}  & \multicolumn{3}{c||}{IV}  &  \multicolumn{1}{c}{}    \\
\cline{2-14}
 \multicolumn{1}{c||}{}  &  0  &  1  &  2  &  0 &  1 & 2   & 0 & 1 & 2 & 0 & 1 & 2   &  ${\mathbb C}$  \\
  \hline
  \hline
  I  & 4494   & 0    & 0 &  \cellcolor{green}2317   &   \cellcolor{green}0  &  \cellcolor{green}0 &  \cellcolor{green}10222  &  \cellcolor{green}0  &  \cellcolor{green}0 &  \cellcolor{green}217 &  \cellcolor{green}0 &  \cellcolor{green}0 & $P_C$ \\
  \cline{2-14}
    & 4494   &  0 & 0 &  \cellcolor{green}348 &  \cellcolor{green}729  &  \cellcolor{green}1240   &  \cellcolor{green}62 &  \cellcolor{green}5482 &  \cellcolor{green}4678 &  \cellcolor{green}217 &  \cellcolor{green}0 &  \cellcolor{green}0 & $P_W$ \\ 
   \hline
   \hline
   II   &  \cellcolor{red}62  &  \cellcolor{red}70   &  \cellcolor{red}36  & 63  &  64 & 51 &  \cellcolor{green}184 &  \cellcolor{green} 168  &  \cellcolor{green}128  & 2 & 1 & 2 & $P_C$\\
  \cline{2-14}
    &  \cellcolor{red}168  &  \cellcolor{red} 0 &  \cellcolor{red}0   & 21 & 53 & 104  &  \cellcolor{green} 4 &  \cellcolor{green}211 &  \cellcolor{green}265  & 5 & 0 & 0 & $P_W$\\ 
    \hline
    \hline
   III   &  \cellcolor{red}2   &  \cellcolor{red}20   &  \cellcolor{red}15  &  \cellcolor{red}4 &   \cellcolor{red}9 &  \cellcolor{red}13 & 7 & 138  & 106  &  \cellcolor{red}0 &  \cellcolor{red}1 &  \cellcolor{red}0 & $P_C$ \\
  \cline{2-14}
     &  \cellcolor{red}37  &   \cellcolor{red}0 &  \cellcolor{red}0  &  \cellcolor{red}3 &  \cellcolor{red}7  &  \cellcolor{red}16  & 0 & 150 & 101 &  \cellcolor{red}1 &  \cellcolor{red}0 &  \cellcolor{red}0 & $P_W$\\
    \hline
    \hline
    IV   &  \cellcolor{red}17 &  \cellcolor{red}0 &  \cellcolor{red}0  & 3  & 0   &  0 &  \cellcolor{green}38 &  \cellcolor{green}0 &  \cellcolor{green}0  & 2  & 0 & 0  & $P_C$ \\
  \cline{2-14}
    &  \cellcolor{red}17   &  \cellcolor{red} 0 &  \cellcolor{red}0  & 0 & 1  & 2  &  \cellcolor{green}0 &  \cellcolor{green}25 &  \cellcolor{green}13  & 2 & 0 & 0 & $P_W$\\  
   \hline
   \hline
\end{tabular}
}
\caption{\label{table:tsub_cut_m2cc} 
The same as Table \ref{table:tsub_cut}, but for the case of CMAOS1($b\ell$;$m_t$), i.e., the transverse momenta of the neutrinos
are now fixed from the minimization of $M_{2CC}^{(b\ell)}$ instead of $M_{T2}^{(b\ell)}$, and then the longitudinal components 
are again obtained from the on-shell conditions for the top quarks on each side. Here the quadrants are defined in terms 
of $M_{2CC}^{(b\ell)}$ and $m_{b\ell}$ and correspond to the right $4\times 4$ table in Table~\ref{table:mt2_m2cc_mbl_cuts}. }
\end{table}
%
\begin{table}[t]
\centering
\scalebox{1.0}{
\begin{tabular}{|c||c|c|c||c|c|c||c|c|c||c|c|c||c|}
\cline{2-13}
 \multicolumn{1}{c||}{}   &\multicolumn{3}{c||}{I}     &  \multicolumn{3}{c||}{II}        & \multicolumn{3}{c||}{III}  & \multicolumn{3}{c||}{IV}  &  \multicolumn{1}{c}{}  \\
\cline{2-14}
 \multicolumn{1}{c||}{}  &  0  &  1  &  2  &  0 &  1 & 2   & 0 & 1 & 2 & 0 & 1 & 2 & ${\mathbb C}$\\
  \hline
  \hline
  I   & 3840   & 497    & 157  &  \cellcolor{green}1810   &  \cellcolor{green} 385  &  \cellcolor{green}122 &  \cellcolor{green}9089  &  \cellcolor{green}931  &  \cellcolor{green}202 &  \cellcolor{green}214 &  \cellcolor{green}3 &  \cellcolor{green}0  & $P_C$\\
  \cline{2-14}
     & 3652   &  612 & 230 &  \cellcolor{green}397  &  \cellcolor{green}879 &  \cellcolor{green}1041   &  \cellcolor{green} 3005  &  \cellcolor{green}3132 &  \cellcolor{green}4085 &  \cellcolor{green}217 &  \cellcolor{green}0 &  \cellcolor{green}0 & $P_W$\\ 
   \hline
   \hline
   II   &  \cellcolor{red}52   &  \cellcolor{red}66   &  \cellcolor{red}50  & 65 &  84 & 29 &  \cellcolor{green}228 &   \cellcolor{green}194  &  \cellcolor{green}58  & 1 & 2 & 2 & $P_C$ \\
  \cline{2-14}
     &  \cellcolor{red}113   &  \cellcolor{red} 37 &  \cellcolor{red}18  & 30 & 81  & 67  &  \cellcolor{green}121 &  \cellcolor{green}143 &  \cellcolor{green}216 & 5 & 0 & 0 & $P_W$\\ 
    \hline
    \hline
   III  &  \cellcolor{red}14   &  \cellcolor{red}13   &  \cellcolor{red}10  &  \cellcolor{red}11   &  \cellcolor{red} 9 &  \cellcolor{red}6 & 101 & 76  & 74  &  \cellcolor{red}1 &  \cellcolor{red}0 &  \cellcolor{red}0  & $P_C$\\
  \cline{2-14}
     &  \cellcolor{red}33  &   \cellcolor{red}3 &  \cellcolor{red}1  &  \cellcolor{red}10 &  \cellcolor{red}11  &  \cellcolor{red}5 & 90 & 88 & 73  &  \cellcolor{red}1 &  \cellcolor{red}0 &  \cellcolor{red}0 & $P_W$\\
    \hline
    \hline
    IV   &  \cellcolor{red}17  &  \cellcolor{red}0 &  \cellcolor{red}0  & 3  & 0   &  0 &  \cellcolor{green}38 &  \cellcolor{green}0 &  \cellcolor{green}0  & 2  & 0 & 0  & $P_C$ \\
  \cline{2-14}
     &  \cellcolor{red}17   &  \cellcolor{red} 0 &  \cellcolor{red}0  & 1 & 0  & 2  &  \cellcolor{green}18 &  \cellcolor{green}12 &  \cellcolor{green}8  & 2 & 0 & 0 & $P_W$\\  
   \hline
   \hline
\end{tabular}
}
\caption{\label{table:wsub_cut_m2cc} 
The same as Table \ref{table:wsub_cut}, but for the case of CMAOS4($b\ell$;$m_W$), i.e., 
the transverse momenta of the neutrinos are now obtained from the minimization of $M_{2CC}^{(b\ell)}$,
then the longitudinal components are computed from the on-shell conditions for the $W$-bosons as before.
}
\end{table}
%
But before we discuss these options, it will be useful to understand the results from Table~\ref{table:tsub_cut} from 
a physics point of view. A careful inspection of Table~\ref{table:tsub_cut} reveals that its content can be summarized as follows:
for any partition $P_k$, quadrants I and IV produce only real solutions, while quadrants II and III lead to only complex solutions.
This means that the existence of complex solutions is correlated with the $x$-axis variable of Fig.~\ref{fig:quadrant}
($M_{T2}^{(b\ell)}$), and not with the $y$-axis variable ($m_{b\ell}$). This is easy to understand: 
$m_{b\ell}$ is formed from visible particle momenta only, and is not directly related to any invisible momenta.
Therefore, a violation of the $m_{b\ell}$ kinematic endpoint by itself does not imply unphysical invisible momenta.
On the other hand, the physical meaning of $M_{T2}^{(b\ell)}$ is the lowest possible mass of the parent particle, in this case the top quark.
If the value of $M_{T2}^{(b\ell)}$ strictly violates the kinematic endpoint $m_t$, i.e., $M_{T2}^{(b\ell)}>m_t$, then enforcing 
the on-shell condition for the top quark will necessarily result in unphysical (complex) values for the momenta.
In particular, quadrants II and III, in which the $M_{T2}^{(b\ell)}$ endpoint is violated by definition, 
will always produce complex momenta. 

While the above logic helps to understand the results from Table~\ref{table:tsub_cut}, it does not carry over directly to the
case of MAOS4($b\ell$;$m_W$) shown in Table~\ref{table:wsub_cut}, since now we are enforcing an on-shell condition 
for a different (relative) particle. The trends which we previously observed in Table~\ref{table:tsub_cut} are still noticeable,
but they are not so clear cut. Nevertheless, if we focus on the unresolved events after Step I (the unshaded cells in Table~\ref{table:wsub_cut}), 
we again see that Step II does not do particularly great on those events. The largest effect is in the $(I,I)$ cell, where
the efficiency for resolving the correct partition is $57.4\%$, which is slightly better than a coin flip. 
Adding up the results from all previously unresolved cells, we find that if we were to perform Step II with the MAOS4($b\ell$;$m_W$) 
version of the method instead of the MAOS1($b\ell$;$m_t$) option used in Ref. \cite{Choi:2011ys}, the overall efficiency would increase to 84.5\%, 
which is still worse than the result (85.3\%) found in section~\ref{sec:step1} with the improvements in Step I alone,
taking advantage of $M_{2CC}^{(b\ell)}$ (see the right $4\times 4$ table in Table~\ref{table:mt2_m2cc_mbl_cuts}).

Tables~\ref{table:tsub_cut_m2cc} and \ref{table:wsub_cut_m2cc} show results from two similar exercises where we use the
corresponding CMAOS methods, i.e., fixing the transverse components of the invisible momenta from $M_{2CC}^{(b\ell)}$
instead of  $M_{T2}^{(b\ell)}$, then applying the on-shell conditions for the parent (top quark) or relative ($W$-boson) particle.
As shown previously in Figs.~\ref{fig:momentum1d}-\ref{fig:momentum2d}, the $M_{2CC}^{(b\ell)}$ variable generally
provides a more accurate estimate of the individual transverse momentum components for the invisible particles, and one might
hope that incorporating $M_{2CC}^{(b\ell)}$ somehow into the Step II algorithm would improve the performance.
However, Table~\ref{table:tsub_cut_m2cc} shows that the improvement in the case of CMAOS1($b\ell$;$m_t$)
is very marginal --- the efficiency increases from 85.3\% after Step I to 85.4\% after Step II.
The effect is slightly better in Table~\ref{table:wsub_cut_m2cc}, which uses the CMAOS4($b\ell$;$m_W$) option ---
there the efficiency increases from 85.3\% after Step I to 85.9\% after Step II. However, even this increase is too small to justify 
the presence of Step II --- as we shall see later on, there exist other, much more effective techniques.
Therefore, as a final summary of this subsection, we conclude that Step II can be safely dropped altogether, since 
its results are largely correlated with Step I.

\subsection{Step III and possible variations}
\label{sec:step3}

In this subsection we shall discuss different possible options for the third step of the method and investigate their performance.
To recap the situation: when we used $M_{T2}^{(b\ell)}$ and $m_{b\ell}$ at Step I, we ended up with 
$N_C=11,920$ correctly resolved events, $N_W=106$ incorrectly resolved events, and $N_U=6,430$ unresolved events,
for an efficiency of 82\% (see the top of the middle column in Table~\ref{table:efficiency}).
\begin{table}[t]
\centering
\def\arraystretch{1.2}
\scalebox{0.92}{
\begin{tabular}{||c||c|c|c||c|c|c||}
\hline
                 & \multicolumn{6}{c||}{Algorithm for Step I} \\ \cline{2-7}  
                & \multicolumn{3}{c||}{$M_{T2}^{(b\ell)}$ and $m_{b\ell}$} & \multicolumn{3}{c||}{$M_{2CC}^{(b\ell)}$ and $m_{b\ell}$}   \\ \cline{2-7}  
                & Correct  & Wrong & Unresolved & Correct  & Wrong & Unresolved \\ \hline
Step I: Quadrant counts & 11,920  & 106 & 6,430 & 13,274  & 249 & 4,933  \\ \cline{2-7}  
efficiency   & \multicolumn{3}{c||}{$\varepsilon=82.0\%$} & \multicolumn{3}{c||}{$\varepsilon=85.3\%$} \\ \hline\hline
Remaining unresolved events &  \multicolumn{3}{c||}{$6,430$} & \multicolumn{3}{c||}{$4,933$}  \\ \hline\hline
$\Delta T_i$ method &  3,445  &   1,573  &   1,412  &     2,820  &  1,462   &  651  \\  \cline{2-7}  
cumulative efficiency  &  \multicolumn{3}{c||}{$\varepsilon=87.1\%$} & \multicolumn{3}{c||}{$\varepsilon=89\%$} \\ \hline\hline
$\Delta T_2$ type cut alone &  3,160  &   3,270  &   ---  &     2,426  &  2,507   &   ---  \\  \cline{2-7}  
cumulative efficiency  &  \multicolumn{3}{c||}{$\varepsilon=81.7\%$} & \multicolumn{3}{c||}{$\varepsilon=85.1\%$}\\ \hline\hline
$\Delta T_3$ and $\Delta T_4$ alone &  4,101  &   1,624  &   705  &     2,868  &  1,328   &  737  \\  \cline{2-7}  
cumulative efficiency  &  \multicolumn{3}{c||}{$\varepsilon=88.7\%$} & \multicolumn{3}{c||}{$\varepsilon=89.5\%$} \\ \hline\hline
\cellcolor{yellow}$m_t^{reco}$ with MAOS4($b\ell$;$m_W$)  &  2,392  &   1,371  &   2,667  &   2,212 & 1,238    &  1,483  \\  \cline{2-7}  
cumulative efficiency &  \multicolumn{3}{c||}{$\varepsilon=84.8\%$} & \multicolumn{3}{c||}{$\varepsilon=87.9\%$}\\ \hline
\cellcolor{yellow}$m_t^{reco}$ with MAOS1($\ell$;$m_W$) &  4,186  &   2,188  &   56  &   3,033 & 1,860    &  40  \\  \cline{2-7}  
cumulative efficiency   &  \multicolumn{3}{c||}{$\varepsilon=87.4\%$} & \multicolumn{3}{c||}{$\varepsilon=88.5\%$}\\ \hline
\cellcolor{yellow}$m_t^{reco}$ with CMAOS4($b\ell$;$m_W$) &  2,262  &   1,307  &   2,861  &   2,122 & 1,228    &  1,583  \\  \cline{2-7}  
cumulative efficiency   &  \multicolumn{3}{c||}{$\varepsilon=84.6\%$} & \multicolumn{3}{c||}{$\varepsilon=87.7\%$}\\ \hline
\cellcolor{yellow}$m_t^{reco}$ with CMAOS1($\ell$;$m_W$)  &  2,995  &   1,790  &   1,645  &   2,565  & 1,617   & 751   \\  \cline{2-7}  
cumulative efficiency  &  \multicolumn{3}{c||}{$\varepsilon=85.3\%$} & \multicolumn{3}{c||}{$\varepsilon=87.9\%$}\\ \hline\hline
\cellcolor{yellow}$m_W^{reco}$ with MAOS1($b\ell$;$m_t$)&  4,344  &   1,856  &   230  &   3,044 & 1,546    &  343  \\  \cline{2-7}  
cumulative efficiency   &  \multicolumn{3}{c||}{$\varepsilon=88.8\%$} & \multicolumn{3}{c||}{$\varepsilon=89.3\%$}\\ \hline
\cellcolor{yellow}$m_W^{reco}$ with MAOS4($\ell$;$m_t$)  &  1,627  &   792  &   4,011  &   1,505 & 760    &  2,688  \\  \cline{2-7}  
cumulative efficiency    &  \multicolumn{3}{c||}{$\varepsilon=84.3\%$} & \multicolumn{3}{c||}{$\varepsilon=87.3\%$}\\ \hline
\cellcolor{yellow}$m_W^{reco}$ with CMAOS1($b\ell$;$m_t$)  &  3,328  &   1,583 &   1,519  &   3,044 & 1,459    &  430 \\  \cline{2-7}  
cumulative efficiency  &  \multicolumn{3}{c||}{$\varepsilon=86.7\%$} & \multicolumn{3}{c||}{$\varepsilon=89.6\%$}\\ \hline
\cellcolor{yellow}$m_W^{reco}$ with CMAOS4($\ell$;$m_t$) &  1,922  &   947 &   3,561 &   1,870 & 928    &  2,135  \\  \cline{2-7}  
cumulative efficiency  &  \multicolumn{3}{c||}{$\varepsilon=84.7\%$} & \multicolumn{3}{c||}{$\varepsilon=87.3\%$}\\
\hline\hline
\end{tabular}
}
\caption{\label{table:efficiency} 
Efficiencies for selecting the correct partitioning (as defined in Eq.~(\ref{efficiency})) for several different procedures which 
use the known values of the top mass $m_t$ or the $W$-boson mass $m_W$. 
} 
\end{table}
If, on the other hand, we choose to use $M_{2CC}^{(b\ell)}$ and $m_{b\ell}$ at Step I, 
we obtain $N_C=13,274$ correctly resolved events, $N_W=249$ incorrectly resolved events, and $N_U=4,933$ unresolved events,
for an efficiency of 85.3\% (see the top of the right column in Table~\ref{table:efficiency}).
Then in section~\ref{sec:complex} we showed that Step II does not add much and can be ignored.
This brings us to Step III, whose purpose is to further classify the remaining unresolved events after Step I (6,430 and 4,933, respectively)
on a statistical basis, using suitable discriminating variables. 

We begin by reviewing the method suggested in Ref.~\cite{Choi:2011ys}, which introduced several kinematic variables, $T_i$, $i=1,\ldots,4$.
These variables were designed so that their values tend to be {\em larger} for the case of the wrong partition, i.e.
\beq
T_i (P_W) > T_i(P_C).
\label{Tihierarchy}
\eeq
While it is not guaranteed that (\ref{Tihierarchy}) will be true in every single event, if it holds for the majority of the events, one can 
attempt to identify the correct partition $P_C$ by declaring \cite{Choi:2011ys}
\beq
P_C = \left\{ 
\begin{array}{l} 
P_1,\ \text{if}\ \Delta T_i(P_2,P_1) >0;  \\
P_2,\ \text{if}\ \Delta T_i(P_2,P_1) <0;
\end{array}
\right. 
\label{pick}
\eeq
where 
\beq
\Delta T_i(P_2,P_1) \equiv T_i(P_2)- T_i(P_1).
\label{DeltaTidef}
\eeq 
In the case of several good variables $T_i$, one can generalize (\ref{pick}) by choosing 
the correct partition $P_C$ to be the partition $P_1$ ($P_2$) if the {\em majority} of the 
quantities $\Delta T_i(P_2,P_1)$ are positive (negative). In the following, we shall refer to this procedure as the 
``$\Delta T_i$ method" \cite{Choi:2011ys}.

The $T_i$ variables considered in Ref. \cite{Choi:2011ys} were the following: 
\begin{eqnarray}
T_1 (P_k) &\equiv& \max_j\{m^{(j)}_{b\ell}\} (P_k) \, ,   \label{T1def}\\ [2mm]
T_2 (P_k) &\equiv& M_{T2}^{(b\ell)} (P_k) \, , \label{T2def}\\ [2mm]
T_3 (P_k)  &\equiv& \sum_{\substack{j=1,2; \\ \alpha=+,-}} | m_{t}^{reco} (j, \alpha) - m_t| (P_k) \, , \label{T3def}\\ [2mm]
T_4 (P_k)  &\equiv& \sum_{\substack{j=1,2; \\ \alpha=+,-}} | m_{W}^{reco} (j, \alpha) - m_W| (P_k) \, , \label{T4def} 
\end{eqnarray}
where $P_k$ is one of two partitions and the index $j$ labels the two decay chains in Fig.~\ref{fig:decaysubsystem}.
The variable $m_{t}^{reco}$ ($m_{W}^{reco}$) is the reconstructed mass of the top quark (the $W$-boson) 
with a MAOS-type method which uses the $W$-boson mass (the top mass) as an input. 
Since the longitudinal invisible momenta are obtained from a quadratic equation, in general there are 
two solutions, labelled by $\alpha=\pm$, corresponding to the two signs in front of the discriminant.
Thus in each event one can obtain four reconstructed top quark masses, $m_{t}^{reco} (j, \alpha)$,
and four reconstructed $W$-boson masses, $m_{W}^{reco} (j, \alpha)$. The idea behind the 
$T_3$ and $T_4$ variables in (\ref{T3def}) and (\ref{T4def}) is to compare those reconstructed values 
to the true values $m_t$ and $m_W$, respectively. For the correct partition $P_C$, on average one might expect
to find the reconstructed values closer to the true ones, in agreement with (\ref{Tihierarchy}). 

In Table \ref{table:efficiency} we test several options for discrimination variables which can be applied at Step III. 
Our benchmark is the $\Delta T_i$ method of Ref.~\cite{Choi:2011ys}, which made use of
only three variables, $T_2$, $T_3$ and $T_4$, since the fourth one, $T_1$, was found to be significantly correlated  with $T_2$.
For consistency, whenever the quadrants from Step I are defined in terms of $M_{2CC}^{(b\ell)}$ (right column in Table~\ref{table:efficiency}),
we shall replace (\ref{T2def}) with $T_2=M_{2CC}^{(b\ell)}$. Table \ref{table:efficiency} reports results from both 
versions of the $\Delta T_i$ method --- we see that the overall efficiency can be further improved to 87.1\% and 89\%, respectively.
The observed improvement at Step III is due to correctly categorizing (at the rate of about 2:1) the majority of the remaining unresolved events --- see
Table~\ref{table:effciency_cut}, which gives the breakdown among the individual ``unresolved" cases from Table~\ref{tab:decision}.
\begin{table}[t]
\centering
\scalebox{1.05}{
\begin{tabular}{||c||c|c|c|c||c|c|c|c||}
\hline
Quadrants & \multicolumn{8}{c||}{  Algorithm for Step I } \\ \cline{2-9}  
with unresolved             & \multicolumn{4}{c||}{$M_{T2}^{(b\ell)}$ and $m_{b\ell}$} & \multicolumn{4}{c||}{$M_{2CC}^{(b\ell)}$ and $m_{b\ell}$}   \\ \cline{2-9}  
events & Total & C & W & U  & Total & C & W & U \\ \hline \hline
(I, I) & 6,194 & 3,266 & 1,521 &  1,407 & 4,494 &  2,471 & 1,308  & 715    \\\hline
(II, II) & 31 & 26 & 5 & -  & 178  & 147  & 26  &  5          \\ \hline
(III, III) & 191 & 149 & 37 & 5 & 251  &  185 & 55  & 11 \\ \hline
(IV, IV) & 6 & 1 & 5 & - &  2 &  0 & 2  &  -  \\ \hline
(II, IV) $\oplus$ (IV, II) & 8 & 3 & 5   & -  & 8   &  3 & 3    & 2 \\ \hline \hline
Total after Step III  &  6,430  & 3,445  &  1,573  & 1,412 & 4,933 & 2,820   & 1,462 & 651 \\ \hline
\end{tabular}
}
\caption{\label{table:effciency_cut} Breakdown of the events from Table~\ref{table:efficiency} which remained unresolved after Step I (6,430 and 4,933, respectively). The table shows the effect of applying the $\Delta T_i$ method at Step III, for the case of  
$T_2=M_{T2}^{(b\ell)}$ (left) and $T_2=M_{2CC}^{(b\ell)}$ (right). 
The resulting cumulative efficiencies are 87.1\% and 89\%, as shown in Table \ref{table:efficiency}.
} 
\end{table}
In spite of this progress, we also notice that a certain number of events (1,412  and 651, correspondingly) still remain unresolved. 
At first glance, this seems odd, since the $\Delta T_i$ method uses an odd number of variables, so for each event,
there should be a clear winner between the two candidate partitions $P_1$ and $P_2$. However, recall that the longitudinal momentum reconstruction 
sometimes results in complex solutions, in which case the corresponding variable $T_3$ or $T_4$ is undefined.\footnote{This will 
become more evident when inspecting the normalization of the plots in Fig.~\ref{fig:step3} below.}
Thus the remaining unresolved events after Step III are those where only two of the three $T_i$ variables were 
calculated, and each preferred a different partition $P_k$.

Table~\ref{table:effciency_cut} demonstrates that Step III was relatively successful. Nevertheless, in the remainder of this section
we shall investigate whether further improvements at the level of Step III are still possible. Let us begin by studying the benefit from
each individual variable, $T_2$, $T_3$ and $T_4$, used in the $\Delta T_i$ algorithm. Following Ref.~\cite{Choi:2011ys},
in Fig.~\ref{fig:step3} we show distributions of the ``ordered" differences 
\beq
\Delta T_i (P_W, P_C) \equiv  T_i (P_W) - T_i (P_C)
\label{DeltaTiordered}
\eeq
for the three variables $T_2$ (left panels), $T_3$ (middle panels) and $T_4$ (right panels), 
where we use MC truth information to make sure that we subtract the variables in the order indicated in (\ref{DeltaTiordered}).
\begin{figure}[t]
\centering
\includegraphics[width=5cm]{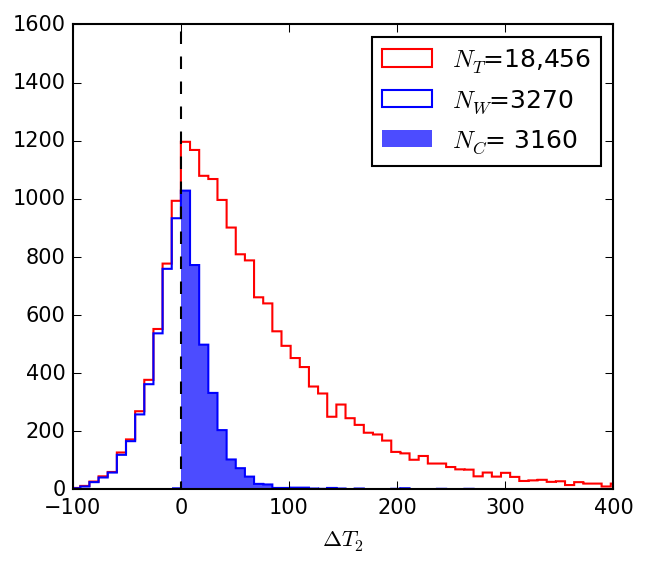} 
\includegraphics[width=5cm]{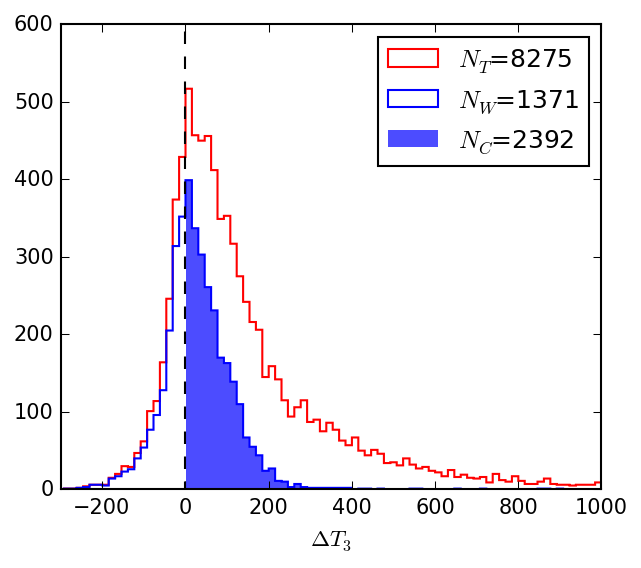} 
\includegraphics[width=5cm]{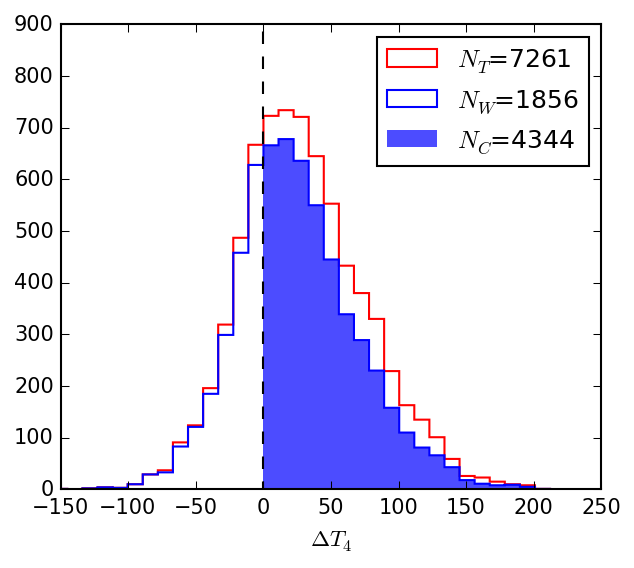} \\
\includegraphics[width=5cm]{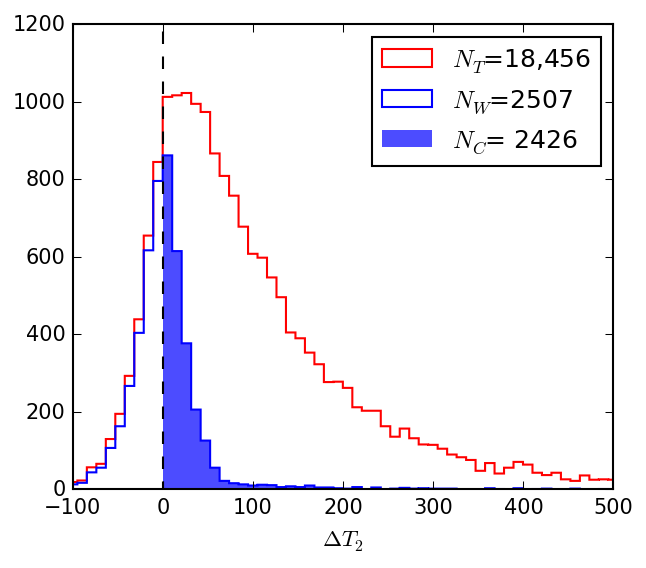} 
\includegraphics[width=5cm]{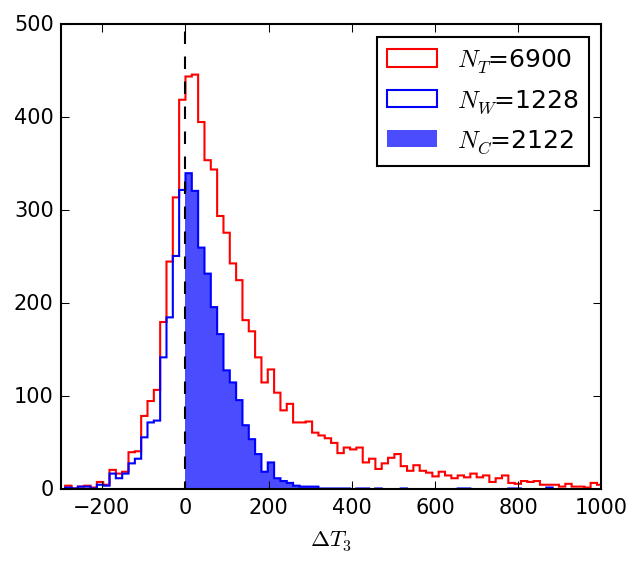} 
\includegraphics[width=5cm]{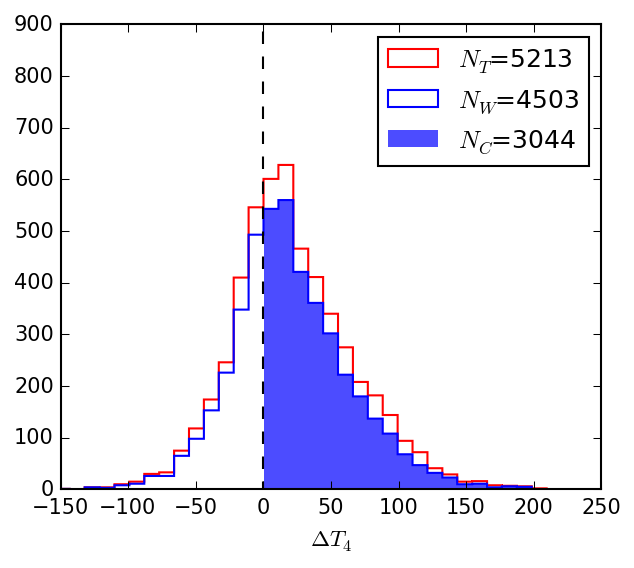} 
\caption{\label{fig:step3} Distributions of the ordered differences (\ref{DeltaTiordered}) for  
$T_2$ (left), $T_3$ (middle) and $T_4$ (right). The red distributions show all events while the blue distributions
only include the events which remain unresolved after Step I.
For the MAOS-type plots in the top row, the transverse invisible momenta were obtained with the help of $M_{T2}^{(b\ell)}$,
while for the CMAOS-type plots in the bottom row the transverse invisible momenta were fixed from $M_{2CC}^{(b\ell)}$.
In the middle and right panels we only plot events with real solutions for the longitudinal momenta.
The shaded (unshaded) portions of the histograms represent events which will be correctly (incorrectly) resolved 
with that particular $\Delta T_i$ variable alone.}
\end{figure}
For the plots in the top (bottom) row of Fig.~\ref{fig:step3}, the transverse invisible momenta were obtained with the help of the
$M_{T2}^{(b\ell)}$ ($M_{2CC}^{(b\ell)}$) variable. Plotting in terms of the ordered difference (\ref{DeltaTiordered}) is very useful, 
since it allows us to see how often the expected relationship (\ref{Tihierarchy}) holds: the difference  (\ref{DeltaTiordered}) 
is positive (negative) if (\ref{Tihierarchy}) is satisfied (violated). Thus, by applying the prescription (\ref{pick}) for a given variable $T_i$, 
we shall correctly resolve all events with positive values of  $\Delta T_i (P_W, P_C)$ (the shaded portions of the distributions in
Fig.~\ref{fig:step3}), and we shall wrongly resolve the events with negative values of  $\Delta T_i (P_W, P_C)$ 
(the unshaded portions of the distributions in Fig.~\ref{fig:step3}). By comparing the areas of the shaded and unshaded
portions of each distribution, we can judge the discrimination power of each variable. For example, the left panels in
Fig.~\ref{fig:step3} show that when considering the whole event sample (red histograms), $T_2$ appears to be a good variable, 
since as many as 79\% of the events have positive values of $\Delta T_2 (P_W, P_C)$ \cite{Choi:2011ys}. 
Unfortunately, we find that this conclusion is invalidated after the application of Step I --- for the remaining unresolved events after Step I 
(the blue histograms in Fig.~\ref{fig:step3}), it is actually more likely to find a negative value of $\Delta T_2 (P_W, P_C)$ instead, 
thus obtaining the wrong answer for $P_C$. This observation reveals that the results from Step I and Step III are also somewhat correlated --- 
the events which are easy to analyze with $T_2$ at Step III would already be correctly resolved at Step I. 
Using $\Delta T_2$ alone (without $\Delta T_3$ or $\Delta T_4$) in fact lowers the cumulative efficiency, 
as shown in Table~\ref{table:efficiency}. This motivates us to drop $\Delta T_2$ from further consideration and 
repeat Step III {\it without} $\Delta T_2$, i.e., with only $\Delta T_3$ and $\Delta T_4$. As shown in Table~\ref{table:efficiency},
this leads to a slight improvement of the cumulative efficiency, to 88.7\% and 89.5\%, respectively, indicating that 
$T_3$ and $T_4$ retain some discrimination power even after Step I (this can also be deduced from the blue histograms in the 
middle and right panels of Fig.~\ref{fig:step3}).


Given the large variety of MAOS and CMAOS methods described in section~\ref{sec:momentum} (see Table~\ref{tab:methods}), next we 
check if there exist alternative versions of the $T_3$ and $T_4$ variables which are better suited for our purpose. 
In the remainder of Table~\ref{table:efficiency} we study the effect on the cumulative efficiency if the reconstruction is performed
with one of the eight yellow-shaded methods from Table~\ref{tab:methods}. Taking one variable at a time, we apply Step III
as in (\ref{pick}) and quote the resulting cumulative efficiency in the last 8 rows of Table~\ref{table:efficiency}.  
The results are rather illuminating, and indicate that ``not mixing" the subsystems provides the best option for constructing a useful $T_i$ variable.
For example, when the transverse invisible components are fixed with the help of the $(b\ell)$ subsystem, in which the top quark is the parent particle, 
then we are better off applying the on-shell condition on the {\em same} particle in order to reconstruct the longitudinal invisible momenta.
Similarly, if the transverse momenta are obtained from the $(\ell)$ subsystem, in which the $W$-boson is the parent particle, 
then it is preferable to apply the on-shell condition on the $W$-boson as well. For both MAOS and CMAOS reconstructions, 
Table~\ref{table:efficiency} shows that the highest efficiencies are obtained in the case of $(b\ell, m_t)$ and $(\ell, m_W)$ type variables.
It is interesting to note that the performance of the modified method above (which used only $\Delta T_3$ and $\Delta T_4$) can be matched
and even slightly exceeded by using a single additional variable at Step III, provided that we pick the right one:
the MAOS1($b\ell$;$m_t$) option gives 88.8\% (compare to 88.7\% before),
while CMAOS1($b\ell$;$m_t$) yields 89.6\% (compare to 89.5\%).

Another interesting point is that in all cases, the cumulative efficiencies in the right column of Table~\ref{table:efficiency}
are higher than those in the middle column, thus reinforcing the idea of using the constrained $M_{2CC}$ variables. 
In section~\ref{sec:step1} we already established that it is beneficial to redefine Step I in terms of $M_{2CC}$ variables, 
and now we see that this advantage is retained after Step III as well. 

Finally, one may wonder if there is any additional benefit in {\em combining} at Step III two or more of the variables 
considered in Table~\ref{table:efficiency}. We test this idea with the following exercise. Let us revisit Step III, 
again dropping the $T_2$ variable (\ref{T2def}) from consideration, while for the definition of $T_3$ and $T_4$ 
let us choose the two best performing CMAOS variables from the right column of Table~\ref{table:efficiency}, namely 
$m_t^{reco}$ with CMAOS1($\ell$;$m_W$) (87.9\%) and $m_W^{reco}$ with CMAOS1($b\ell$;$m_t$) (89.6\%).  
In this modified scheme, we obtain a cumulative efficiency of $\varepsilon=89.2\%$ with $N_C=2,483$ correctly identified, 
$N_W=1,023$ wrongly identified and $N_U=1,427$ unresolved events. 
Similarly, choosing the two best MAOS options in the middle column of Table~\ref{table:efficiency},
namely $m_t^{reco}$ with MAOS1($\ell$;$m_W$) (87.4\%) and $m_W^{reco}$ with MAOS1($b\ell$;$m_t$) (88.8\%),
we find a final efficiency of  $\varepsilon=88.3\%$ (with $N_C=3,493$ correctly identified, 
$N_W=1,176$ wrongly identified and $N_U=1,761$ unresolved events). 
In both exercises, the final efficiency is slightly worse than what would be obtained with the single best variable alone, 
although the number of unresolved events decreased. These two exercises indicate that there exist non-trivial correlations 
between the different variables and an improvement of the efficiency is not guaranteed by simply merging or combining different methods. 
This is one of the reasons why we kept the results for the different methods separate in Table \ref{table:efficiency}.

Up to this point, in Steps II and III we have used reconstruction methods which require the knowledge of 
a particle mass ($m_t$ or $m_W$). However, when the method is being applied in studies of new physics, 
such information may not be immediately available. Therefore, it is prudent to consider modifications of Steps II and III, 
where one uses methods from Table~\ref{tab:methods} which do not rely on any mass information. 
\begin{table}[t]
\centering
\def\arraystretch{1.2}
\scalebox{0.92}{
\begin{tabular}{||c||c|c|c||c|c|c||}
\hline
                 & \multicolumn{6}{c||}{Algorithm for Step I} \\ \cline{2-7}  
               & \multicolumn{3}{c||}{$M_{T2}^{(b\ell)}$ and $m_{b\ell}$} & \multicolumn{3}{c||}{$M_{2CC}^{(b\ell)}$ and $m_{b\ell}$}   \\ \cline{2-7}  
                & Correct  & Wrong & Unresolved & Correct  & Wrong & Unresolved \\ \hline
Step I: Quadrant counts& 11,920  & 106 & 6,430 & 13,274  & 249 & 4,933  \\ \cline{2-7} 
efficiency  & \multicolumn{3}{c||}{$\varepsilon=82.0\%$} & \multicolumn{3}{c||}{$\varepsilon=85.3\%$} \\ \hline\hline
Remaining unresolved events &  \multicolumn{3}{c||}{$6,430$} & \multicolumn{3}{c||}{$4,933$}  \\ \hline\hline
\cellcolor{orange}$m_t^{reco}$ with MAOS3($b\ell$;$b\ell$)   &  3,527  &   2,903  &   -  &   2,980 & 1,953    &  -  \\  \cline{2-7}  
cumulative efficiency    &  \multicolumn{3}{c||}{$\varepsilon=83.7\%$} & \multicolumn{3}{c||}{$\varepsilon=88\%$}\\ \hline
\cellcolor{orange}$m_t^{reco}$ with CMAOS3($b\ell$;$b\ell$)   &  3,657  &   2,773  &   -  &   2,914 & 2,019    &  -  \\  \cline{2-7}  
cumulative efficiency  &  \multicolumn{3}{c||}{$\varepsilon=84.4\%$} & \multicolumn{3}{c||}{$\varepsilon=87.7\%$}\\ \hline
\cellcolor{orange}$m_t^{reco}$ with M${}_2$A($b\ell$) &  3,495  &   2,935  &   -  &      2,759 & 2,174    &  -  \\  \cline{2-7}  
cumulative efficiency  &  \multicolumn{3}{c||}{$\varepsilon=83.5\%$} & \multicolumn{3}{c||}{$\varepsilon=86.9\%$}\\ \hline
\cellcolor{orange}$m_t^{reco}$ with M${}_2$A($\ell$) &  3,719  &   2,711  &   - &   2,699 & 2,234    &  -  \\  \cline{2-7}  
cumulative efficiency &  \multicolumn{3}{c||}{$\varepsilon=84.7\%$} & \multicolumn{3}{c||}{$\varepsilon=88.6\%$}\\ \hline\hline
\cellcolor{orange}$m_W^{reco}$ with MAOS3($b\ell$;$b\ell$)  &  3,877  &   2,553  &   -  &   2,783 & 2,150    &  -  \\  \cline{2-7}  
cumulative efficiency &  \multicolumn{3}{c||}{$\varepsilon=85.6\%$} & \multicolumn{3}{c||}{$\varepsilon=87\%$}\\ \hline
\cellcolor{orange}$m_W^{reco}$ with CMAOS3($b\ell$;$b\ell$)  &  3,628  &   2,802  &   -  &   2,622 & 2,311    &  -  \\  \cline{2-7}  
cumulative efficiency  &  \multicolumn{3}{c||}{$\varepsilon=84.2\%$} & \multicolumn{3}{c||}{$\varepsilon=86.1\%$}\\ \hline
\cellcolor{orange}$m_W^{reco}$ with M${}_2$A($b\ell$)   &  3,769  &   2,661  &   -  &   2,658 & 2,275    &  -  \\  \cline{2-7}  
cumulative efficiency  &  \multicolumn{3}{c||}{$\varepsilon=85\%$} & \multicolumn{3}{c||}{$\varepsilon=86.3\%$}\\ \hline
\cellcolor{orange}$m_W^{reco}$ with M${}_2$A($\ell$)   &  3,448  &  2,982  &   - &   2,529 & 2,404    &  -  \\  \cline{2-7}  
cumulative efficiency  &  \multicolumn{3}{c||}{$\varepsilon=83.3 \%$} & \multicolumn{3}{c||}{$\varepsilon=85.6\%$}\\ \hline
\hline
\end{tabular}
}
\caption{\label{table:efficiencyM2CC} 
The same as Table~\ref{table:efficiency}, but for a few representative methods from Table~\ref{tab:methods} which do not use mass information.
} 
\end{table}
To be specific, we focus on the four methods listed in the orange-shaded cells of Table~\ref{tab:methods}
and show the corresponding results in Table~\ref{table:efficiencyM2CC}, which is the direct analogue of Table~\ref{table:efficiency}.
Once again, it turns out that Step II is unnecessary, albeit for a different reason --- this time in all cases the solutions for the 
invisible momenta are found to be real and therefore invisible momentum reconstruction is always possible for both partitions.
It also follows that there will be no unresolved events after Step III, since the relevant kinematic variables can always be computed 
and compared for the two partitions. The results in Table~\ref{table:efficiencyM2CC} help identify the most promising 
variables for Step III --- $m_W^{reco}$ with MAOS3($b\ell$;$b\ell$) for the middle column ($\varepsilon=85.6\%$) and
$m_t^{reco}$ with M${}_2$A($\ell$) for the right column ($\varepsilon=88.6\%$). In both cases, the use of a single variable 
at Step III leads to an improvement in the efficiency found after Step I of more than 3\%.

In conclusion of this section, we summarize our main findings from the review of the method of Refs.~\cite{Baringer:2011nh,Choi:2011ys}.
\begin{enumerate}
\item The efficiency after Step I is increased if we define the quadrants of Fig.~\ref{fig:quadrant} 
in terms of $M_{2CC}^{(b\ell)}$ instead of $M_{T2}^{(b\ell)}$. 
\item Step II does not lead to any appreciable effect after Step I, and can be safely omitted from the algorithm.
\item The use of the variable $T_2$ in Step III is counterproductive, thus $T_2$ can be safely dropped from consideration.
\item The use of a single optimal variable at Step III (as opposed to a combination of variables) 
is generally sufficient to produce the desired result.
\end{enumerate}

\section{A few ideas for further improvement}
\label{sec:newmethod}

In the previous section, we considered the partitioning method as defined in Refs.~\cite{Baringer:2011nh,Choi:2011ys}
and found that with a few slight tweaks the efficiency (\ref{efficiency}) can reach over 89\% (88\%)
with (without) mass information. In this section we shall consider a few more serious departures from
the original algorithm, which can potentially further increase the efficiency. Some of the changes are simply quantitative 
(as in section~\ref{sec:quadrants}, where we increase the number of variables used at Step I), others are qualitative 
(as in sections~\ref{sec:M2Ct} and \ref{sec:smintheta}).

\subsection{Generalizing the quadrant counts}
\label{sec:quadrants}

Recall that the main idea at Step I was to use two kinematic variables, in this case $m_{b\ell}$ and $M_{T2}^{(b\ell)}$, 
which have clear kinematic endpoints for the case of the correct partition $P_C$. The resulting efficiency after Step I was 
82\%, and when we replaced $M_{T2}^{(b\ell)}$ with $M_{2CC}^{(b\ell)}$, the efficiency increased to 85.3\%. 
But one should be able to do even better at Step I. The main point is that in the event topology of Fig.~\ref{fig:decaysubsystem}
there are not two, but {\em three} independent kinematic endpoints (they allow for a complete measurement of the
mass spectrum \cite{Burns:2008va,Cho:2014yma}). Therefore, one can expect that the addition of a third variable
at Step I, i.e., generalizing the plane of Fig.~\ref{fig:quadrant} to a three-dimensional parameter space divided into eight octants,
would further improve the performance of Step I.

\begin{figure}[t]
\centering
\includegraphics[width=7cm]{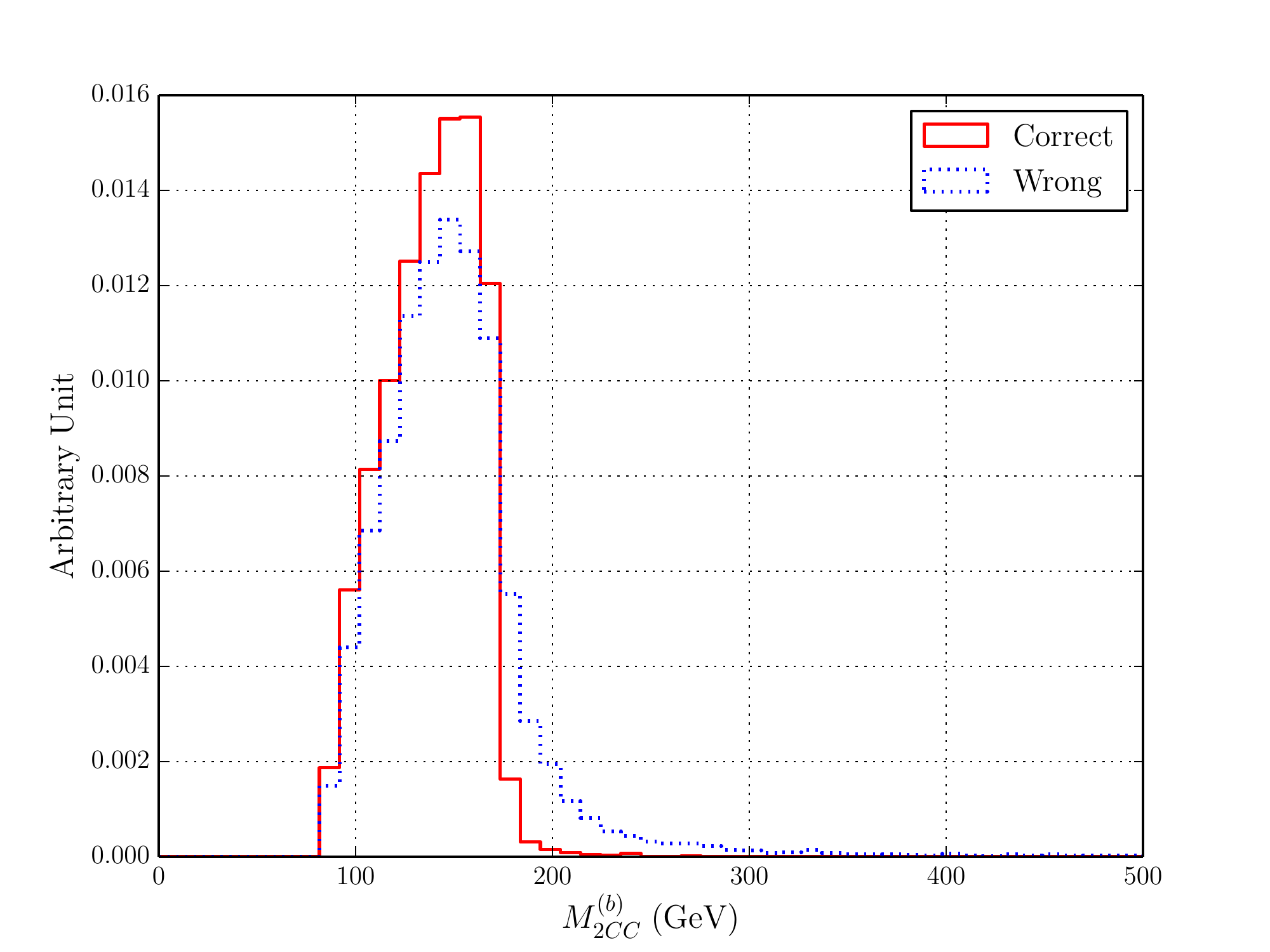}  \hspace*{-0.1cm}
\includegraphics[width=7cm]{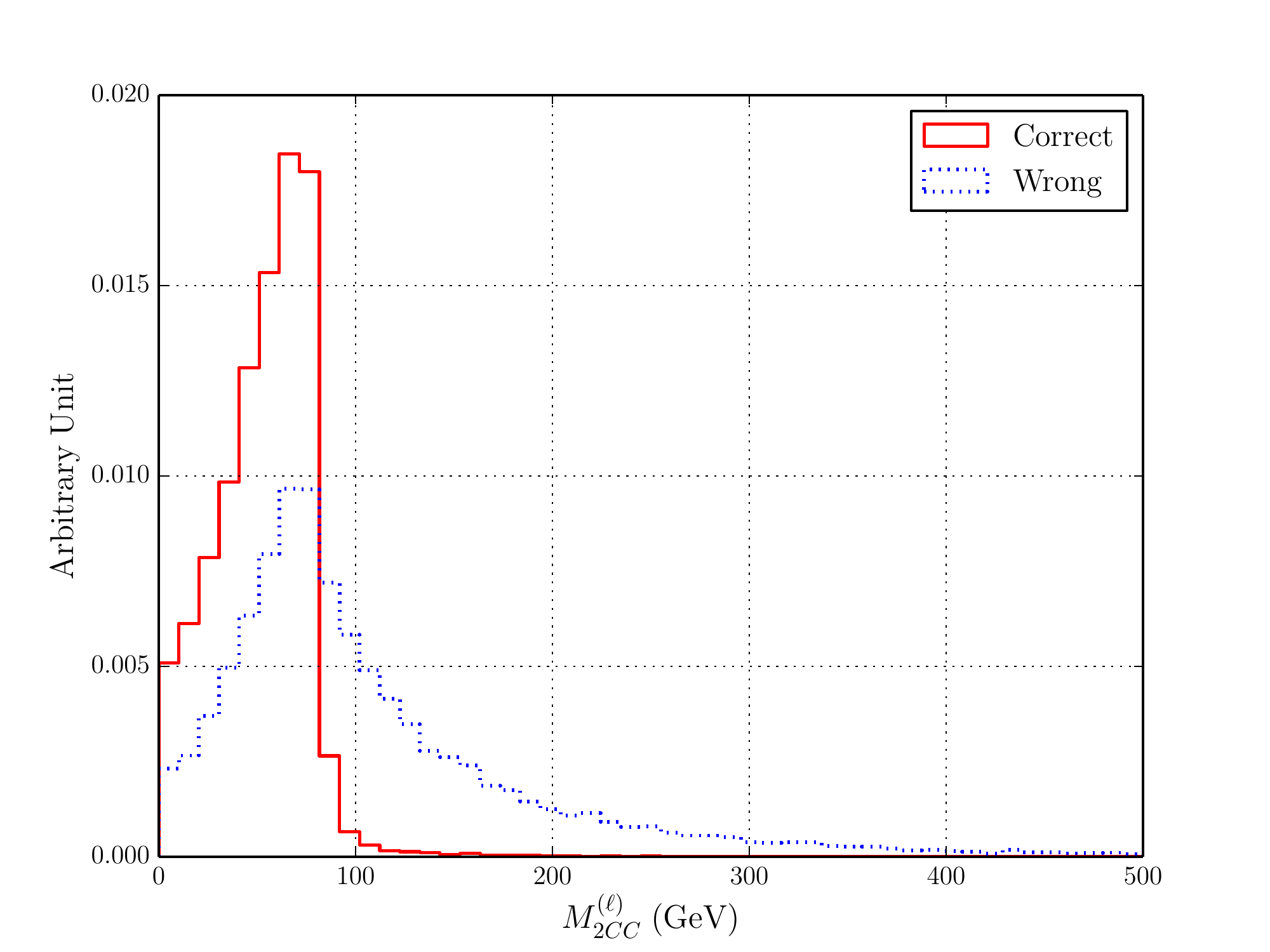}\hspace*{-0.1cm}  \\
\includegraphics[width=7cm]{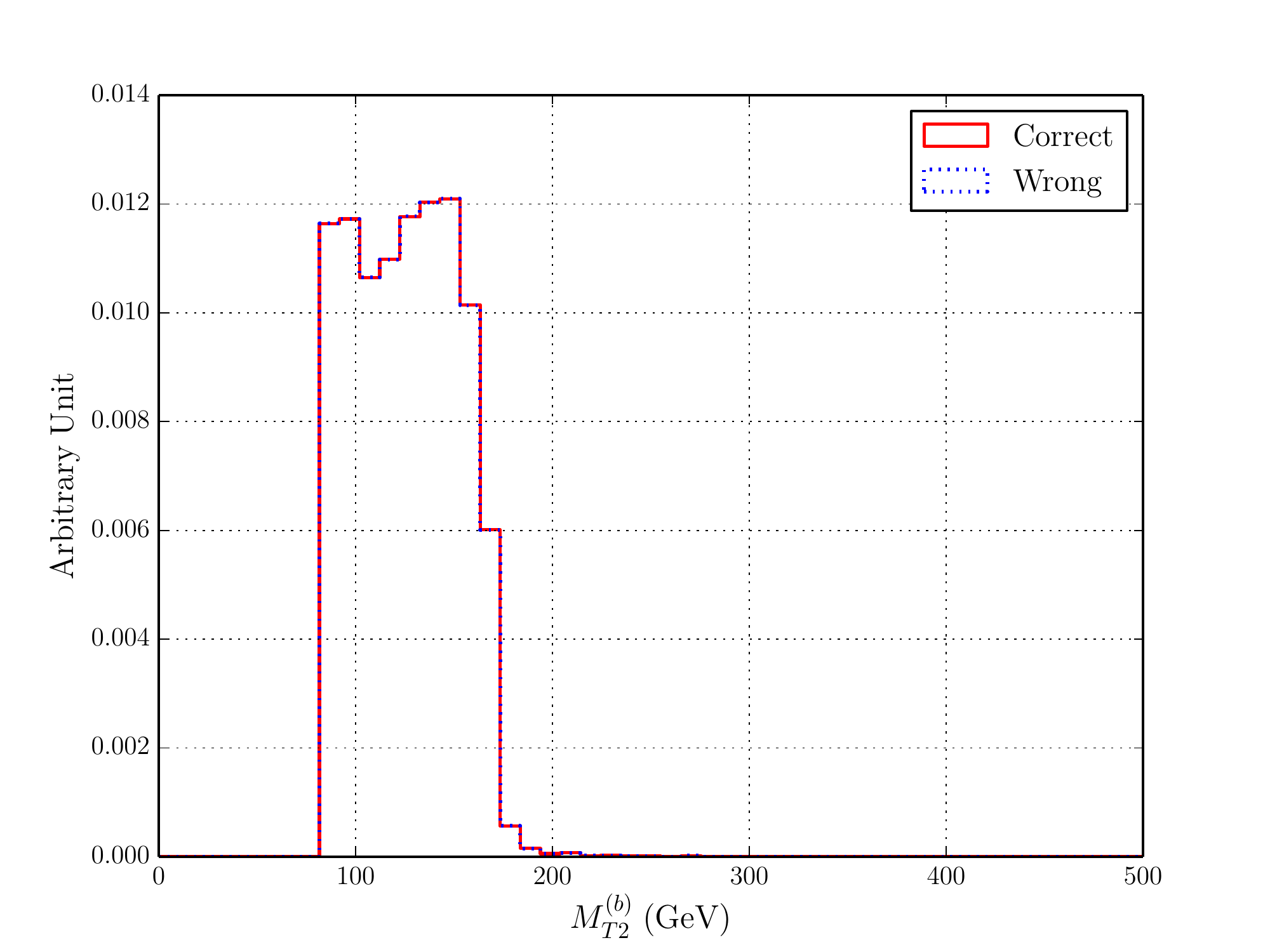}  \hspace*{-0.1cm}
\includegraphics[width=7cm]{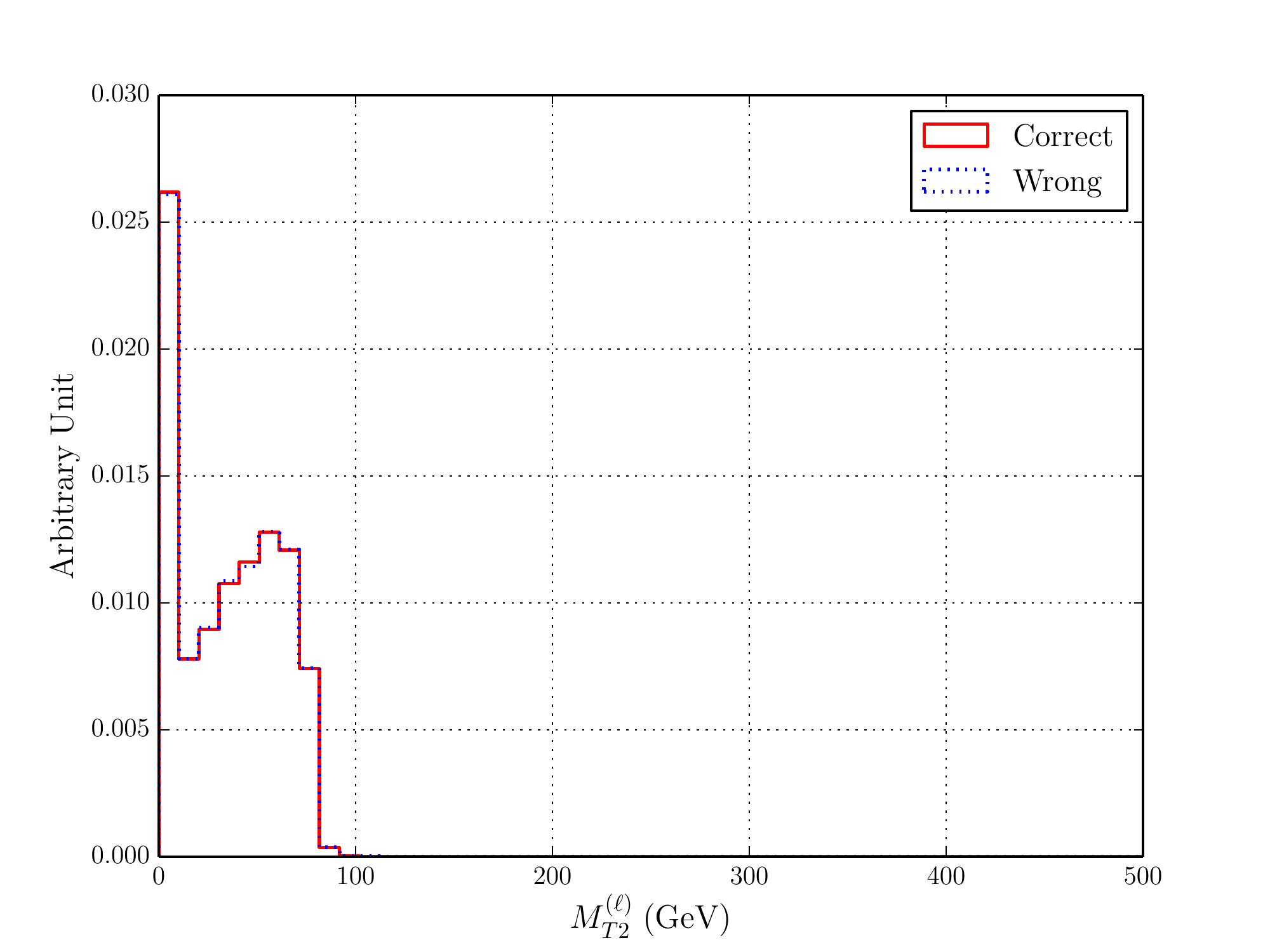}\hspace*{-0.1cm} 
\caption{\label{fig:subsystem} Distributions of $M_{2CC}$ (top row) and $M_{T2}$ (bottom row) for the case of subsystem $(b)$ (left panels) 
or subsystem $(\ell)$ (right panels). The result for the correct (wrong) partition is shown in red (blue). 
}
\end{figure}

In order to test this idea, we need to pick a suitable third variable to go along with our original two. We focus on the 
$M_{T2}$ and $M_{2CC}$ variables in the remaining two subsystems, $(b)$ and $(\ell)$, and show
their distributions in Fig.~\ref{fig:subsystem}.
The plots in the bottom row indicate that the variables $M_{T2}^{(b)}$ and $M_{T2}^{(\ell)}$ 
are not suitable for our purpose since they do not depend on the chosen partition --- 
the distributions for $P_C$ (red solid lines)
exactly coincide with the distributions for $P_W$ (blue dotted lines).
On the other hand, the distributions of $M_{2CC}^{(b)}$ (top left panel) and 
$M_{2CC}^{(\ell)}$ (top right panel) by construction depend on the partition via the mass shell constraint for the relative particle.
The differences are more pronounced in the case of $M_{2CC}^{(\ell)}$, which we shall choose as our third variable to go along
with the previous two, $m_{b\ell}$ and $M_{2CC}^{(b\ell)}$.

We can now generalize our previous discussion of the quadrant counts in Fig.~\ref{fig:quadrant} by 
populating our events in the three-dimensional space 
\beq
\left(x,y,z\right) \equiv
\left(   \,
m_{b\ell}^{max}-\max_j\{m^{(j)}_{b\ell}\}, \,
m_t - M_{2CC}^{(b\ell)}, \,
m_W - M_{2CC}^{(\ell)} \,
\right).
\label{setofthree}
\eeq 
As before, we expect that for the correct partition $P_C$, the events will be populating predominantly the first octant, where
$(\sign(x), \sign(y), \sign(z))=(+,+,+)$, while for the wrong partition $P_W$, the events will be more randomly distributed 
throughout the eight octants. These expectations are tested in Table \ref{table:mbl_m2cc_bl_l},
which generalizes the right table in Table~\ref{table:mt2_m2cc_mbl_cuts} by
additionally incorporating the subsystem variable $M_{2CC}^{(\ell)}$.
\begin{table}[t]
\centering
\scalebox{0.85}{
\begin{tabular}{||c||c|c|c|c|c|c|c|c||}
\hline
C$\, \backslash$W       & $(+,+,+)$ & $(+,+,-)$ & $(+,-,+)$ & $(-,+,+)$ & $(+,-,-)$ & $(-,+,-)$ & $(-,-,+)$ & $(-,-,-)$ \\ \hline \hline
$(+,+,+)$ &  3,697    &  621      &  615      &  216     & 1,581    &    0      &  3,325   &   6,611        \\ \hline
$(+,+,-)$  & 108        &   68       &   8         &    1        &   113     &    0      &    76      &   210        \\ \hline 
$(+,-,+)$  &    70       &   21       &   40       &     3       &    80      &    0      &     90     &    243       \\ \hline
$(-,+,+)$  &    17       &  0          &     1       &     2       &     2       &   0       &     14     &    24       \\ \hline
$(+,-,-)$   &    52       &  25         &   12      &      2      &   46       &   0       &    38     &    109       \\ \hline
$(-,+,-)$   &      0       &      0       &        0   &        0    &        0    &       0   &     0      &       0    \\ \hline
$(-,-,+)$   &    10       &    5         &    6       &    0        &     7      &     0      &    48     &      55     \\ \hline
$(-,-,-)$    &   19        &    3         &    8       &     1      &     5        &   0       &   46      &    102         \\ \hline
\end{tabular}
}
\caption{\label{table:mbl_m2cc_bl_l} 
Octant counts utilizing the set of three variables (\ref{setofthree}), where
rows (columns) are labelled by the sign signature $(\sign(x), \sign(y), \sign(z))$ for the correct (wrong) partition.} 
\end{table}
In Table~\ref{table:mbl_m2cc_bl_l}, each of the eight octants of the parameter space (\ref{setofthree})
is labelled by its corresponding sign signature $(\sign(x), \sign(y), \sign(z))$, and 
rows (columns) correspond to the correct (wrong) partition. Just like Table~\ref{table:mt2_m2cc_mbl_cuts} earlier,
Table~\ref{table:mbl_m2cc_bl_l} reveals that the endpoint violations are more likely to occur 
in the case of the wrong partition. We note that the entries in the $(-,+,-)$ row and column 
are identically zero --- this is due to the fact that the values of $m_{b\ell}^{(j)}$ are direct input to the calculation of
$M_{2CC}^{(b\ell)}$, so that once the $m_{b\ell}$ endpoint is violated in both decay chains ($j=1$ and $j=2$), 
it is very difficult to satisfy the endpoint of $M_{2CC}^{(b\ell)}$, unless the invisible momenta are rather soft 
(in the $b\ell$ rest frame). Note that under those circumstances the $M_{2CC}^{(\ell)}$ endpoint would be satisfied 
as well, which explains the nonzero entries in the $(-,+,+)$ row and column.

\begin{table}[t]
\centering
\scalebox{1}{
\begin{tabular}{||c||c|c|c|c||}
\hline
C$\, \backslash$W  & 0        &     1    &    2     &     3        \\ \hline \hline
0       & 3,697 & \cellcolor{green}1,452 & \cellcolor{green}4,906 & \cellcolor{green}6,611    \\ \hline 
1       & \cellcolor{red}195    & 144    &  \cellcolor{green}375   & \cellcolor{green}477   \\ \hline 
2       & \cellcolor{red}62      & \cellcolor{red}50      &  139   & \cellcolor{green}164   \\ \hline 
3       & \cellcolor{red}19      & \cellcolor{red}12      & \cellcolor{red}51      & 102    \\ \hline  
\multicolumn{2}{c}{} 
\end{tabular}
%
}
\caption{\label{table:mbl_m2cc_bl_b_l2} 
Event counts summarizing the number of endpoint violations in Table~\ref{table:mbl_m2cc_bl_l}.
Unresolved events are in the diagonal entries, while events above (below) the diagonal 
are correctly (wrongly) resolved. The resulting final efficiency after Step I is 86.8\%.
} 
\end{table}

We are now in position to evaluate the effect of Step I in the presence of the additional variable --- we simply
compare the number of endpoint violations (0, 1, 2 or 3) for each partition $P_k$, and declare the 
winner $P_C$ to be the case with fewer endpoint violations, as illustrated in Table~\ref{table:mbl_m2cc_bl_b_l2}.
The events for which both partitions give the same number of endpoint violations, remain unresolved ---
adding up the diagonal (unshaded) cells of Table~\ref{table:mbl_m2cc_bl_b_l2}, we find 
$N_U=4,082$. The events above (below) the diagonal cells are correctly (wrongly) resolved, 
giving $N_C=13,985$ and $N_W=389$, for a final efficiency of 86.8\%.\footnote{We have checked 
that adding a fourth variable, $M_{2CC}^{(b)}$, and repeating the procedure in four dimensions, 
leads to a very marginal improvement: $N_U=3,962$, $N_C=14,076$ and $N_W=418$, and efficiency of 87\%.}
This result should be contrasted with the 85.3\% efficiency found in the right column of Table~\ref{table:efficiency}.
Thus the benefit from adding a third variable to Step I can be quantified as an extra 1.5\% in the efficiency.


 \subsection{Utilizing the variables $M_{2CW}^{(b\ell)}$ and $M_{2Ct}^{(\ell)}$}
 \label{sec:M2Ct}

In this section we shall have in mind situations where full mass information is available, such as polarization 
studies in dilepton top events. Given the known values of $m_t$ and $m_W$, the relevant question is
whether they have been optimally utilized in the algorithm. With the MAOS method, this mass information
is not used at all during the reconstruction of the {\em transverse} invisible components.
One way to incorporate the mass information from the very beginning is to 
consider the additionally constrained variables $M_{2CW}^{(b\ell)}$ and $M_{2Ct}^{(\ell)}$
defined in (\ref{eq:m2CWdef}) and (\ref{eq:m2Ctdef}), respectively \cite{Kim:2017awi}.
We have already seen that those variables show the best performance in terms of 
reconstructing the neutrino momenta (see Figs.~\ref{fig:momentum1d} and \ref{fig:momentum2dmass}).
%
\begin{table}[t]
\centering
\scalebox{0.85}{
\begin{tabular}{||c||c|c|c|c|c|c|c|c||}
\hline
C$\, \backslash$W       & $(+,+,+)$ & $(+,+,-)$ & $(+,-,+)$ & $(-,+,+)$ & $(+,-,-)$ & $(-,+,-)$ & $(-,-,+)$ & $(-,-,-)$ \\ \hline \hline
$(+,+,+)$ &  2,388    &   500     &   353     &     0      & 2,526     &     0     &   187  &   8,662        \\ \hline
$(+,+,-)$  &    69      &      41    &     8       &     0       &  124      &    0      &      6     &    352        \\ \hline 
$(+,-,+)$  &    157     &    26      &     65     &     0       &    156     &    0      &     23     &   809        \\ \hline
$(-,+,+)$  &   0         &      0       &   0         &     0       &     0       &     0     &    0      &     0      \\ \hline
$(+,-,-)$   &    174     &     70      &    15     &     0       &     485    &     0     &   6      &   879        \\ \hline
$(-,+,-)$   &    0         &     0        &    0       &   0         &     0       &      0    &      0     &     0      \\ \hline
$(-,-,+)$   &   11        &      1       &    5       &    0        &     3       &     0      &     2    &    38       \\ \hline
$(-,-,-)$    &    21       &       2      &    6       &    0       &     34        &   0       &    1     &   251          \\ \hline
%
\end{tabular}
}
\caption{\label{table:mbl_m2ct_m2cw} The same as Table \ref{table:mbl_m2cc_bl_l}, but 
using the set of variables (\ref{setofthreeWt}).} 
\end{table}
%
\begin{table}[t]
\centering
\scalebox{1}{
\begin{tabular}{||c||c|c|c|c||}
\hline
C$\, \backslash$W  & 0        &     1    &    2     &     3        \\ \hline \hline
0       & 2,277 & \cellcolor{green}887 & \cellcolor{green}2,663 & \cellcolor{green}8,425    \\ \hline 
1       & \cellcolor{red}260    & 152    &  \cellcolor{green}357   & \cellcolor{green}1,376   \\ \hline 
2       & \cellcolor{red}191      & \cellcolor{red}95      &  508   & \cellcolor{green}949   \\ \hline 
3       & \cellcolor{red}21      & \cellcolor{red}8      & \cellcolor{red}35      & 252    \\ \hline  
\multicolumn{2}{c}{} 
\end{tabular}
}
\caption{\label{table:mbl_m2ct_m2cw2} The same as Table \ref{table:mbl_m2cc_bl_b_l2} but based on the 
results from Table~\ref{table:mbl_m2ct_m2cw}.
The final efficiency is 88.1\%.} 
\end{table}
%
This motivates us to perform Step I as in the previous section~\ref{sec:quadrants},
but in terms of the alternative parameter space
\beq
\left(x,y,z\right) \equiv
\left(   \,
m_{b\ell}^{max}-\max_j\{m^{(j)}_{b\ell}\}, \,
m_t - M_{2CW}^{(b\ell)}, \,
m_W - M_{2Ct}^{(\ell)} \,
\right)
\label{setofthreeWt}
\eeq 
instead of (\ref{setofthree}). The corresponding results are shown in Tables~\ref{table:mbl_m2ct_m2cw} and 
\ref{table:mbl_m2ct_m2cw2}. We use the {\sc OPTIMASS} package \cite{Cho:2015laa} to compute the values of $M_{2CW}^{(b\ell)}$
and $M_{2Ct}^{(\ell)}$. It should be noted that in certain cases {\sc OPTIMASS} is unable to find a viable solution, 
since all constraints cannot be simultaneously satisfied. 
This can happen, e.g., when we consider the wrong partition of an event, or if some particles are produced off-shell. 
For the purposes of tabulating the results in Tables~\ref{table:mbl_m2ct_m2cw} and \ref{table:mbl_m2ct_m2cw2},
such cases are assigned a ``minus" sign.

Table~\ref{table:mbl_m2ct_m2cw2} shows that when Step I is performed in terms of the alternative variables (\ref{setofthreeWt}), 
the efficiency of Step I alone is as high as 88.1\%. This is comparable to the results with several versions of the full algorithm 
(including Step III) which were considered in the previous section. Thus we conclude that in cases where the masses of the 
intermediate particles are known, it is best to perform Step I in terms of the parameters (\ref{setofthreeWt}) which use the variables
$M_{2CW}^{(b\ell)}$ and $M_{2Ct}^{(\ell)}$.

 \subsection{Using reconstructed event kinematics: the $\sqrt{\hat s}-\cos\theta$ method}
 \label{sec:smintheta}

In this section, we shall use the fact that once we choose an ansatz for the invisible momenta via one of the methods from
Table~\ref{tab:methods}, the full event kinematics is completely fixed as well. This means that when it comes to 
discriminating the unresolved events after Step I, we are not limited to only invariant mass variables, but we have the full 
set of kinematics tools at our disposal. In particular, we can study angular variables, as well as global inclusive variables, 
whose definition does not rely on partitioning the event. 

An example of the latter type of variables is the total invariant mass in the event, $\sqrt{\hat{s}}$, where 
\beq
\hat{s}\equiv \left( \sum_{j=1}^2 \left(p_{a_j}+p_{b_j}+ q_j\right)\right)^2.
\eeq
It has been shown that a preselection cut on $\sqrt{\hat{s}}$ improves the efficiency at the cost of lowering the statistics \cite{Choi:2011ys}.
This suggests that $\sqrt{\hat{s}}$ can potentially be a useful variable for categorizing the unresolved events after Step I.
\begin{figure}[t]
\centering
\includegraphics[width=7cm]{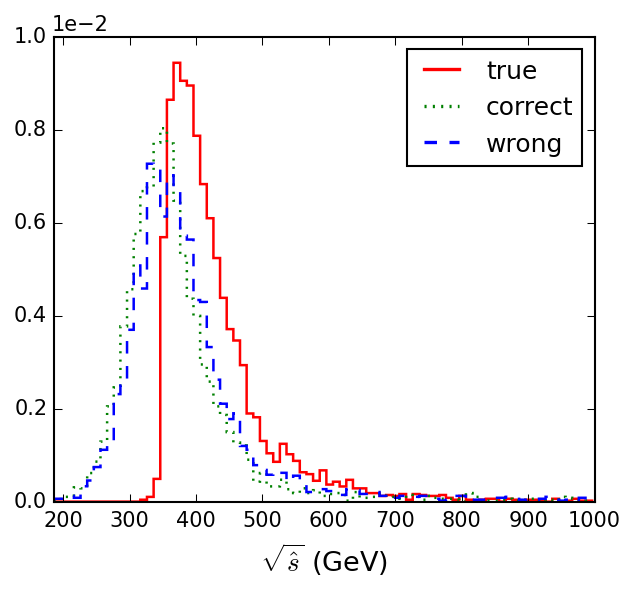} 
\includegraphics[width=7cm]{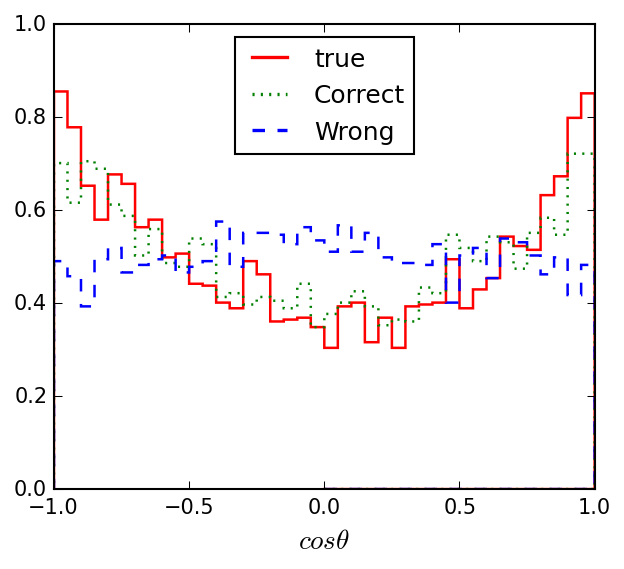} \\
\includegraphics[width=7cm]{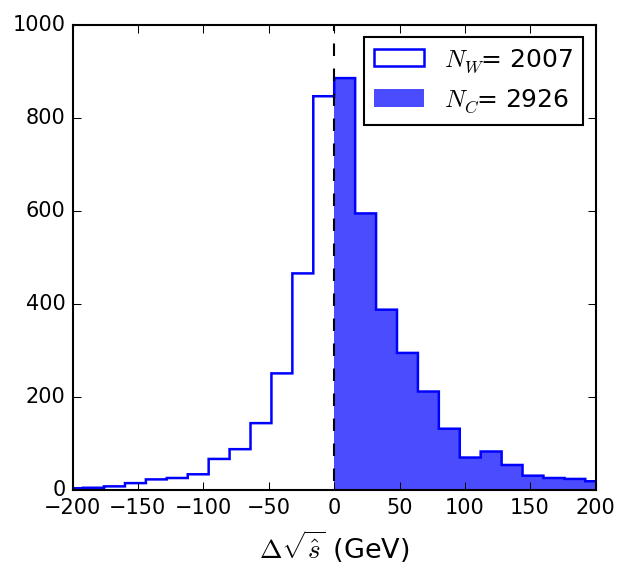} 
\includegraphics[width=7cm]{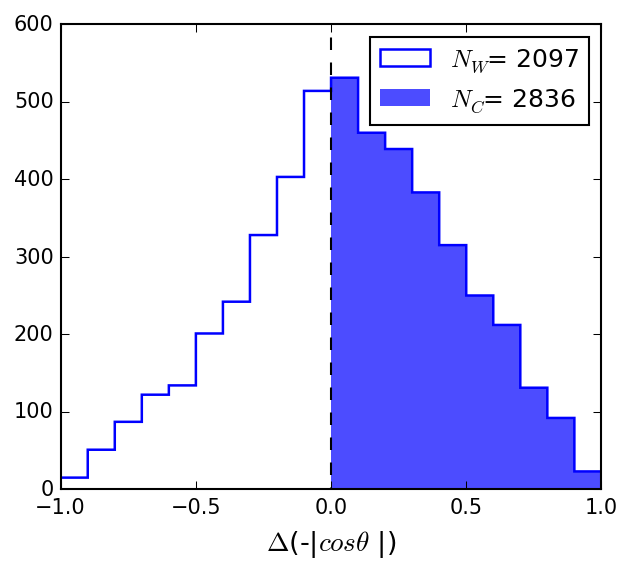} 
\caption{\label{fig:roots_costheta} Upper panels: unit-normalized distributions of the reconstructed variables
$\sqrt{\hat{s}}$ (left) and $\cos\theta$ (right) for the unresolved events after Step I.
The green dotted (blue dashed) lines correspond to results obtained with the correct (wrong) partition.
The solid red line shows the MC truth distribution. Bottom plots: distributions of the 
ordered differences (\ref{DeltaRootsordered}) (left panel) and 
(\ref{DeltaCosordered}) (right panel).
}
\end{figure}
The idea is tested in the left panels of Fig.~\ref{fig:roots_costheta}. The upper left panel shows $\sqrt{\hat{s}}$ distributions
obtained with the M${}_2$A($b\ell$) method from Table \ref{tab:methods}, for the $N_U=4,933$ unresolved events arising after Step I 
when it is done in terms of $M_{2CC}^{(b\ell)}$ and $m_{b\ell}$ (see the right columns in Tables~\ref{table:efficiency} and \ref{table:efficiencyM2CC}).
The green dotted (blue dashed) line shows the case of the correct (wrong) partition. For reference, the solid red line gives the true
$\sqrt{\hat{s}}$ distribution. As previously observed in Ref. \cite{Konar:2008ei}, the reconstructed $\sqrt{\hat{s}}$ distribution peaks at 
threshold ($2m_t$). We also notice that the distribution for the wrong partition $P_W$ is slightly harder, which suggests to 
investigate the ordered difference 
\beq
\Delta \sqrt{\hat{s}}(P_W, P_C) \equiv  \sqrt{\hat{s}}(P_W) - \sqrt{\hat{s}}(P_C)
\label{DeltaRootsordered}
\eeq 
in analogy to (\ref{DeltaTiordered}). The distribution of the variable (\ref{DeltaRootsordered}) is shown in the lower left panel of Fig.~\ref{fig:roots_costheta}.
If we attempt to resolve events by applying the condition (\ref{pick}) to the variable $\sqrt{\hat{s}}$, we find $N_C=2,926$ correctly resolved events 
(the shaded portion of the distribution in the lower left panel of Fig.~\ref{fig:roots_costheta})
and $N_W=2,007$ wrongly resolved events (the unshaded portion of the distribution).
The overall efficiency is then increased from 85.3\% after Step I to 87.8\% (see Table~\ref{tab:efficiency}).

An alternative handle to sort the unresolved events is provided by the angular distribution of the parent particles at production.
For concreteness, in the upper right panel of Fig.~\ref{fig:roots_costheta} we compare the distributions of the scattering angle $\theta$ 
of the top quarks in their center of mass frame, for the case of the correct partition (green dotted line), the wrong partition
(blue dashed line) and the MC truth (solid red line). We notice that in reality, the top quarks are produced predominantly 
in the forward direction, due to the presence of $t$-channel and $u$-channel diagrams. When the momenta are 
reconstructed using the correct partition, this tendency is retained, while the distribution obtained with the wrong partition 
is mostly flat in $\cos\theta$. This motivates us to consider the corresponding ordered difference
\beq
\Delta (-|\cos\theta|)(P_W, P_C) \equiv  (-|\cos\theta|)(P_W) - (-|\cos\theta|)(P_C)
\label{DeltaCosordered}
\eeq
whose distribution is shown in the lower right panel of Fig.~\ref{fig:roots_costheta}. 
As before, events with positive\footnote{This is why in the definition (\ref{DeltaCosordered}) we 
chose to consider the function $-|\cos\theta|$ instead of simply $|\cos\theta|$.} 
values of the ordered difference (\ref{DeltaCosordered}) will be correctly resolved, and
they represent the shaded portion of the distribution. As summarized in Table~\ref{tab:efficiency},
by applying the condition (\ref{pick}) to the $\cos\theta$ variable, we obtain $N_C=2,836$ ($N_W=2,097$) correctly (wrongly) 
resolved events and an overall efficiency of 87.3\%.
\begin{table}[t]
\centering
\scalebox{1.05}{
\begin{tabular}{||c||c|c|c|c||}
\hline
method & $N_C$ & $N_W$ & $N_U$  & $\varepsilon$\\ \hline
$\sqrt{\hat s}$ & 2926 & 2007 & - & 87.8 \%\\\hline
$-|\cos\theta|$ & 2836 & 2097 & - & 87.3 \%\\ \hline
${\cal P}(\sqrt{\hat s}, \cos\theta)$ & 3199 & 1720 & 14 & 89.2 \% \\ \hline
\end{tabular}
}
\caption{\label{tab:efficiency} Results from alternative methods for classifying the $N_U=4,933$ unresolved events 
remaining after Step I when done in terms of $M_{2CC}^{(b\ell)}$ and $m_{b\ell}$.
The two partitions are compared based on the resulting values of $\sqrt{\hat{s}}$, of
$(-|\cos\theta|)$, or probabilistically based on the templates in Fig.~\ref{fig:roots_costheta2d}.}
\end{table}

\begin{figure}[t]
\centering
\includegraphics[width=7cm]{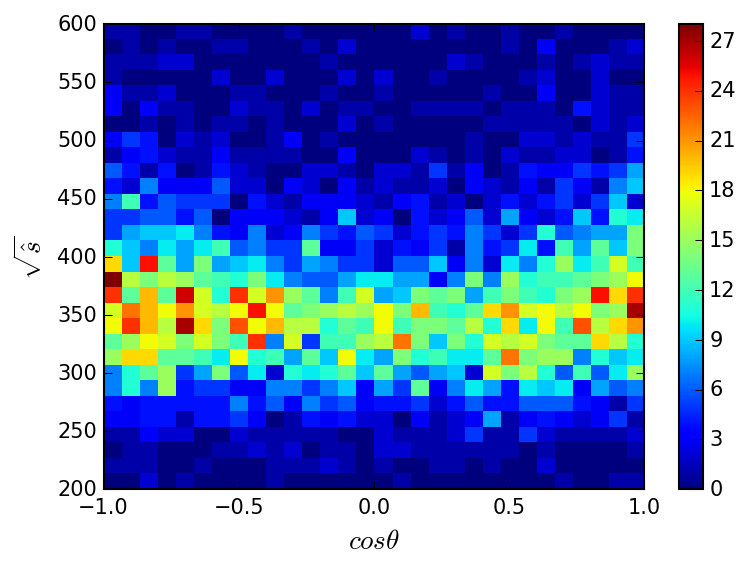}
 \includegraphics[width=7cm]{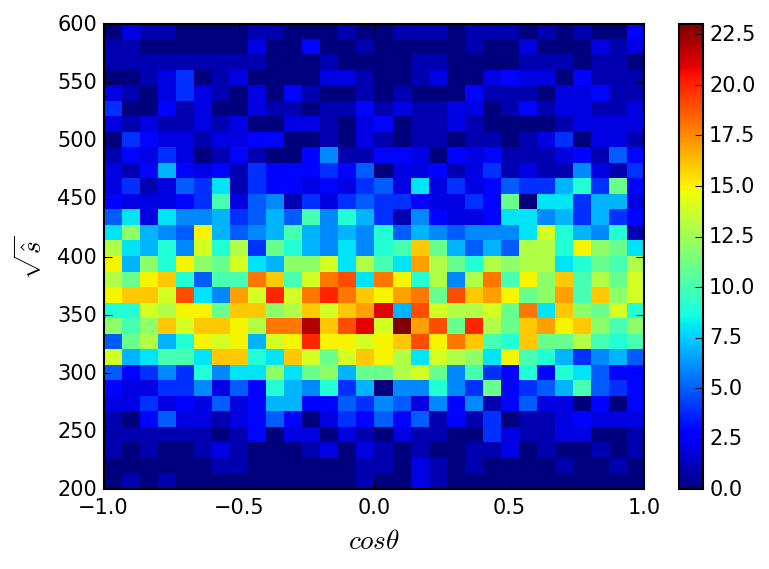}
\caption{\label{fig:roots_costheta2d} The two-dimensional templates ${\cal P}_C$ (left) and ${\cal P}_W$ (right)
in the $(\cos\theta,\sqrt{\hat{s}})$ plane.}
\end{figure}

Table~\ref{tab:efficiency} demonstrates that both variables $\cos\theta$ and $\sqrt{\hat{s}}$ are useful 
in categorizing the unresolved events, if applied separately. We now check whether they can be combined into a single method
leading to an even higher efficiency. Since it is not possible to derive analytical expressions for the expected 2D distribution in
$(\cos\theta,\sqrt{\hat{s}})$, we shall use a template method. Fig.~\ref{fig:roots_costheta2d} shows the relevant 
two-dimensional distributions using only unresolved events. 
In the left (right) plot the invisible momenta were reconstructed with the correct (wrong) partition. 
These two plots define two probability distributions, ${\cal P}_C$ and ${\cal P}_W$. 
Each of the two possible partitions $P_k$ in an event comes with its own values of $\cos\theta$ and $\sqrt{\hat{s}}$,
say $\cos\theta_k$ and $\sqrt{\hat{s}_k}$. Since we do not know whether $P_1$ corresponds to $P_C$ or $P_W$, we try it both ways, 
and select
\beq
P_C = \left\{ 
\begin{array}{l} 
P_1,\ \text{if}\ {\cal P}_C(P_1) {\cal P}_W(P_2)>{\cal P}_C(P_2) {\cal P}_W(P_1);  \\ [2mm]
P_2,\ \text{if}\ {\cal P}_C(P_2) {\cal P}_W(P_1)>{\cal P}_C(P_1) {\cal P}_W(P_2).
\end{array}
\right.
\label{P1P2}
\eeq
As shown in Table~\ref{tab:efficiency}, the prescription (\ref{P1P2}) results in 
$N_C=3,199$ correctly resolved events and $N_W=1,720$ wrongly resolved events,\footnote{The $N_U=14$ unresolved events seen in Table~\ref{tab:efficiency}
are due to our finite binning --- for those events, the two points $(\cos\theta_k,\sqrt{\hat{s}_k}), k=1,2,$ happened to fall within the same bin, which resulted in a tie. 
Since the templates of Fig.~\ref{fig:roots_costheta2d} are built from Monte Carlo, in principle one can use
more statistics for their generation, and correspondingly smaller bin sizes, which will make such ties increasingly rare.}
and the overall efficiency is increased to 89.2\%, which ranks among the best results we have found so far at the end of Step I.

\section{Finding the correct partition without any mass or endpoint information}
\label{sec:generalcase}

The potential improvements considered in the previous section relied on some prior knowledge about the 
masses of the particles involved in the decay chains. However, in new physics applications of our 
event topology from Fig.~\ref{fig:decaysubsystem}, such information may be difficult to get at first.
This is why in this section we shall consider scenarios where no mass information is available, i.e.,
the masses of the particles $A_i$, $B_i$ and $C_i$ are a priori unknown, and furthermore, 
kinematic endpoint measurements are also unavailable or have very large uncertainties to be useful.

Let us now revisit the method under those assumptions. Steps I and II do require mass information
(for defining the quadrants of Fig.~\ref{fig:quadrant} and for longitudinal momentum reconstruction, respectively)
and therefore cannot be used. Similarly, two of the variables from Step III, namely $\Delta T_3$ and $\Delta T_4$,
also need mass inputs for their calculation. For the moment, this leaves us with only $\Delta T_1$ and $\Delta T_2$ at our disposal.
Since for the correct partition the values of $T_1$ and $T_2$ are in principle limited from above by a kinematic endpoint, 
we can still expect (\ref{Tihierarchy}) to be mostly true. 
In fact, it is known that with the simple prescription (\ref{pick}) using $\Delta T_1$ ($\Delta T_2$) alone,
the efficiency is 80\% (79\%) over all events \cite{Choi:2011ys}.
As before, we can combine the two variables $\Delta T_1$ and $\Delta T_2$
and assign the correct partition $P_C$ to be the one chosen by both variables. In order 
to estimate the resulting efficiency, in the first row of Table~\ref{table:delta12} we list the number of events 
with a given sign signature $(\sign(\Delta T_1(P_W,P_C)), \sign(\Delta T_2(P_W,P_C)))$.
The events with signature $(+,+)$ (green-shaded cells) will be correctly identified,
the events with signature $(-,-)$ (red-shaded cells) will be wrongly identified, while the events with signatures
$(+, -)$ or $(-,+)$ (unshaded cells) will remain unresolved at this point. 
The resulting efficiency of this combined $\Delta T_1\oplus\Delta T_2$ method is 81.8\%. 

We emphasize that the $\Delta T_1\oplus\Delta T_2$ method does not use any mass information:
we simply compare the two possible values for $T_1(P_k)$ (as well as the two possible values for $T_2(P_k)$) 
for $k=1$ and $k=2$, and choose the larger (the smaller) to indicate the wrong (correct) partition.
However, a potential problem with the method is that the variables $T_1$ and $T_2$ were found to be 
correlated \cite{Choi:2011ys}. This motivates us to look for an alternative set of variables.
Since we saw previously that $M_{2CC}^{(b\ell)}$ is more efficient than $M_{T2}^{(b\ell)}$ 
(compare the middle and right panels in Fig.~\ref{fig:step1}), we can try to replace $T_2$ with 
\beq
T_5 (P_i) \equiv M_{2CC}^{(b\ell)} (P_i).
\label{T5def}
\eeq
The efficiency of the resulting $\Delta T_1\oplus\Delta T_5$ method is 83.6\%, as shown in the second row of Table~\ref{table:delta12}. 
One can go one step further and add a third variable to the mix, e.g., $M_{2CC}^{(\ell)}$, as was done in section~\ref{sec:quadrants}:
\beq
T_6 (P_i) \equiv M_{2CC}^{(\ell)} (P_i).
\eeq
The resulting combined method $\Delta T_1\oplus\Delta T_5\oplus \Delta T_6$ involves an odd number of variables, 
thus each event will be resolved based on the sign signature. This is illustrated in Table~\ref{table:delta156}, which does not contain any unshaded cells. 
The corresponding efficiency of the method is 85.1\%, which is a noticeable improvement over the results in Table~\ref{table:delta12}.

\begin{table}[t]
\centering
\scalebox{0.95}{
\begin{tabular}{||c||c|c|c|c||c||}
\hline
 & $(+,+)$	& $(+,-)$   &   $(-,+)$  & $(-,-)$ & efficiency    \\   \hline
$(\sign(\Delta T_1(P_W,P_C)), \sign(\Delta T_2(P_W,P_C)))$ & \cellcolor{green}14,477 & 693 & 559& \cellcolor{red} 2,727 & 81.8\%         \\ \hline 
$(\sign(\Delta T_1(P_W,P_C)), \sign(\Delta T_5(P_W,P_C)))$ & \cellcolor{green}14,677 & 493 & 1,003& \cellcolor{red}2,283 & 83.6\%         \\ \hline 
\end{tabular}
}
\caption{\label{table:delta12} 
Categorizing events by their sign signature for $(\Delta T_1,\Delta T_2)$ or $(\Delta T_1,\Delta T_5)$.} 
\end{table}


\begin{table}[t]
\centering
\scalebox{0.95}{
\begin{tabular}{||c|c|c|c|c|c|c|c||}
\hline
\multicolumn{8}{||c||}{$(\sign(\Delta T_1(P_W,P_C)), \sign(\Delta T_5(P_W,P_C)), \sign(\Delta T_6(P_W,P_C)))$} \\
\hline
 $(+ + +)$ &$(+ + -)$  &  $(+ - +)$ &  $(- + +)$ & $(+ - -)$ & $(- + -)$ & $(- - +)$ & $(- - -)$      \\  \hline
\cellcolor{green} 12,301 	 & \cellcolor{green}2,376     & \cellcolor{green}175	          &	\cellcolor{green}856	      &	\cellcolor{red} 318       &  \cellcolor{red}   147      &  \cellcolor{red}  1,017         &\cellcolor{red}  1,266\\ \hline 
\end{tabular}
}
\caption{\label{table:delta156} 
The same as Table~\ref{table:delta12}, but for the combined $\Delta T_1\oplus\Delta T_5\oplus \Delta T_6$ method. The resulting efficiency is 85.1\%.} 
\end{table}

\begin{figure}[t]
\centering
\includegraphics[width=5.1cm]{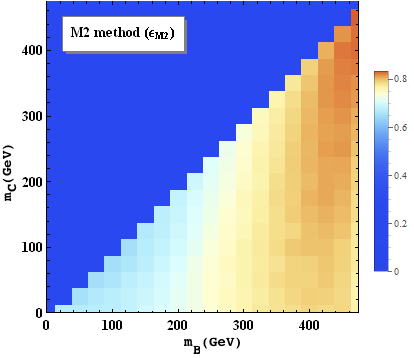}
\includegraphics[width=5.1cm]{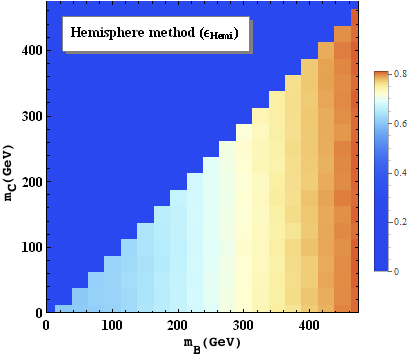} 
\includegraphics[width=5.1cm]{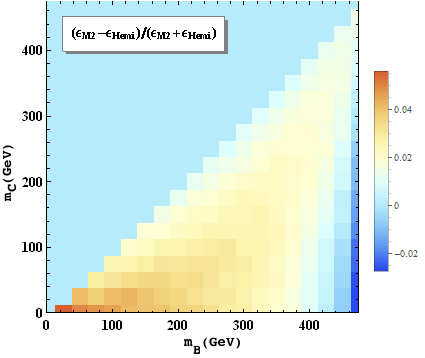}
\caption{\label{fig:methods} Efficiencies for choosing the correct partitioning 
for the $\Delta T_1\oplus\Delta T_5\oplus \Delta T_6$ method (left panel)
and the hemisphere (i.e., $\Delta T_1$) method (middle panel), as a function of the mass spectrum, for fixed $m_A=500$ GeV and for
$m_B>m_C$.
The right panel compares the efficiencies of the two methods. 
}
\end{figure}

Until now, we have tested the efficiencies of the methods with a SM sample of dilepton $t\bar{t}$ events,
i.e., the masses of the particles $A_i$, $B_i$ and $C_i$ were respectively the top mass $m_t$, 
the $W$-boson mass $m_W$, and the neutrino mass $m_\nu$. One may wonder how sensitive 
the results are to the choice of a benchmark study point. This issue is investigated in Fig.~\ref{fig:methods}, 
where we fix the mass of particle $A$ to $m_A=500$ GeV, and then freely vary the other two masses $m_B$ and $m_C$.
The left panel in Fig.~\ref{fig:methods} shows the efficiency of the $\Delta T_1\oplus\Delta T_5\oplus \Delta T_6$
method considered above. The efficiency varies noticeably throughout the mass parameter space, and seems to be correlated 
mostly with $m_B$ and less with $m_C$. The highest efficiency is obtained in the region where the spectrum becomes 
relatively degenerate --- in that case the visible decay products $a_i$ and $b_i$ are highly correlated 
with the direction of the parent particle $A_i$. 
 
For completeness, in Fig.~\ref{fig:methods} we also show results for the standard hemisphere method 
\cite{Ball:2007zza,Matsumoto:2006ws,Cho:2007dh,Nojiri:2008hy}
when applied to our event topology. In the hemisphere method, one clusters the visible particles
into two groups trying to keep the invariant mass of each cluster to a minimum. It is not difficult to see that 
in our language this is nothing but the $\Delta T_1$ method. The corresponding efficiency is shown in the
middle panel of Fig.~\ref{fig:methods} and it exhibits the same qualitative behavior. The right panel of Fig.~\ref{fig:methods}
compares the two methods by plotting the fractional difference of their efficiencies. We see that throughout 
most of the parameter space, the efficiency of the $\Delta T_1\oplus\Delta T_5\oplus \Delta T_6$ method 
is higher by 2-5\%.

\section{Summary and outlook}
\label{sec:conclusions}

The combinatorial problem is a very important issue in experimental particle physics. Identifying the correct
event topology on an event by event basis is a key element of many analyses which attempt to 
measure particle properties such as spin, couplings, CP quantum numbers, etc. A successful method 
which can avoid combinatorial ambiguities, especially in jetty events, is bound to improve the sensitivity 
of new physics searches as well.



In this paper, we revisited some of the existing methods \cite{Baringer:2011nh,Choi:2011ys} 
for resolving the combinatorial ambiguity in the dilepton $t\bar t$ event topology 
of Fig.~\ref{fig:decaysubsystem}. To summarize our main findings:
\begin{enumerate}
\item The efficiency after Step I can be increased if a) the quadrants of Fig.~\ref{fig:quadrant} 
are defined in terms of $M_{2CC}^{(b\ell)}$ instead of $M_{T2}^{(b\ell)}$ and b) if the quadrants are generalized to ``octants" 
as discussed in section~\ref{sec:quadrants}.
\item Step II does not lead to any appreciable effect after Step I, and can be safely omitted from the algorithm.
\item The use of the variable $T_2$ in Step III is counterproductive, and $T_2$ can also be dropped from consideration.
\item The use of a single optimal variable at Step III (as opposed to a combination of variables) 
is generally sufficient to produce the desired result.
\item The efficiency is also increased if the available mass information is incorporated as early as possible, e.g., 
by utilizing the variables $M_{2CW}^{(b\ell)}$ and $M_{2Ct}^{(\ell)}$, as discussed in section~\ref{sec:M2Ct}.
\item We investigated further improvements of the algorithm, by invoking other types of variables, including 
a global inclusive variable like $\sqrt{\hat{s}}$ and an angular variable like $\cos\theta$, see section~\ref{sec:smintheta}.
\item In section~\ref{sec:generalcase} we discussed a more general approach which does not rely on any mass information. 
\end{enumerate}
One should keep in mind that the efficiency can always be further improved at the cost of statistics. 
For instance, a cut on $\sqrt{\hat{s}}_{min}$ reduces the number of signal events, but the resulting efficiency 
can be increased beyond $90\%$ \cite{Choi:2011ys}.

Our results are directly applicable to any studies of final states containing $b\bar{b} W^+W^-$. 
In searches for new physics, dilepton $t\bar t$ would be the dominant background and our results should help 
in reducing it and increasing the sensitivity. In addition, there are several interesting physics scenarios where
a similar combinatorial problem plagues the signal itself:
\begin{itemize}
\item resonant di-Higgs production in the $b{\bar b}W^+W^-$ channel \cite{Huang:2017jws};
\item direct CP measurement of the Higgs-top coupling \cite{Buckley:2015vsa};
\item constraining new resonant physics with top spin polarization information \cite{Englert:2017uec};
\item triple Higgs boson production \cite{Papaefstathiou:2015paa};
\item multi-boson production processes such as $W^\pm W^\mp H$ or $W^\pm W^\mp H H$ \cite{Englert:2017gdy}; 
\item studies of anomalous triple and quartic gauge coupling such as $W^+ W^- \gamma$, $W^+ W^- Z$, $\gamma\gamma W^+ W^-$
and $\gamma Z W^+ W^-$ \cite{Kunkle:2015swy,Aaboud:2016ffv,Khachatryan:2014sta}.
\end{itemize}

Looking ahead, there are several directions in which the study presented here can be evolved.
\begin{enumerate}
\item Following the previous literature, in this initial investigation we considered a relatively simple
situation, where there were only two possible alternatives, and we had to pick one of them. 
As we increase the number of indistinguishable objects in the final state, things get much more complicated. 
We intend to tackle this more difficult problem in the very near future.
\item As the number of jets in the signature is increased, it becomes important to cross-check the 
parton-level results with more detailed simulations including detector effects, initial and final state radiation, etc.
\item The kinematic variables which we used here were designed for event topologies with 2 missing particles.
It would be interesting to generalize our analysis to event topologies with more than 2 missing particles, where one would
have to use a different set of $M_2$ variables suitably adapted for that case.
\end{enumerate}

\acknowledgments
This work is supported in part by a US Department of Energy grant DE-SC0010296 and DE-FG02-12ER41809. 
DD acknowledges support from the University of Florida Informatics Institute in the form of a
Graduate Student Fellowship. 
DK is supported by the Korean Research Foundation (KRF) through the CERN-Korea Fellowship program.
\dk{JHK is supported in part by the University of Kansas General Research Fund allocation 2302091.}

\bibliographystyle{JHEP}
\bibliography{draft}

\end{document}